\documentclass[fleqn,usenatbib]{mnras}

\usepackage{newtxtext,newtxmath}

\usepackage{figsize}
\usepackage{url}
\usepackage{graphicx}	
\usepackage{amsmath}	
\usepackage{longtable}
\usepackage{multirow}
\usepackage{supertabular,booktabs}
\usepackage{lscape}
\usepackage{rotate} 
\usepackage{hyperref}
\usepackage{float}

\usepackage{ulem}

\usepackage[T1]{fontenc}

\DeclareRobustCommand{\VAN}[3]{#2}
\let\VANthebibliography\thebibliography
\def\thebibliography{\DeclareRobustCommand{\VAN}[3]{##3}\VANthebibliography}

\title[The 2019/2021 X-ray monitoring campaigns]{The lively accretion disk in NGC 2992. II. The 2019/2021 X-ray monitoring campaigns.}

\author[R. Middei ]{R. Middei$^{1,2}$\thanks{E-mail: riccardo.middei@ssdc.asi.it}, A. Marinucci$^{3}$, V. Braito$^{4,5}$, S. Bianchi$^{6}$, B. De Marco$^{7}$, A. Luminari $^{1,8}$, 
\newauthor G. Matt$^{6}$, E. Nardini$^{9}$, M. Perri$^{1,2}$, J. N. Reeves$^{4,5}$, F. Vagnetti$^{8,10}$
	\\
	$^{1}$ INAF - Osservatorio Astronomico di Roma, Via Frascati 33, 00078, Monte Porzio Catone (Roma), Italy.\\
	$^{2}$ Space Science Data Center - ASI, Via del Politecnico s.n.c., 00133 Roma, Italy\\
	$^{3}$ ASI - Italian Space Agency, Via del Politecnico snc, 00133, Rome, Italy\\
    $^4$Center for Space Science and Technology, University of Maryland Baltimore County, 1000 Hilltop Circle, Baltimore, MD 21250, USA\\
    $^5$INAF - Osservatorio Astronomico di Brera, Via Bianchi 46 I-23807 Merate (LC), Italy\\
    $^6$Dipartimento di Matematica e Fisica, Universit\`a degli Studi Roma Tre, via della Vasca Navale 84, 00146 Roma, Italy\\
    $^7$Departament de F\`isica, EEBE, Universitat Politècnica de Catalunya, Av. Eduard Maristany 16, E-08019 Barcelona, Spain\\
    $^{8}$INAF - Istituto di Astrofisica e Planetologia Spaziali, Via del Fosso del Cavaliere, 00133, Roma, Italy\\
    $^{9}$INAF - Osservatorio Astrofisico di Arcetri, Largo Enrico Fermi 5, 50125 Firenze, Italy\\
$^{10}$Dipartimento di Fisica, Universit\`a degli Studi di Roma “Tor Vergata”, Via della Ricerca Scientifica 1, 00133,Roma, Italy \\
}

\date{Accepted XXX. Received YYY; in original form ZZZ}

\pubyear{2022}

\begin{document}
	\label{firstpage}
	\pagerange{\pageref{firstpage}--\pageref{lastpage}}
	\maketitle
	\begin{abstract}
		We report on the short and long term X-ray properties of the bright nearby Seyfert 2 galaxy NGC 2992, which was extensively observed with \textit{Swift}, \textit{XMM-Newton} and \textit{NuSTAR}. \textit{Swift} targeted the source more than 100 times between 2019 and 2021 in the context of two monitoring campaigns. Both time-averaged and time-resolved analyses are performed, and we find that the short-to-long term spectral properties of NGC 2992 are dominated by a highly variable nuclear continuum. The source varied in the 2-10 keV energy band from 0.6 to 12 $\times$ 10$^{-11}$ erg cm$^{-2}$ s$^{-1}$ during the two year long {\it Swift} monitoring. The fastest 2-10 keV flux change (by a factor of $\sim60\%$) occurred on a timescale of a few hours. The overall emission spectrum of the source is consistent with a power law-like continuum ($\Gamma=1.69\pm0.01$) absorbed by a constant line-of-sight column density N$_{\rm H}=(7.8\pm0.1)\times$ 10$^{21}$  $\rm cm^{-2}$. The reflected emission is likely due to matter with an average column density N$_{\rm H}=(9.6\pm2.7)\times$ 10$^{22}$  $\rm cm^{-2}$, thus NGC 2992 appears to have a globally Compton-thin circumnuclear medium. This scenario is fully supported by an independent analysis of the fractional variability and by {\it XMM-Newton} multi-year spectra.
	\end{abstract}
	
	
	
	
\section{Introduction}
\indent Active galactic nuclei (AGNs) are extragalactic sources that emit across the whole electromagnetic spectrum. Such systems are composite and each sub-structure has its own role in shaping the emerging spectrum  \cite[see][for a comprehensive review]{Padovani2017}. It is ubiquitously accepted that  the X-ray emission originates in the very inner regions of AGNs, near the central supermassive black hole (SMBH).  Accretion of matter infalling onto the SMBH is responsible for  the enormous amount of optical-UV photons, a fraction of which can be further energised via inverse-Compton \cite[][]{Sunyaev1980} off thermal electrons \cite[the so-called hot corona:][]{haar91,haar93,Zdziarski1995,Madejski1995} up to the X-rays. The maximum energy gain for these seed photons is mainly set by the hot plasma's temperature, and, to a lower extent, by its opacity \cite[e.g.][]{Rybi79,Belo99,Middei2019}. In fact, the X-ray continuum in AGNs is well modelled by a power law with a high energy roll-over \cite[e.g.][]{Perola2002,Dadina07,Molina09,Molina13,Mali14,Ricci18,Fabi15,Fabi17,Tortosa2018}. AGN X-ray spectra may show additional features due to reprocessing of the primary X-ray emission by the circumnuclear material. A fluorescence emission line from the Fe K-shell is commonly observed as the most prominent feature \cite[e.g.][]{Bian09} and its analysis carries a wealth of information on the physics of the reflecting material. This emission line has an intrinsically narrow profile that can undergo distortions, such as broadening, due to special and general relativistic effects. In particular, the closer to the SMBH the reflectors, the more distorted (i.e. the broader) the neutral or ionised Fe line profile \cite[e.g.][]{Fabian1995}. On the contrary, at larger distance, these effects are negligible, thus the Fe K$\alpha$ shape is consistent with a narrow profile. Additionally, in the case of Compton-thick reflectors (i.e. N$_{\rm H}\gtrsim1.5\times10^{24}$ cm$^{-2}$) the X-ray spectra show a typical emission excess around 30 keV, the so-called Compton-hump \cite[e.g.][]{Matt93}. The effect of any absorbing matter crossing our line of sight can significantly attenuate the observed number of photons, especially in the soft X-rays \cite[e.g.][]{Cappi1999,Awaki2000,Matt2002,Bian09,Middei2021}.\\
\indent The X-ray emission of AGNs is also well-known to be variable in spectral shape and amplitude. Nearby Seyfert galaxies as well as distant quasars show a typical softer-when brighter behaviour \cite[e.g.][]{Sobolewska2009,Serafinelli2017}, where softer spectral states, characterised by a photon index $\Gamma>2$, generally correspond to higher flux states. Variability is also commonly witnessed in terms of flux changes that occur on different time intervals. Changes from months to decades are common \cite[e.g.][]{Papadakis2008,Vagnetti2011,Vagnetti2016,Falocco2017,Paolillo2017} and, in the X-rays, variations are also observed down to kiloseconds timescales \cite[e.g][]{Uttley2002,Ponti12}. The origin of such a rapid variability cannot be solely ascribed to the X-ray band merely mimicking the variations of the disc optical-UV photons \cite[][]{Nandra2001}. Moreover, a tight relation between short timescales variations and SMBH mass \cite[e.g.][]{Papadakis2004,O'Neill2005,McHardy2006,Ponti12} is well established. \\
\indent Multi-epoch high S/N spectral and timing data provide compelling pieces of information to shed light onto the physics behind X-ray variability and its tight link with the emerging X-ray spectrum. In this context, we report on the X-ray spectral properties of  NGC 2992, a nearby highly inclined spiral galaxy \cite[z=0.00771,][]{Keel1996} classified as a Seyfert 1.5-1.9 galaxy \cite[][]{Trippe2008}. This source was the target of two consecutive XMM-Newton orbits in 2019, the second of which had a simultaneous  but shorter \textit{NuSTAR} exposure. Multiple transient Fe K emission lines between 5-7 keV were found, originating from several flaring sectors of the accretion disk \cite[][]{Marinucci2020}. Moreover, variable absorption structures above 9 keV were also detected, associated to an intermittent disk wind (Luminari et al, submitted). The general trend of the source \cite[already reported in][]{Yaqoob2007,Shu2010,Marinucci2018}, where relativistic emission lines are observed at high flux levels, was therefore confirmed. In this paper, we report on the analysis of all the Neil Gehrels Swift Observatory (hereafter Swift) data, most of which were taken in the context of two monitoring campaigns. Then, we  present a detailed timing and spectral analysis of the {\it XMM-Newton} and {\it NuSTAR} 2019 observations.

\section{Data reduction and science products extraction}

\begin{table}
	\centering
	\setlength{\tabcolsep}{1.pt}
	\caption{\small{The observation log for the \textit{XMM-Newton} and \textit{NuSTAR} data is presented. The \textit{NuSTAR} exposure was simultaneously taken during the second orbit of \textit{XMM-Newton}.}\label{obslog}}
	\begin{tabular}{c c c c c}
		\hline
		\\
		Satellite& Detector Obs.& ID Obs.& Net exposure& Start-date\\
		\textit{XMM-Newton}&pn&0840920201&92.6 ks&2019-05-07\\
		\textit{XMM-Newton}&pn&0840920301&92.8 ks&2019-05-09\\
		\textit{NuSTAR}&FPMA/B&90501623002&57.4 ks&2019-05-10\\
	\end{tabular}
\end{table}

 \begin{table}
	\centering
	\setlength{\tabcolsep}{1.5pt}
	\caption{\small{Extraction regions properties as a function of the \textit{XRT} observed rate.}\label{pileup1}}
	\begin{tabular}{c c c}
		\hline
		\\
		Region's shape&Radius(Inner Radius)&Rate\\
		&pixel&cts/s\\
		circle&20&<0.6\\
		annulus&(2)&>0.6\\
		annulus&(3)&>1.4\\
		annulus&(4)&>1.7\\
		annulus&(5)&>2.8\\
		annulus&(6)&>3.2\\
	\end{tabular}
\end{table}

The present paper focuses on NGC 2992 observations (longer than 100 seconds) taken in the context of  \textit{Swift} monitoring campaigns. In the first one, \textit{Swift-XRT} monitored NGC 2992 throughout 2019 (from March 26 to December 14), with the aim of triggering a deep, high flux observation of the source. On May 6, 2019, the triggering flux threshold was met (F$_{\rm 2-10}$=7.0$\times$10$^{-11}$ erg cm$^{-2}$ s$^{-1}$) and \textit{XMM-Newton} started observing the source on 2019 May 7 for two consecutive orbits (ObsIDs: 0840920201, 0840920301). \textit{NuSTAR} \cite[][]{Harr13} observed NGC 2992 on May 10, 2019 for $\sim$120 ks, simultaneously with the second \textit{XMM-Newton} orbit. In this paper, we consider the same data set presented in \cite{Marinucci2020} and Luminari et al., sub, (see also Table \ref{obslog}) and we address the reader to these papers for details on the data reduction.
Then, we also consider exposures obtained during a novel 2021 \textit{Swift} monitoring aimed at keeping track of the extreme variability of NGC 2992.\\
\indent The extraction of the high level science products of each \textit{Swift-XRT} exposure resulted from an automatic process that downloads and reduces raw data taken via photon counting acquisition mode. The procedure is based on the standard pipelines \textit{xrtpipeline} and \textit{xrtproducts} described in Capalbi et al.,~(2005)\footnote{\url{https://swift.gsfc.nasa.gov/analysis/xrt\_swguide\_v1\_2.pdf}}. The regions used to extract the source and background spectra and light curves are selected taking into account any pile-up affecting that specific observation. In particular, after computing the source net count rate with \textit{ximage}, the procedure selected a circular region or an annulus in the case of a rate $<$0.6 cts/s or $>$0.6 cts/s, respectively. Then, the inner radius of the annulus is determined on the basis of the observed count rate. In Table ~\ref{pileup1} we list the rates limits for each inner radius of the extracting annulus. The region used to extract the source always has an outer radius of 50 arcsec, regardless of its circular or annular shape. On the other hand, the background is always extracted using an annular region centered on the source. A difference of 25 pixels ($\sim$60 arcsec) is set between the inner and outer radii of such a region. The outer radii of the source and background regions are always $\sim$60 arcsec apart. Spectra were then binned requiring a minimum of 5 counts per bin and were fitted adopting the Cash statistic \citep[][]{Cash1979}.
\\
\indent Finally,  we relied on a similar automatic procedure to extract science products for each UVOT exposure. In particular, we checked that two regions, one circular and centered on the source (radius=6 arcsec) and a concentric annulus ($\Delta$radius=7 arcsec) were free of any other sources or spurious detection. The returned count rates for the UVOT filters are not corrected for Galactic nor intrinsic reddening. We notice that correcting for such effects would not modify our results, instead it would lead to a shift of the rates. Our computations showed that considering a reddening of E(B-V)=0.0519 \cite[][]{Schlafly2011} would decrease the rates by $\sim$13\% or $\sim$30\% for the V and the UVW2 filters, respectively.

\section{Timing properties}
The \textit{Swift}, \textit{XMM-Newton} and \textit{NuSTAR} observations provide a compelling dataset to shed light onto the temporal properties of NGC 2992 at different timescales.\\
\begin{figure*}
	\centering
	\includegraphics[width=0.99\textwidth]{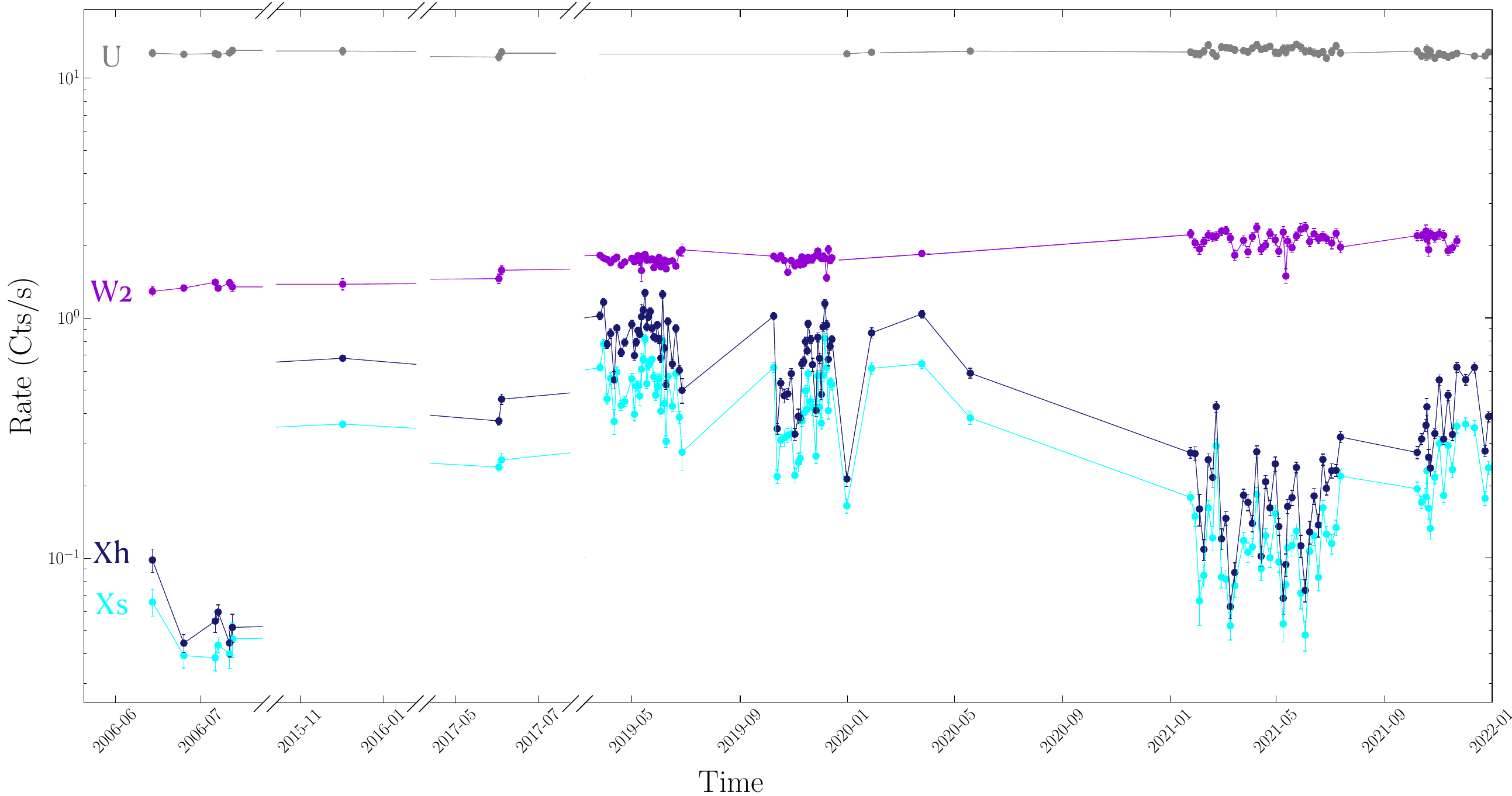}
	\caption{\small{Swift-XRT and -UVOT light curves from the 123 observations. Remarkable variability from short to long timescales can be observed in the X-rays, while the optical-UV light curves (not corrected for the extinction) are characterised by a more constant behaviour. UVOT filters are labelled in the plot while Xs and Xh account for X-rays in the  0.3-2 and 2-10 keV energy ranges. We notice that for visual purposes, each segment of the x-axis has a different length.}}
	\label{swift_lc}
\end{figure*}
\begin{figure}
	\centering
	\includegraphics[width=\columnwidth]{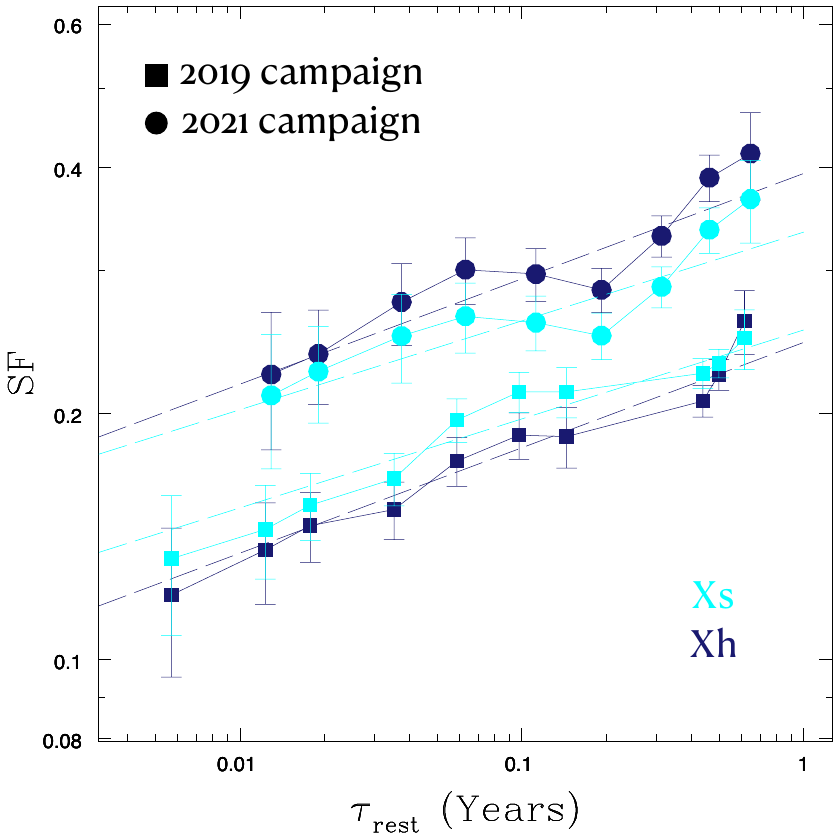}
	\vspace{-0.5cm}
	\caption{\small{Structure functions for the soft and hard X-rays computed for the two monitoring campaigns. Dashed lines account for the weighted linear regression describing the SF. Light curves in 2021 were more variable than in 2019 and both the hard and soft X-rays varied of the same amount within each monitoring.}}
	\label{SF}
\end{figure}
\begin{figure}
	\centering
	\includegraphics[width=\columnwidth]{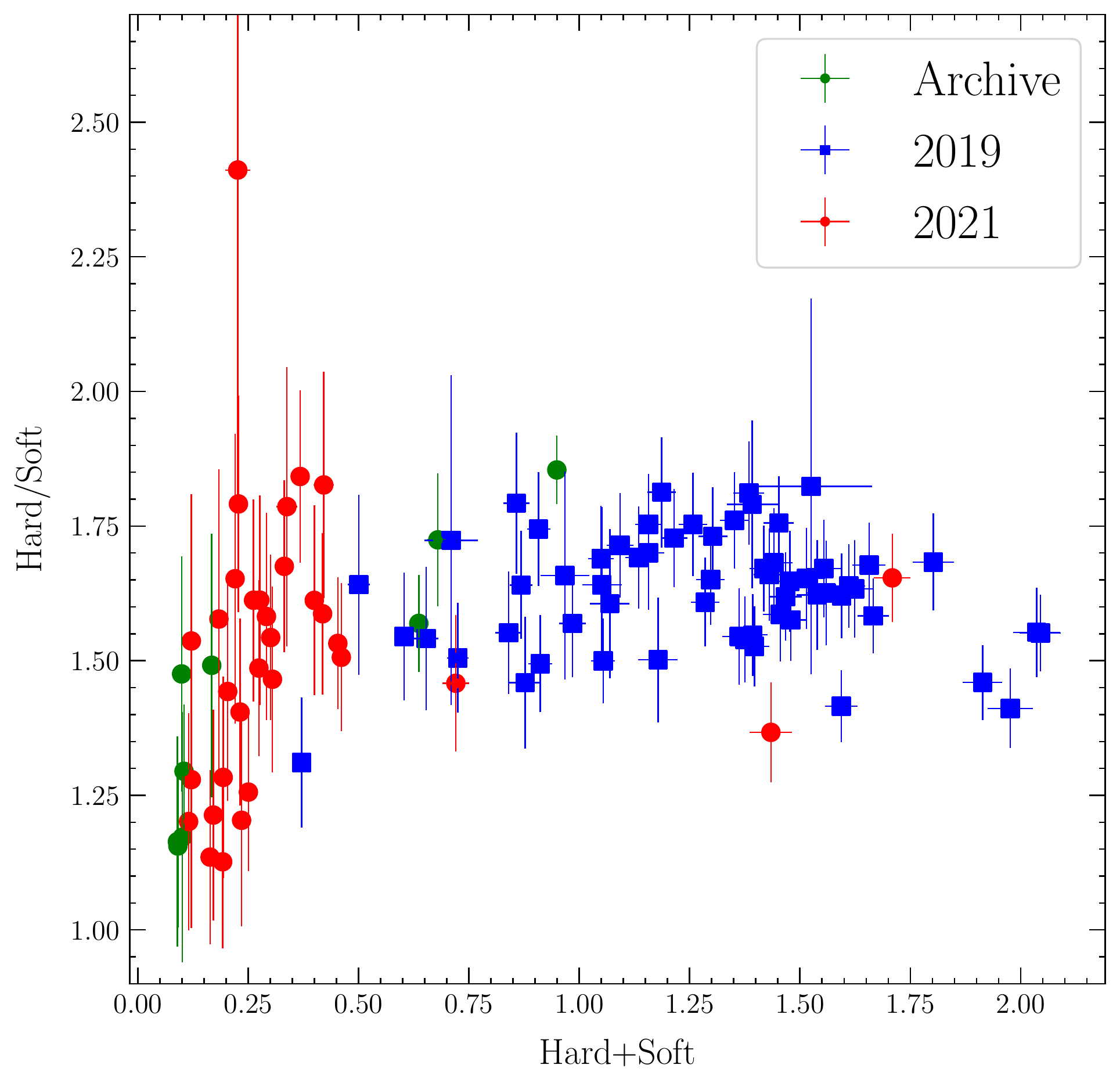}
	\vspace{-0.5cm}
	\caption{\small{NGC 2992 hardness ratios as a function of the total collected counts in the 0.3-10 keV band. Different colours identify data from the 2019 (blue), 2021 (red) and those already in the archive (green).}}
	\label{swift_ratio}
\end{figure}
\begin{figure}
	\centering
	\includegraphics[width=\columnwidth]{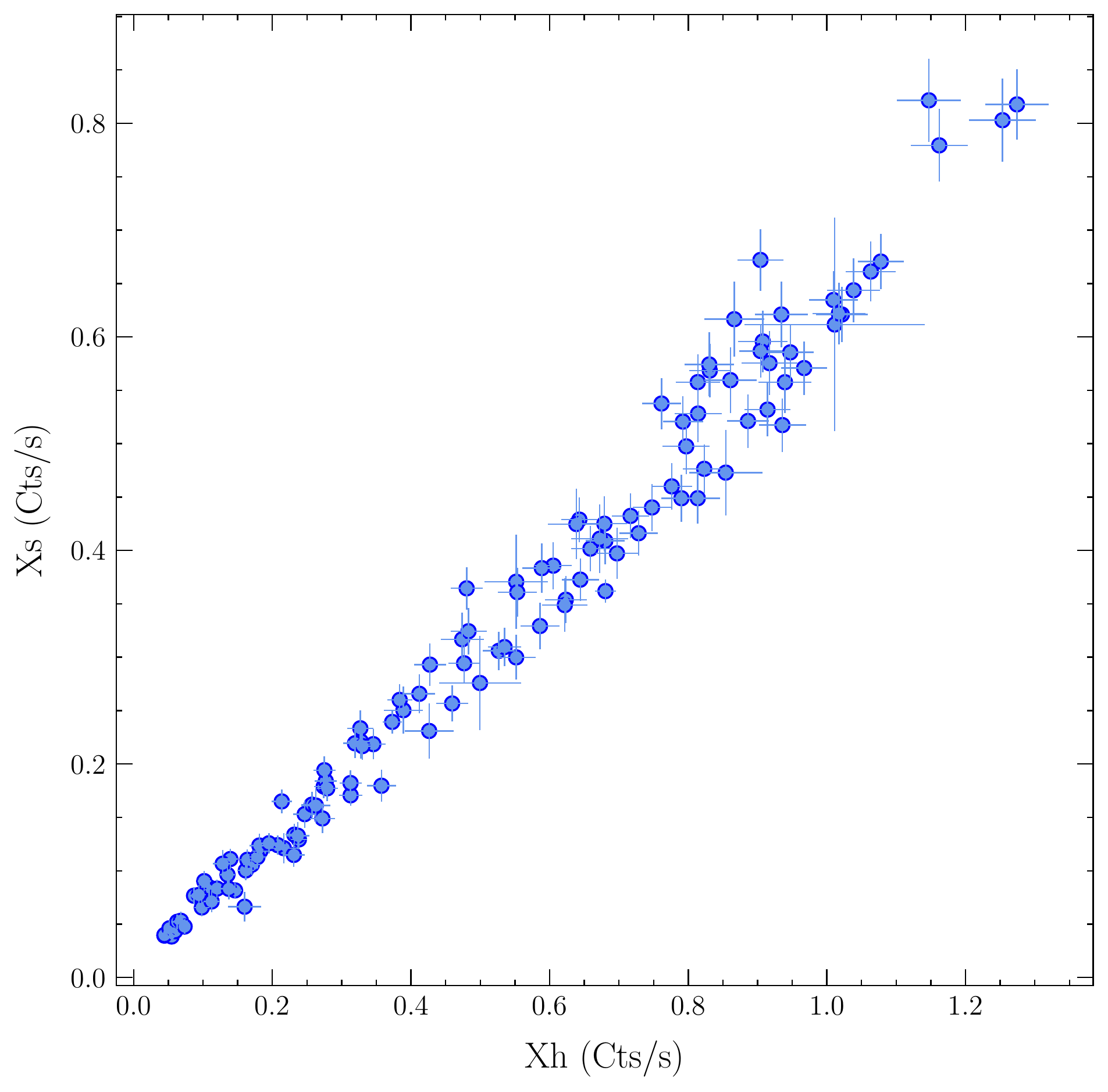}
	\caption{\small{Correlation between soft and hard count rates (0.3-2 and 2-10 keV, respectively). This suggests that the power-law component dominates the $\sim$1-10 keV spectrum of NGC 2992}}
	\label{correlation}
\end{figure}
\begin{figure}
	\centering
	\includegraphics[width=\columnwidth]{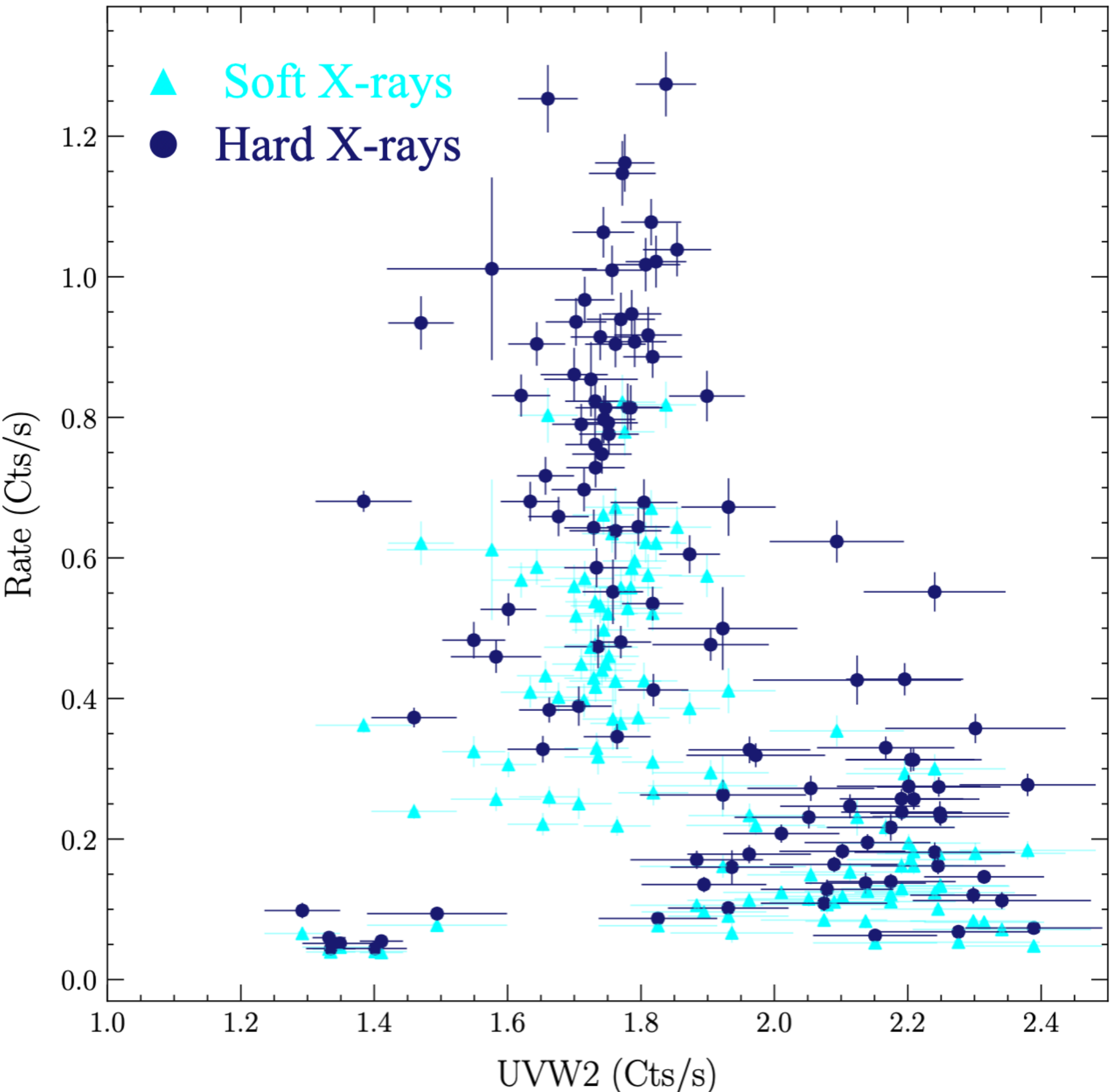}
	\vspace{-0.5cm}
	\caption{\small{Lack of correlation between the X-rays and ultraviolet UVW2 filter. Cyan triangles account for the soft X-rays while blue circles are used for the hard band. Pearson cross-correlation coefficients of P$_{\rm cc}$=-0.21 P(<r)=2\% and P$_{\rm cc}$=-0.19 P(<r)=3\% are found for the Xs vs UVW2 and Xh vs UVW2, respectively.}.}
	\label{nocorrelation}
\end{figure}
\begin{figure}
	\centering
	\includegraphics[width=\columnwidth]{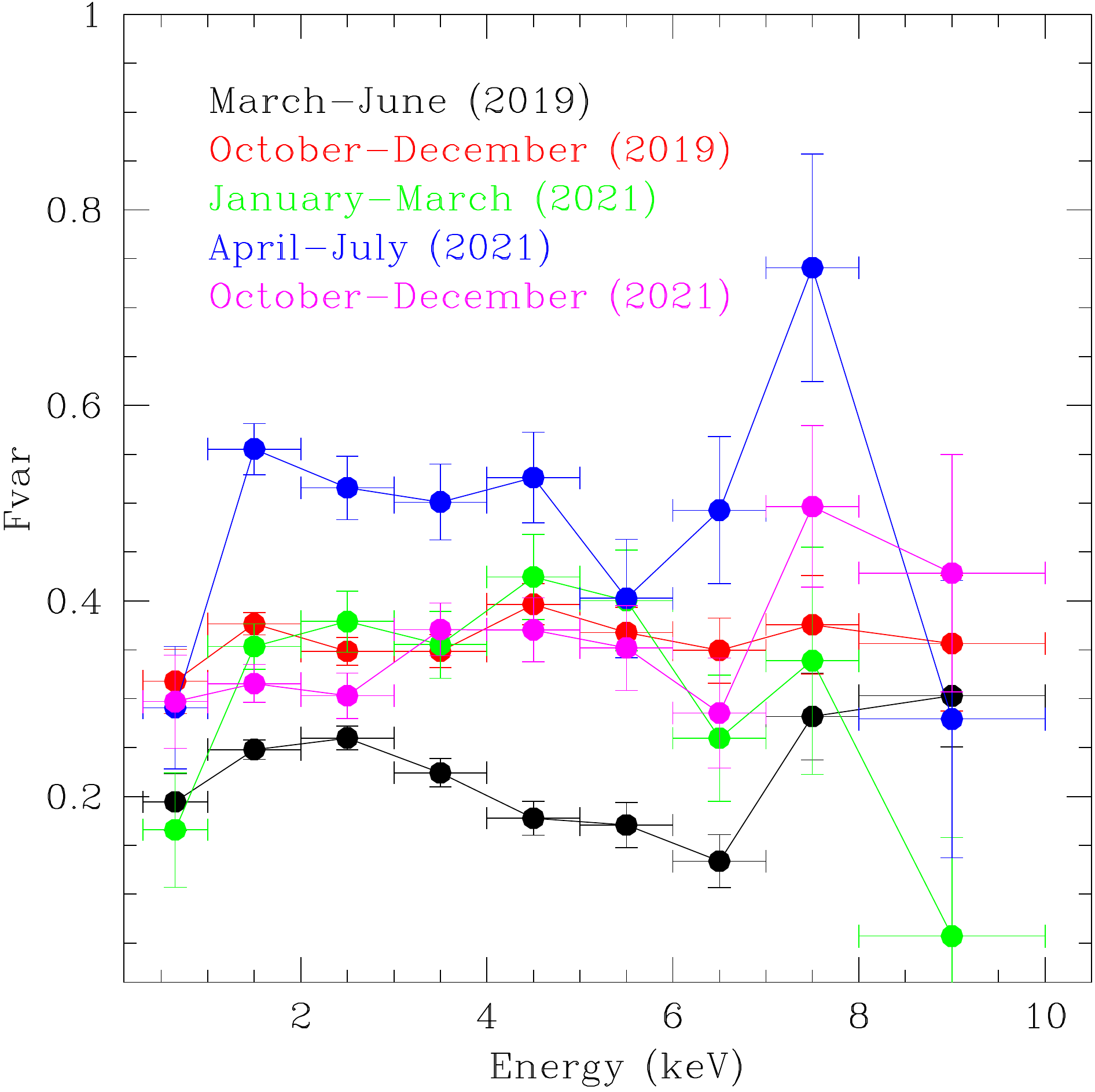}
	\vspace{-0.5cm}
	\caption{\small{Fractional variability spectra derived for the five Swift's subsamples described in Sect. 3. Excess variance spectra clearly have different normalisations suggesting that the source varied by different amounts over monthly timescales. Error bars only account for the Poissonian noise.}}
	\label{fvar}
\end{figure}
\subsection{Daily to yearly variations}
\indent \textit{Swift} extensively observed NGC 2992. From 2006 to present days more than one hundred observations are available, and the bulk of the  exposures belong to two distinct monitoring campaigns, one held during 2019 from which 60 $\sim$2 ks exposures were derived, and a novel one covering 2021.\\
\indent Multi-epoch and multi-wavelength light curves are showed in Fig. \ref{swift_lc}, where remarkable variations are observed from daily to yearly timescales. The lowest X-ray state corresponded to the archival observations from 2006 and a maximal variation larger than a factor of 10 is found.
To the fast X-ray variations correspond fairly flat optical and UV time series. Such a constant behaviour, is clearly observed in the optical filters U (with both V and B showing the same trend) and UVW2, although in this ultraviolet band we can see a marginal long term increase. The flat shape of the optical-UV light curves for NGC 2992 can be straightforwardly accounted for by the obscured nature of this source, implying that even if the AGN may contribute, it does not dominate the emission in these bands.\\
\indent The well sampled X-ray light curves obtained from the monitoring campaigns allow us to compute the corresponding structure functions (SFs).
The SF has been widely adopted in different bands of the electromagnetic spectrum  \cite[][]{Trevese1994,DeVries2005,Bauer2009} and quantifies the amount of variability giving a measure of the mean change between two observations separated by a time lag $\tau$. It has been used for ensemble studies \cite[see][for details]{Vagnetti2011,Vagnetti2016} as well as for single AGN \cite[e.g.][]{Gallo2018,Laurenti2020}. Different mathematical formulations for such an estimator have been proposed \cite[e.g.][]{Simonetti1985,diClemente1996}, and we here use the one described in Sect. 3 of \cite{Middei2017}. In Fig.~\ref{SF} we show the soft (0.3-2 keV) and hard (2-10 keV) SFs for the two \textit{Swift-XRT} campaigns. Flux changes in both bands increase with the rest-frame time lags, SF$_{\rm hard}\sim\tau^{0.12\pm0.02}$ and SF$_{\rm soft}\sim\tau^{0.10\pm0.02}$, with these slopes being compatible with the one found from ensemble studies focusing on the average variability of AGNs \cite[SF$_{\rm ensemble}\sim\tau^{0.121\pm0.004}$,][]{Vagnetti2016}. The normalisations of the SFs account for two different variability levels, the larger the variability, the higher the SF. Interestingly, larger variability is measured during 2021 corresponding to a lower flux state of the source.\\
\indent We then searched for correlations possibly connected with variations in the column density of the obscurer. For this reason we computed the ratio between the hard and soft X-rays and we studied it as a function of the total counts. In accordance with the plot in Fig.~\ref{swift_ratio} no correlation holds between the hard/soft X-ray ratio and full band rate, and a marginal trend can be only observed for a full band rate below 0.5 cts/s. The lack of a noticeable trend is consistent with a fairly constant column density of the obscurer. On the other hand, the soft and hard X-rays are strongly correlated (P$_{\rm cc}$=0.99, P(<r)<0.01\%), see Fig.~\ref{correlation}, suggesting that both the soft and hard X-rays are produced by the very same spectral component.

Finally, we further stress that neither the soft nor the hard X-ray bands are correlated with the ultraviolet emission, see Fig.~\ref{nocorrelation}. Such a correlation has been commonly observed in samples of unobscured AGN \cite[e.g.][]{Edelson2002,Lusso2010,Edelson2015,Lusso2017} and the lack of correlation has been associated either to an incumbent changing look process \citep[e.g.][]{Ricci2020,Ricci2021,Laha2022arXiv} or to intervening absorbing matter. Due to the Seyfert 2 nature of NGC 2992, the lack of a correlation between the disc and coronal emissions can be ascribed to the predominantly non-nuclear nature of the UV emission observed with
\textit{Swift-UVOT}.\\

\indent We further quantified the variability properties of NGC 2992 computing the so-called fractional  variability (F$_{\rm var}$). This estimator \cite[e.g.][]{Edelson2002,Vaughan2003,Ponti2004,Ponti2006} provides a direct measure of a light curve variability and is defined as the square root of the normalised excess variance. \cite[e.g.][]{Vaughan2004,Ponti2006,Matzeu2016mnras,Matzeu2017,Alston2019,Parker2020,DeMarco2020,Igo2020}. To derive compatible excess variance spectra, we only considered data taken during the 2019 and 2021 monitoring campaigns. In particular, for a meaningful comparison, we need to compute the F$_{\rm var}$ spectra using light curves of similar length. The 2019 campaign spans a time interval of about 6 months although not performed continuously due to visibility issues. We thus divided each set of observations into subsets roughly covering an about 3 months long time interval. We thus ended up with five different subsamples: sample \textit{a}  (March-June 2019), sample \textit{b} (October-December 2019), sample \textit{c} (January-March 2021) and sample \textit{d} (April-July 2001) and sample \textit{e} (October-December 2021. We then computed the light curves for the samples in different energy intervals and derived the corresponding fractional variability. In Fig.~\ref{fvar} we show the resulting   F$_{\rm var}$ spectra. Aside from some fluctuations above $\sim$7 keV, likely due to background issues and/or low S/N, and the first energy bin that is mostly due to distant scattering, all the spectra have a fairly constant behaviour.  This suggests that the variability is driven by a single variable component, which in our case is the primary continuum emission. Concerning the normalisation of the spectra, we notice that the higher the F$_{\rm var}$ spectrum the lower the flux, which is in agreement with what is commonly observed in other AGNs \cite[e.g.][]{Barr1986,Green1993,Lawrence1993}.

\subsection{Hourly to daily changes}

\begin{figure*}
	\centering
	\includegraphics[width=0.9999\textwidth]{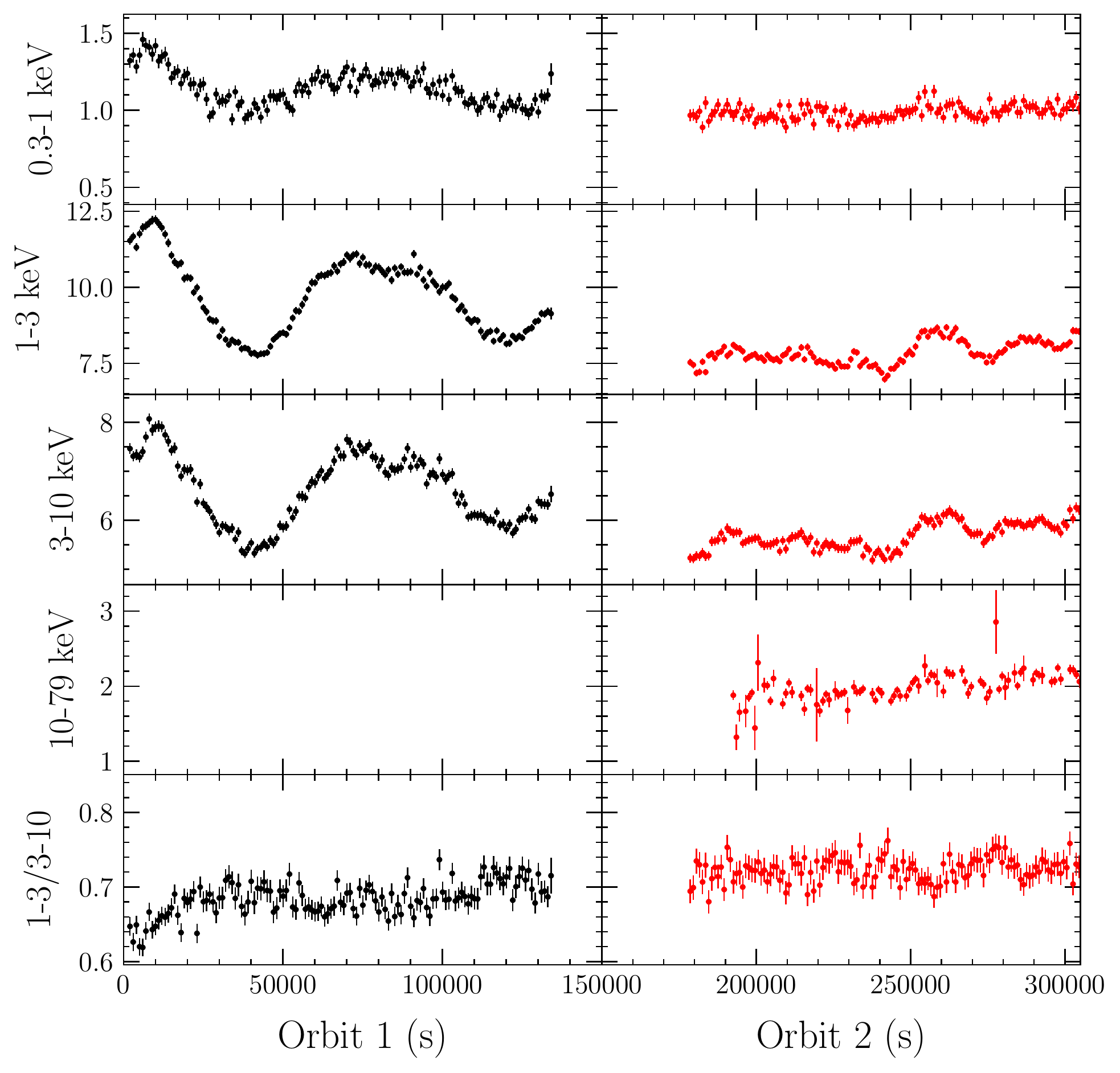}
	\caption{\small{Background subtracted light curves in units of counts/sec are shown for \textit{XMM-Newton} and \textit{NuSTAR} data. Aside from the 0.3-1 keV energy band, light curves below 10 keV show remarkable changes in the first orbit and moderate flux variability in the second one. The \textit{NuSTAR} light curve (10-79 keV) is not completely simultaneous with the second \textit{XMM-Newton} orbit. The 3-10 keV \textit{XMM-Newton} (orbit 2) and the 10-79 keV \textit{NuSTAR} light curves are characterised by a similar amount of variations above 1, which suggests the primary continuum to dominate also the \textit{NuSTAR} energy band}.}
	\label{2992lc}
\end{figure*}
\begin{figure}
	\centering
	\includegraphics[width=0.99\columnwidth]{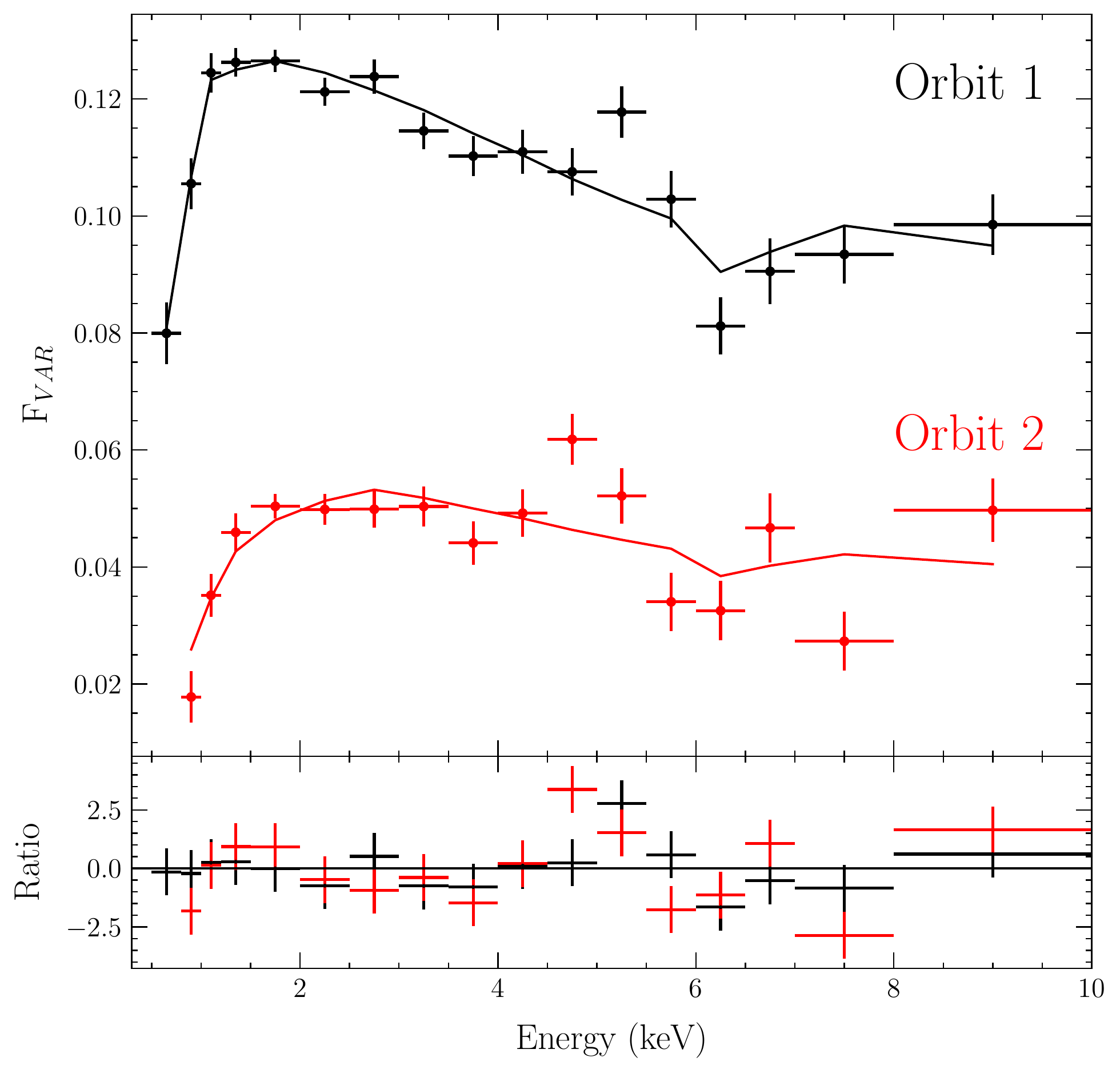}
	\caption{\small{Fractional variability spectra of Orbit 1 and 2 derived from the background subtracted light curves. The adopted temporal binning is set to 1000 seconds. Interestingly, these spectra are characterised by two different amounts of variability. In particular, the higher the flux the larger the amount of variability. We notice that the second orbit spectrum lacks the 0.3-0.5 keV  energy bin as it was found consistent with zero. Fitted F$_{\rm var}$ spectra are showed. In both orbits, substantial residuals can be observed around 5 keV.}}\label{fvarx}
\end{figure}
The two consecutive \textit{XMM-Newton} orbits, one of which also has a simultaneous \textit{NuSTAR} exposure, are extremely suitable for quantifying NGC 2992 variations on very short term. We started deriving the NGC 2992 light curves, see Fig.~\ref{2992lc}. The different panels, from top to bottom, account for the 0.3-1, 1-3 and 3-10, 10-79 keV bands, while the last row shows on the ratios between the soft and hard X-rays (1-3 keV/3-10 keV). The constancy of the 0.3-1 light curves is consistent with extra-nuclear, ionised emission.
In orbit 1, \textit{XMM-Newton} caught the source in a higher flux state than in orbit 2, for which data up to 79 keV are also showed.
Variations during the first orbit are about a factor of 60\% in both the 1-3 and 3-10 keV energy bands. Such fast flux changes occur on a timescale of $\sim$40 ks. On the other hand, the amount of variability in orbit 2 is reduced to a few percent for both the soft and the hard X-ray bands. \textit{NuSTAR}'s light curve is consistent with the 3-10 keV XMM-Newton time series and shows a similar amount of variability suggesting for the presence of a weak reflected component and a primary continuum still dominating this high energy band. The hardness ratios reported in the last row of Fig.~\ref{2992lc} are suggestive of a smooth spectral softening over the observed temporal window. In fact, ratios between the 1-3 and 3-10 keV bands increases from 0.65 at the beginning of the first orbit to $\sim$0.75 at the end of the second \textit{XMM-Newton} exposure.\\
\indent Thanks to the high S/N of this dataset, we derived the F$_{\rm var}$ spectra of the two \textit{XMM-Newton} exposures. We used the background subtracted light curves binned every 1000 seconds. The resulting variability spectra are shown in Fig.~\ref{fvarx}. 
The spectrum of orbit 1 has a larger normalisation than orbit 2, in agreement with the light curves in Fig. \ref{2992lc}. During orbit 1, the 1-3 keV X-rays are more variable than the hard X-rays. A moderate drop in the variability spectrum is observed around 6.4 keV, as expected for a constant Fe K$\alpha$ emission. A similar drop is also observed in orbit 2, despite the  $\sim$3 times less variable spectrum. Both spectra show a drop below $\sim$1 keV, as this energy range is dominated by a constant component. Finally, both spectra show an interesting variable feature around 5 keV that can possibly be associated to the transient emission line component discussed in \cite{Marinucci2020}.  In a recent work by \cite{Parker2020}, the authors computed different tables to model \textit{F$_{\rm var}$} spectra with standard spectral fitting packages, such as \textit{Xspec} \cite[][]{Arnaud1996}.
Following \cite{Parker2020} and the prescriptions in the web page\footnote{\url{https://www.michaelparker.space/variance-models}}, we tried to model our excess variance spectra. In particular, we used the following model:
\begin{equation*}
\rm Fvar\_pidamp\_1.fits \times Fvar\_pow.fits \times Fvar\_xildamp.fits .
\end{equation*}
\noindent The first table is based on the \textit{Spex} photoionisation model Pion \cite[][]{Miller2015,Mehdipour2016} and accounts for the drop in variance due to a constant photoionised emission. The model's parameters are \textit{frac} and \textit{$\xi$}, the ratio of the 0.5-10 keV flux of the reflection to the average log primary flux and the disc's ionisation in erg cm s$^{-1}$.
The second table reproduces the variance of a powerlaw-like continuum changing in log(flux). Its parameters, \textit{var} and \textit{corr} define the variance of the logarithmic flux for the primary continuum in the 0.5-10 keV energy range and the correlation between the photon index of the power law and the variable flux. Finally, the last table is needed to account for the reduction of F$_{\rm var}$ due to unblurred reflection. In this case, the \textit{frac$_{xill}$} is the ratio of the 0.5-10 keV flux of the reflection to the average power law flux whose flux is computed in logarithm.
We fitted this very same model to the two spectra finding that the first high flux orbit is well described by the model with $\chi^2$=15 for 13 d.o.f. and that the model works fairly well for the continuum variability of orbit 2, despite the $\chi^2$/d.o.f.=38/12. This poor statistics is indeed mainly due to the unaccounted excess around 5 keV and, more marginally, by scattered data above 7 keV.

\begin{table}
	\centering
	\setlength{\tabcolsep}{1.5pt}
	\caption{\small{Best-fir parameters for the two excess variance spectra derived from the \textit{XMM-Newton} orbits 1 and 2. The fit statistics are $\chi^2$/d.o.f.=15/13 and $\chi^2$/d.o.f=38/12 for the two spectra, respectively.}\label{pileup}}
	\begin{tabular}{c c c c}
		\hline
		\\
		model&parameer&obs1&obs2\\
		\hline
		Fvar\_pidamp&frac&0.26$^{+0.21}_{-0.09}$&1.0$^{+1.9}_{-0.4}$\\
		&xi& 0.85$^{+0.61}_{-0.40}$&1.6$\pm$0.3\\
		Fvar\_pow&var&$0.059^{+0.002}_{-0.004}$&0.022$^{+0.005}_{-0.003}$\\	
		&corr&<0.24&<0.63\\
		Fvar\_xildamp&frac&0.26$^{+0.03}_{-0.12}$&<0.34\\
	\end{tabular}
\end{table}

\section{Spectral properties}
As for Sect. 3, the multi-epoch broadband data presented in this paper provide a compelling collection of observations suitable to perform both time -average and -resolved spectral analyses.

\subsection{Mid-to-long term spectral properties}
\indent The high flux of NGC 2992 allows us to extract \textit{XRT} spectra for each of the 123 observations presented in this paper. We thus adopted a simple model to determine the basic properties of the primary continuum and the neutral absorbing column of NGC 2992 across the years, by fitting within \textit{Xspec} the following model:

\begin{equation*}
\rm tbabs\times ztbabs \times powerlaw.
\end{equation*}

\begin{figure*}
	\centering
	\includegraphics[width=.99\textwidth]{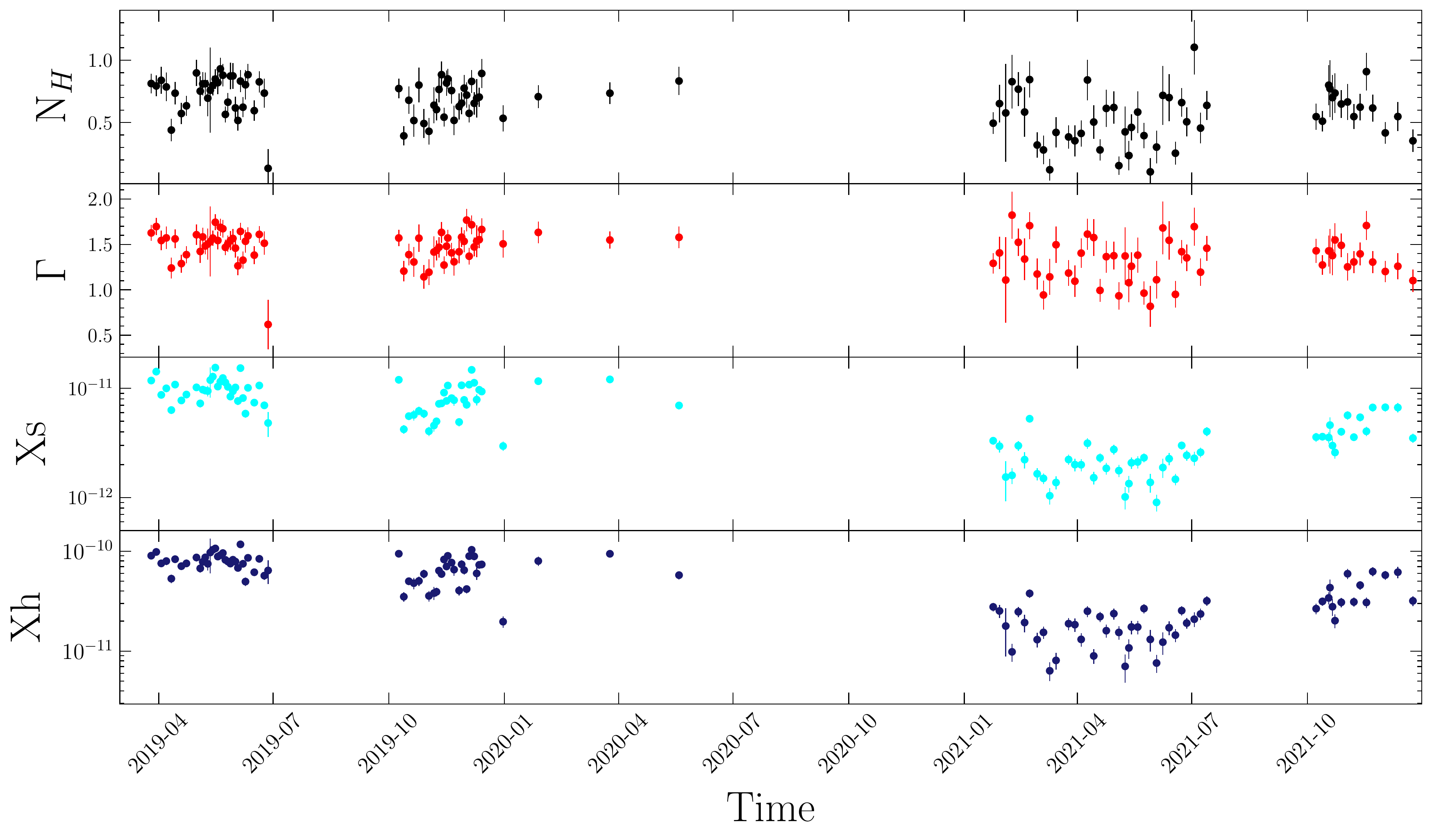}
	\caption{\small{Best-fit parameters derived from the analyses of the XRT exposures. The column density is in units of $\rm \times10^{22}~cm^{-2}$ and fluxes are in units of  erg ~cm$^{-2}$ ~s$^{-1}$. The same results are reported in Table~\ref{swifttable}}.}
	\label{swift_bestfit}
\end{figure*}

\begin{figure}
	\centering
	\includegraphics[width=\columnwidth]{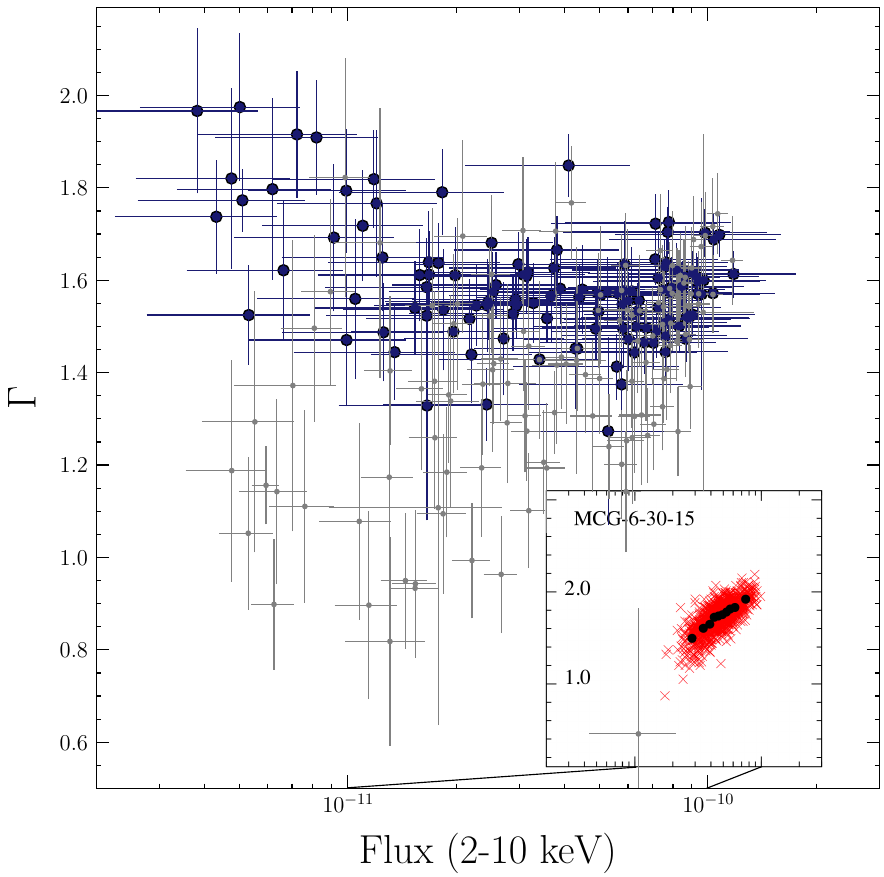}
	\caption{\small{\textit{XRT} photon indices as a function of the 2-10 keV flux (in units of $\times$10$^{-11}$ erg cm$^{-2}$ s$^{-1}$). Blue dots accounts for $\Gamma$ values derived using a fixed absorbing column density N$_{\rm H}$=7.8 $\times$10$^{21}$ cm$^{-2}$. Gray crosses, instead, represent flatter $\Gamma$ values that had been derived with a free to vary N$_{\rm H}$. The inset \citep[taken from][]{Sobolewska2009} shows MCG-6-30-15, which in contrast, does show a softer when brighter behavior more typical of other AGN.}}\label{deltagamma}
\end{figure}

\noindent The power law models the nuclear X-ray emission and both the local and Galactic absorptions are accounted for. 
For each observation, we fitted  the column density of the local absorber, as well as the continuum photon index and its normalisation. 
Via this procedure we determined the best-fit parameters shown in Fig.~\ref{swift_bestfit} and quoted in Table~\ref{swifttable}.
Although the adopted model does not include the soft scattered component, all the spectra are well accounted for, as they all have a Cstat/d.o.f. ratio close to unity. Hard and soft X-ray fluxes show remarkable variations also down to daily timescales, see Fig.~\ref{swift_bestfit}.
It is hard to assess whether there are spectral changes or not. As expected, in fact, the power-law photon index and the obscurer column density are strongly correlated, diluting any intrinsic spectral variability. We thus refitted all the observations keeping N$_{\rm H}$ fixed to its average value of N$_{\rm H}$=7.8$\times$10$^{21}$ cm$^{-2}$ (determined from the 123 \textit{Swift} observations). This new attempt led to steeper values of $\Gamma$, found to have an average value of  $\Gamma$=1.59$\pm$0.04 and covering the range 1.37-1.97,  see blue points in Fig.~\ref{deltagamma}. Interestingly, no correlation holds between these  photon indices and the total flux, so that the source does not obviously show the typical softer when brighter behaviour commonly observed in AGNs \cite[e.g.][see example in the inset of Fig.~\ref{deltagamma}]{Sobolewska2009}.

\subsection{Short term spectral properties: the Fe K$\alpha$ complex}

\begin{figure}
	\centering
	\includegraphics[width=\columnwidth]{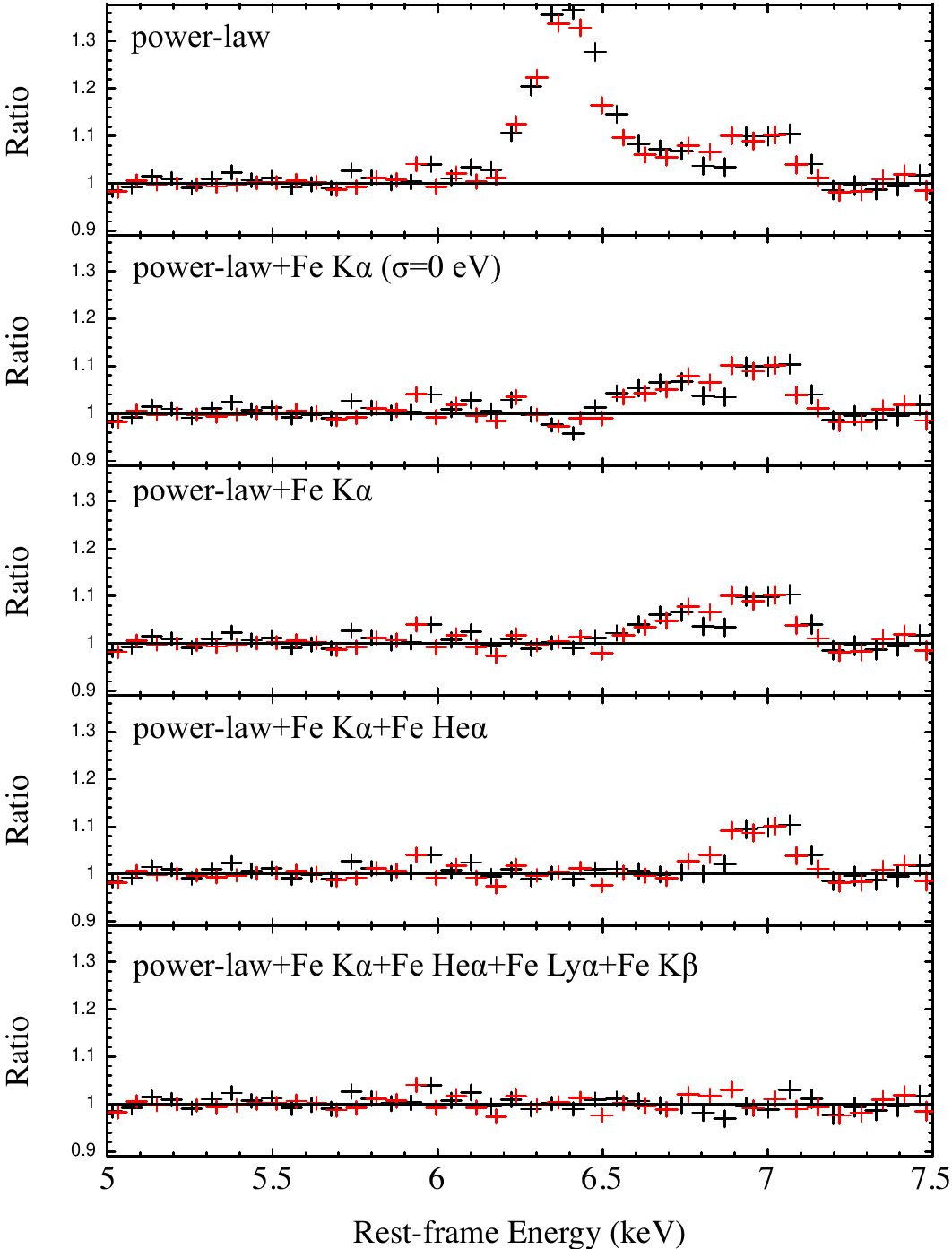}
		\vspace{-0.5cm}
	\caption{\small{Zoom in the 5-7.5 keV energy band of the data to model ratios. The spectrum in black refers to orbit 1 data while the red spectrum accounts for data taken during orbit 2.}}
	\label{line}
\end{figure}

We started focusing on the Fe complex of NGC 2992 testing a simple power law to the \textit{XMM-Newton} spectra. We worked on the 3-8 keV energy range and we fitted the photon index and the normalisation of the continuum for both orbit 1 and 2. In Fig.~\ref{line}, we show zoomed spectra where modelling the sole continuum leaves prominent residuals between 6 and 7 keV. Then we added two Gaussian components with zero width to model the Fe K$\alpha$ and its accompanying Fe K$\beta$. We assumed the lines not to change between the two orbits so that we fitted the Gaussian energy centroid and normalisation for the Fe K$\alpha$. The Fe K$\beta$, had its energy fixed to 7.06 keV and its normalisation was free to vary 
up to 14\% of the Fe K$\alpha$ flux \cite[][]{Molendi2003}. Although this leads to a significant reduction in the fit statistic ($\chi^2$/$\Delta$d.o.f.=264/160), the data are far from being well reproduced. First of all, data between 6.4-7.1 keV are not yet accounted for, suggesting that the Fe K$\alpha$ profile might be the superposition of different components. We then allowed the Fe K$\alpha$ width to vary finding a better fit ($\Delta\chi^2$=-19). The residuals around 6.4 keV are now accounted for. Then, we added two additional Gaussian components for the additional residuals at $\sim$6.7 and $\sim$7 keV. As for the 6.4 keV Fe K$\alpha$ line, we fitted the central energy and normalisation of these two Gaussians (with null width) tying the values between the orbits. In particular, one line models the Fe Ly$\alpha$ emission at 6.96 keV ($\Delta\chi^2$/$\Delta$d.o.f.=-42/-2) while the second accounts for the Fe He$\alpha$ at 6.7 keV ($\Delta\chi^2$/$\Delta$d.o.f.=-47/-2).\\
\indent These steps led us to a best-fit of $\chi^2$=156 for 155 d.o.f. and the inferred parameters are quoted in Table ~\ref{xmmlines}. These tests are consistent with a weakly broad Fe K$\alpha$ that may be the superposition of two different components.\\
\begin{table}
	\centering
	\setlength{\tabcolsep}{1.5pt}
	\caption{\small{Best-fit quantities derived from the \textit{XMM-Newton} exposures fitted in the 3-10 keV energy range. The dagger indicates that the energy centroid of the Fe K$\beta$ was fixed.}\label{xmmlines}}
	\begin{tabular}{c c c c c}
		Component&Parameter &Orbit 1& Orbit 2&Units \\
		\hline
        Tbabs&N$_{\rm H}$&1.0$\pm$0.2&0.9$\pm$0.2&$\times$10$^{22}$ cm$^{-2}$\\
		Pow&$\Gamma$&1.62$\pm$0.03 &1.66$\pm$0.02& \\
		&N$_{\rm pow}$&1.7$\pm$0.1 &2.1$\pm$0.1& $\times$10$^{-2}$ ph.\,keV$^{-1}$\,$\rm cm^{2}$\,s$^{-1}$\\
	    zGauss$_{\rm Fe K\alpha}$&E&6.38$\pm$0.05&&keV\\
 	&$\sigma$&46$\pm$10&& eV\\
	 &EW&90$\pm$5&&eV\\
	  &Norm&8.6$\pm$0.3&&$\times$10$^{-5}$ ph.\,cm$^{-2}$\,s$^{-1}$\\
       zGauss$_{\rm Fe K\beta}\dagger$&E&7.06&&keV\\
	   &EW&<20&&eV\\
       &Norm&<5.9&&$\times$10$^{-6}$ ph.\,cm$^{-2}$\,s$^{-1}$\\
		zGauss$_{\rm Fe Ly\alpha}$&E&6.96$\pm$0.01&&keV\\
       &EW&25$\pm$4&&eV\\
		&Norm&2.0$\pm$0.3&&$\times$10$^{-5}$ ph.\,cm$^{-2}$\,s$^{-1}$\\
		zGauss$_{\rm Fe He\alpha}$&E&6.71$\pm$0.04&&keV\\
		&EW&12$\pm$3&&eV\\
		&Norm&1.2$\pm$0.3&&$\times$10$^{-5}$ ph.\,cm$^{-2}$\,s$^{-1}$\\
		\end{tabular}
	    \end{table}
	    
\subsection{The \textit{XMM-Newton/NuSTAR} 2019 observations: Time averaged spectral properties}

We here derive the NGC 2992 time-average properties in the  0.5-79 keV energy range by fitting the \textit{XMM-Newton} and the \textit{NuSTAR} spectra. Relying on our findings in previous Sect. 4.2 and the phenomenological model by \cite{Marinucci2018}, we built the following model in \textit{Xspec}:

\begin{multline}
\rm tbabs \times(apec + Cloudy + (ztbabs \times po)+ \\
\rm+~zGauss + MyTorusL+ MyTorusS+zGauss+zGauss).
\end{multline}

\begin{figure}
	\centering
	\includegraphics[width=\columnwidth]{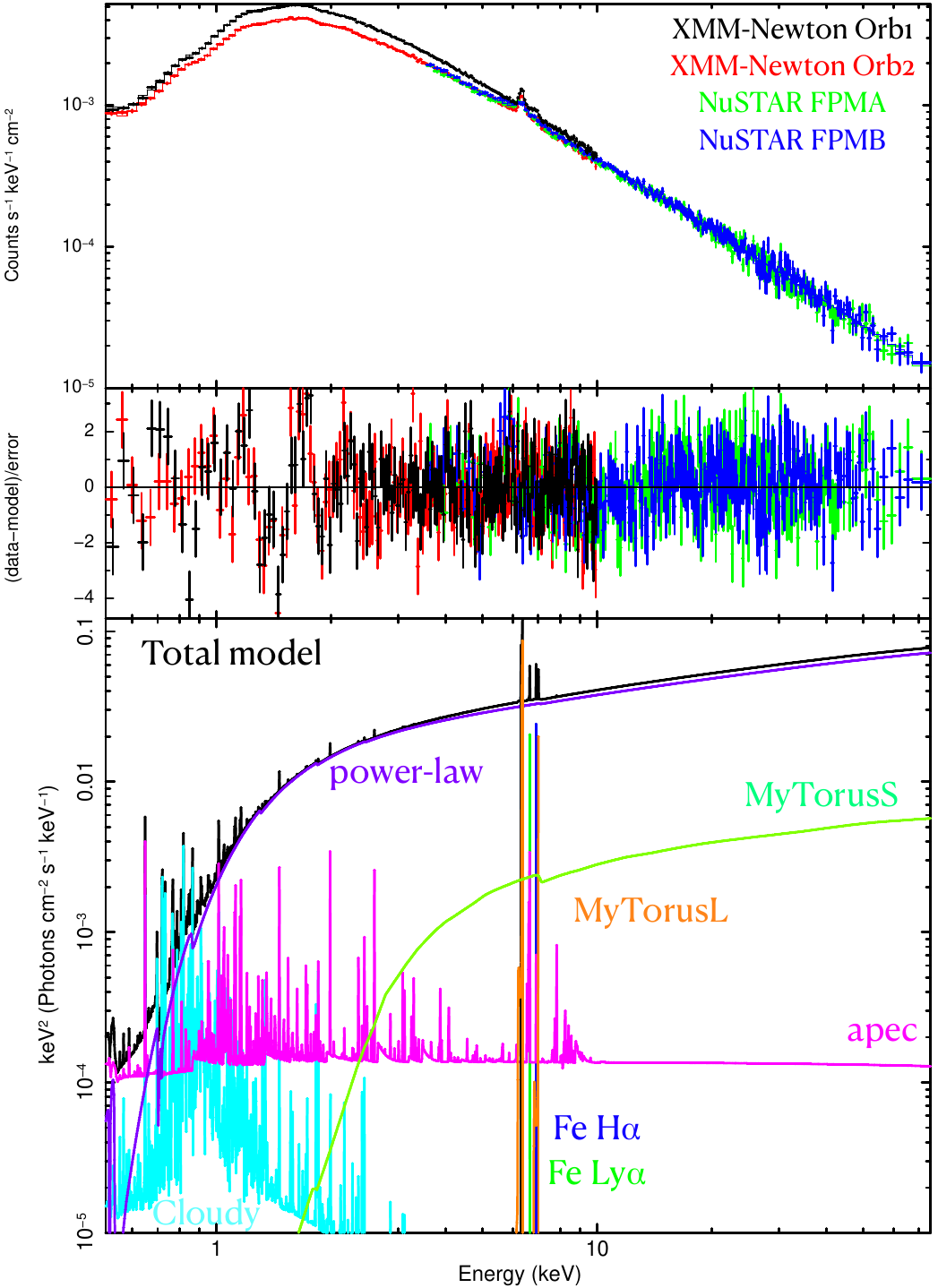}
	\caption{\small{Best-fit for the two \textit{XMM-Newton} orbits and the accompanying \textit{NuSTAR} data. The inferred parameters are reported in Table~\ref{xmmaverage}. The model components shown and labelled in the bottom panel are those derived for the simultaneous    \textit{XMM-Newton}-\textit{NuSTAR} data only.}}
	\label{timeaverage}
\end{figure}

\noindent \textbf{Soft X-rays}: The neutral Galactic and intrinsic absorption is taken into account using the $tbabs$ model. The soft X-rays of type 2 AGNs  are generally dominated by emission lines whose origin is a photonionised gas consistent with the narrow line region \cite[NLR; e.g.][]{Awaki1991,Turner1997a,Turner1997b,Bianchi2006,Guainazzi2007,Laha2020}, thus we accounted for this emission component using a grid model for \textit{Xspec} computed with \textit{CLOUDY} 17 \cite[][]{Ferland2017}. This table,  already presented in past studies \cite[][]{Bianchi2010,Marinucci2011,Marinucci2017}, has two parameters: the ionising flux $\log$ U = [-2.00 : 4.00], with step of 0.25, and cloud column density $\rm\log N_H$ = [19.0 : 23.5], with step of 0.1. Then, the \textit{apec} model was used to reproduce thermal emission from extra-nuclear material observed with \textit{Chandra} \cite[][]{Colbert2005}. 
We fitted the temperature and the normalisation for the \textit{apec} component and the column density, the ionisation and the normalisation of the \textit{Cloudy} one. However, these parameters were tied between the spectra as no variations are expected for this larger scale gas down to the investigated timescales.\\
\textbf{Hard X-rays}: A power law reproduces the nuclear continuum while \textit{MyTorusS} and \textit{MyTorusL}, both additive components, model the reflected emission plus its accompanying Fe\,K$\alpha$, Fe\,K$\beta$ fluorescent emission lines. \textit{MyTorus} \cite[][]{Murphy2009,Yaqoob2012} includes the Compton down-scattering effect and the self-consistent reflected components assuming a fixed geometry of the toroidal X-ray reprocessor, for which the covering factor of the torus corresponds to a fixed half-opening angle of 60$^{\circ}$. Here we assumed: the medium absorbing the primary continuum and the one reflecting it to have different column densities. To set this scenario up we fitted independently the column density of ztbabs\footnote{The NGC 2992 obscurer has a column density N$_{\rm H }$<10$^{22}$ cm$^{-2}$, too low to adopt the multiplicative table \textit{MyTorusZ} commonly used to account for absorption in the line of sight.} and \textit{MyTorus} tables independently i.e. using the so-called decoupled mode to allow the reflector and absorber to have different column densities. Then, we fixed the viewing angle of \textit{MyTorusL} and \textit{MyTorusS} to 0$^{\circ}$, thus implementing the back scattering scenario. In the fits, the photon index of the power law was tied with those of the two \textit{MyTorus} tables and computed for both the orbits . The normalisation of the nuclear emission was computed for both orbits, similar to the one of the \textit{MyTorus} model, which we assumed to be the same between  \textit{MyTorusL} and \textit{MyTorusS}.\\
\textbf{Emission lines}: the 5-7 keV energy range hosts prominent features in emission and we used Gaussian lines to account for all of them except for the Fe K$\alpha$ and the Fe K${\beta}$ lines, which are already included in \textit{MyTorus}. However, in Sect. 4.2 we found evidence for a weakly broad Fe K$\alpha$ line, thus we added  a Gaussian component whose energy centroid was computed tying its value between the spectra and fitting its normalisation in both orbits. Moreover, we fixed the centroid energy of two additional ionised Gaussians to E=6.7 keV, E=6.96 keV, respectively, and assumed a narrow profile ($\sigma$=0 eV) for both of them.\\ 
\indent These steps led to a fit statistic of $\chi^2$/d.o.f.= 1074/731. Residuals between 3 and 5 keV suggest the photon index may not be the same for \textit{XMM-Newton} orbit 2 and \textit{NuSTAR} ($\Delta \Gamma\sim 0.06$). This may be either due to the non-simultaneity of the spectra or due to inter-calibration issues among the detectors as also reported for other observations \cite[e.g.][]{Porquet2018,Laha2021b}. Allowing for different $\Gamma$ values for \textit{XMM-Newton} and \textit{NuSTAR} in orbit 2 yields a better fit statistic of $\chi^2$/d.o.f.=980/730 (see Fig.~\ref{timeaverage}), and we report in Table~\ref{xmmaverage} the corresponding best-fit parameters.
\begin{table}
	\centering
	\setlength{\tabcolsep}{.1pt}
	\caption{\small{Best-fit values for the fit with statistic $\chi^2$=980 for 730 d.o.f. as derived in accordance with Sect. 4.2. The $\dagger$ is used to identify those parameters that have been computed tying the values among the orbits.}\label{xmmaverage}}
	\begin{tabular}{c c c c c}
		Component&Parameter &Orbit 1& Orbit 2+NuSTAR&Units \\
		\hline
		Cloudy$\dagger$&$\log$U&2.66$\pm$0.01&&\\
		&$\log N_{\rm H}$&20.4$\pm$0.1&&\\
        &N&6.1$\pm$0.2&&$\times$10$^{-16}$\\
        Apec$\dagger$&kT&0.68$\pm$0.02&&\\
       &Norm&1.0$\pm$0.4&&$\times$10$^{-4}$\\
        ztbabs&N$_{\rm H}$&0.79$\pm$0.01&0.78$\pm$0.01&$\times$10$^{22}$ cm$^{-2}$\\
		Pow&$\Gamma$&1.70$\pm$0.01 &1.68$\pm$0.01& \\
		&N$_{\rm pow}$&2.1$\pm$0.2 &1.6$\pm$0.1& $\times$10$^{-2}$ ph.\,keV$^{-1}$\,$\rm cm^{2}$\,s$^{-1}$\\
		zGauss$\dagger$&E&6.33$\pm$0.05&& keV\\
		&EW&25$^{+31}_{-10}$&15$^{+16}_{-9}$& eV\\
		&N&1.9$^{+2.2}_{-0.7}$&1.6$^{+1.1}_{-0.6}$& $\times$10$^{-5}$ ph.\,cm$^{-2}$\,s$^{-1}$\\
		MyTorusS&N$_{\rm H}$&9.2$\pm$3.1&10$\pm$2.4& $\times$10$^{22} \rm cm^{-2}$\\
		&N&6.4$\pm$2.0&5.2$\pm$0.9&  $\times$10$^{-2}$  ph.\,keV$^{-1}$\,$\rm cm^{2}$\,s$^{-1}$\\
		zGauss&N$_{\rm6.96~keV}$&1.15$\pm$0.04&0.9$\pm$0.3 &$\times$10$^{-5}$ ph.\,cm$^{-2}$\,s$^{-1}$\\
		zGauss&N$_{\rm6.70~keV}$&1.40$\pm$0.04&1.0$\pm$0.3 &$\times$10$^{-5}$ ph.\,cm$^{-2}$\,s$^{-1}$\\
		\hline
		&F$_{\rm 0.5-2~keV}$&1.3$\pm$0.2 &1.1$\pm$0.3&$\times$10$^{-11}$ erg cm$^{-2}$ s$^{-1}$ \\
		&F$_{\rm 2-10~keV}$&8.6$\pm$0.1 &7.5$\pm$0.01&$\times$10$^{-11}$ erg cm$^{-2}$ s$^{-1}$ \\
	   \end{tabular}
       \end{table}

In accordance with the light curves, the first orbit NGC 2992 showed a higher flux than in orbit 2, accompanied by a spectral shape characterised by a similar photon index of $\Gamma$=1.68$\pm$0.01 and absorbing column density of N$_{\rm H}$=7.8$\pm$0.2 $\times$10$^{21}$ cm$^{-2}$. In a similar fashion, the reflected emission has a compatible flux between the two observations and a rather constant column density N$_{\rm H}\sim$9.6$\times$10$^{22}$ cm$^{-2}$ for the scattered component out of the line of sight. The presence of Compton-thin matter  both along the line of sight and out of the line of sight implies the overall emission spectrum of NGC 2992 being globally Compton-thin. Only upper limits were found for the variable Fe K$\alpha$ red tail while the Fe XXV He-$\alpha$ and Fe XXVI Ly-$\alpha$ are well constrained in both orbits. Despite the fairly acceptable statistics, the model well reproduces the \textit{XMM-Newton} and \textit{NuSTAR} spectra as the high $\chi^2$ is mainly due to residuals between 1 and 2 keV likely resulting from calibration issues.

\subsection{High energy cut-off}
\indent The broadband coverage provided by a simultaneous \textit{XMM-Newton-NuSTAR} exposure is extremely suitable to investigate for the high energy roll-over of the nuclear continuum emission, which provides direct clues on the physical properties of the hot corona. For this reason, here we focus on data belonging to \textit{XMM-Newton} orbit 2 and those from its accompanying \textit{NuSTAR} observation.\\
\indent In Section 4.3, we found the overall emission spectrum of NGC 2992 to be globally Compton thin. We thus further investigate the properties of the X-ray emission in NGC 2992 replacing \textit{MyTorus} with the \textit{Borus} model \cite[][]{2Balokovic18,Balokovic2019,Balokovic2021}. In this model, in fact, the high energy cut-off of the primary continuum is set as a free parameter and is not fixed to 300 keV. In this model, a homogeneous spherical scattering medium is considered to surround the central X-ray source.  Except for Fe, whose relative abundance (A$_{\rm Fe}$) can be derived, a solar abundance is considered. We therefore modelled the simultaneous \textit{XMM-Newton/NuSTAR} orbit 2 replacing the \textit{MyTorus} tables with the \textit{Borus} one (borus02\_v170323a.fits) ending up with the model:

\begin{multline}
\rm tbabs_G \times(apec + Cloudy + (tbabs_z \times cutoffpl) +\\
\rm +Borus+zGauss+zGauss+zGauss).
\end{multline}

\noindent We fit XMM-Newton orbit 2 and NuSTAR data computing the \textit{Borus} normalisation and column density. We assumed the cut-off power law and \textit{Borus} to have the same primary photon index, high energy cut-off and normalisation, hence we tied these parameters between the two models. The iron abundance was set to 1.\\
\indent  These simple steps led to a best-fit of $\chi^2$/d.o.f.=687/563. Based on this model, the primary continuum emission of NGC 2992 has a slope of $\Gamma$=1.67$\pm$0.01 and is absorbed by a column of N$_{\rm H}$=(7.8$\pm$0.1)$\times10^{21}$ cm$^{-2}$. A lower limit for the high energy cut-off is found, E$_{\rm c}$>390 keV, and the reflected emission is due to matter with N$_{\rm H}$=(8.7$\pm$0.4) $\times$ 10$^{22}$ cm$^{-2}$. \textit{Borus} also allows for a Comptonised continuum via the table \textit{borus11\_v190815a.fits}. We thus refitted the spectra adopting this novel table, substituting the cut-off power-law by \textit{nthcomp} \cite[][]{Zdziarski1996,Zycki1999} and tying the equivalent parameters between the two models. The model yielded a best-fit of $\chi^2$=686 for 563 d.o.f, fully compatible with the previous one. Aside from a slightly steeper photon index ($\Gamma$=1.71$\pm$0.01), a very high temperature of kT>115 keV is implied and, under the assumption of a spherical corona, we derived an optical depth of $\tau$<1.2. All the other parameters are consistent with the previous fit.\\
\indent It is worth noticing that irrespectively from the model adopted, matter on/out of the line of sight has a column density smaller than the threshold for Compton-thick regime, this further confirming the emission spectrum of NGC 2992 to be globally Compton-thin.

\section{Very short term spectral properties}
 \cite{Marinucci2020} presented a time resolved spectral analysis on 50 \textit{EPIC-pn} slices (each $\sim$5 ks long). We here perform a further step re-analysing the same spectral chunks adding \textit{NuSTAR} when available and replacing the phenomenological model presented in that paper with the best-fit model discussed in Sect. 4.3. In particular, we seek for short variations of the continuum shape and normalisation and its associated reflected component \textit{MyTorusS}.\\
\begin{figure*}
	\centering
	\includegraphics[width=\textwidth]{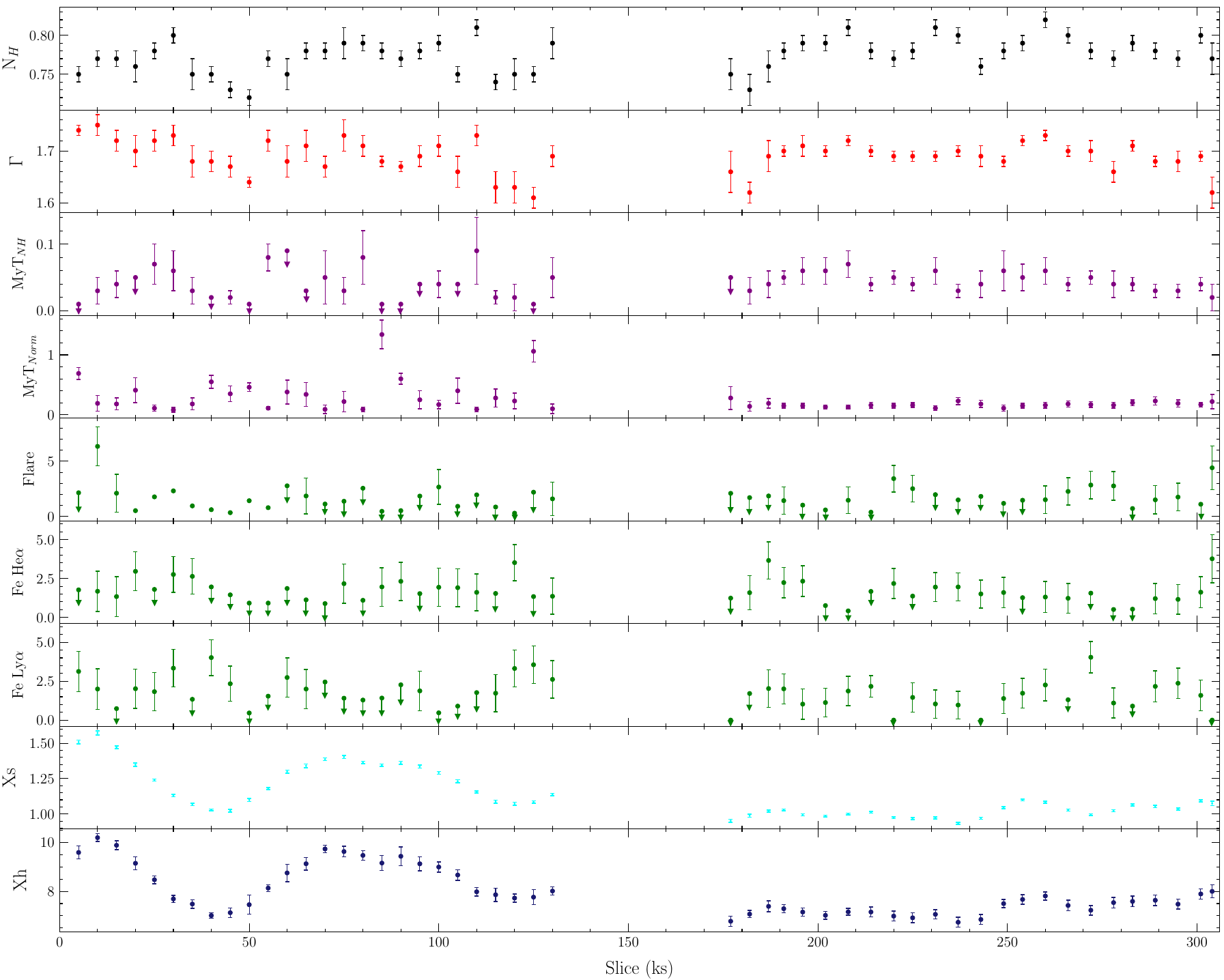}
	\caption{\small{Best-fit for the two \textit{XMM-Newton} orbits and the accompanying \textit{NuSTAR} data. Xs and Xh account for the soft (0.5-2 keV) and hard  (2-10 keV) fluxes, respectively. Units and multiplicative factors for the parameters are given in Table~\ref{xmmtimeresolved}.}}
	\label{xmmtime_resolved}
\end{figure*}
\indent For each spectral chunk, we computed the photon index and the normalisation of the continuum  and, for the reflected component \textit{MyTorus}, the column density and its normalisation. In \cite{Marinucci2020} the presence of the Fe He$\alpha$ and Fe Ly$\alpha$ plus a variable red component has already been presented. To account for these features, we relied on the best-fit model presented in that paper, but we recalculated the line normalisations as we have adopted a different continuum. Finally, for the \textit{apec} and \textit{CLOUDY} components considered the best-fit values from Table~\ref{xmmaverage}, but allowed to vary their normalisations.\\
\indent This fitting strategy provided a good representation for all the spectral slices, see Fig.~\ref{slices1} and Fig.~\ref{slices2}, and allowed us to infer the best fit parameters in Table~\ref{xmmtimeresolved}. The same quantities are also shown in Fig.~\ref{xmmtime_resolved}. From this figure, we notice that prominent flux variability of the soft and hard X-ray bands does not correlate with the absorbing column density, nor the source spectral shape, suggesting that the variability in NGC 2992 is intrinsic, or in other words, it is not driven by absorption changes. We also notice that both $\Gamma$ and N$_{\rm H}$ had a fairly constant value during the observation and, besides a few exceptions, they are consistent with the average values of $\Gamma$=1.68$\pm$0.03 and N$_{\rm H}$=(7.7$\pm$0.2)$\times$10$^{21}$ cm$^{-2}$, respectively. In a similar fashion, the reflected component does not vary significantly between the exposures and is consistent with originating from Compton-thin material, where N$_{\rm H}$=(6.3$\pm$2.4)$\times$10$^{22}$ cm$^{-2}$.

\section{Discussion and conclusions}
\begin{figure*}
	\centering
	\includegraphics[width=\textwidth]{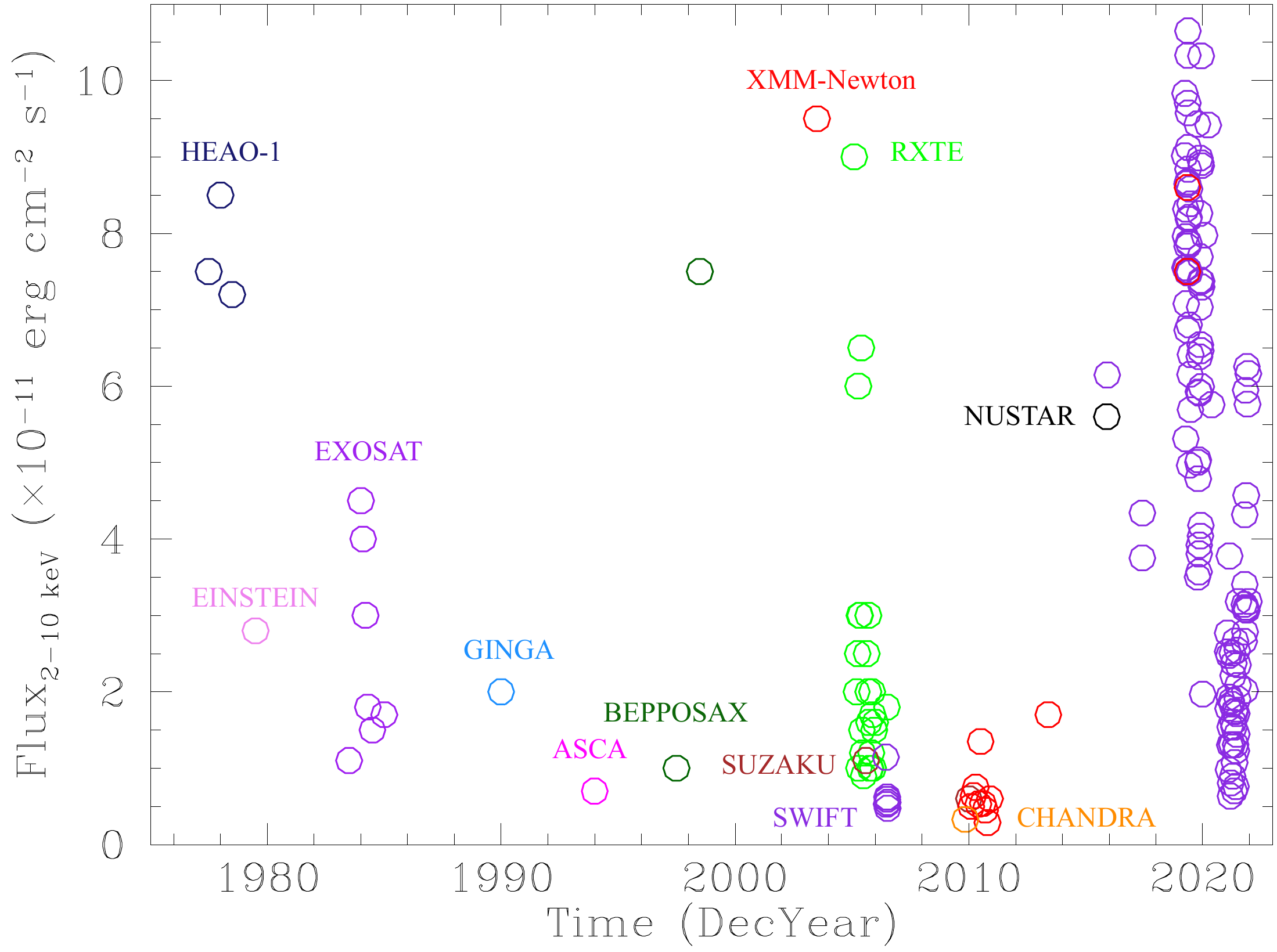}
	\caption{\small{The historical 2-10 keV light curve of NGC 2992 as observed with various X-ray satellites. We address the reader to the paper by \citet{Yaqoob2007} and references therein, for details on data taken before 2008. Data collected after 2008 were presented in \citet[][]{Marinucci2018,Marinucci2020} and this paper. The 2-10 keV flux derived using \textit{NuSTAR} data taken in 2019 is consistent with the one derived from \textit{XMM-Newton} orbit 2, thus we did not included it in this plot.}}
	\label{history}
\end{figure*}

NGC 2992 is a X-ray bright AGN that has been repeatedly observed by all the major X-ray facilities. The hallmark of its emission is the significant variability of the primary continuum that has now a $\sim$40 year long light curve, see Fig. \ref{history}.
In this paper we focused on data taken after 2019 analysing a rich collection of X-ray observations of NGC 2992.
In particular, by exploiting \textit{Swift}, \textit{XMM-Newton} and \textit{NuSTAR} data, we derived both temporal and spectral properties for this source over different timescales, from hours to years. 
In the following, we summarise and discuss our main findings. \\

\begin{figure}
	\centering
	\includegraphics[width=\columnwidth]{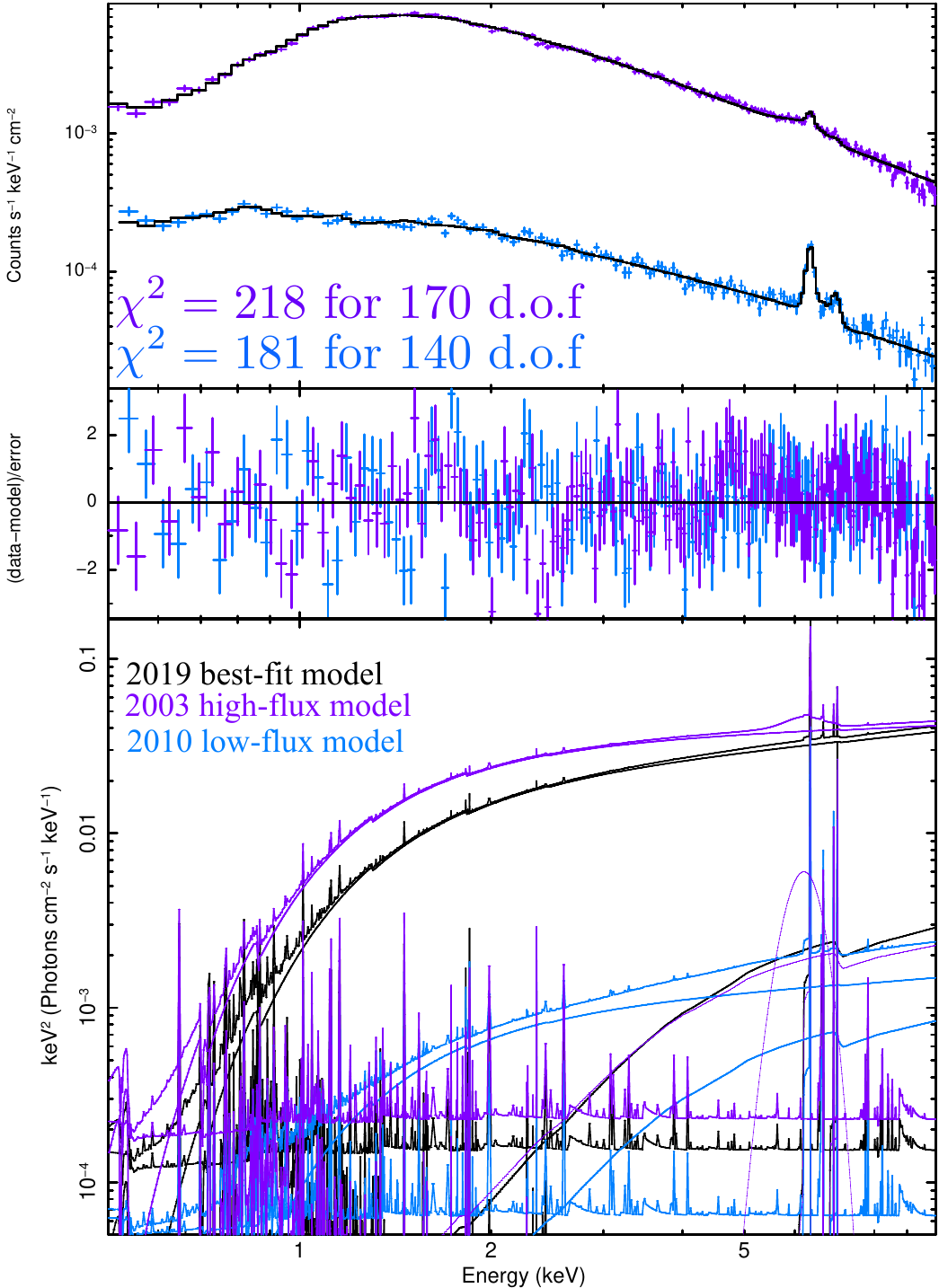}
	\caption{\small{Fit to the high (purple) and low (dodgerblue) flux states data using model 1 presented in Sect. 4.3 and based on the 2019 \textit{XMM-Newton/NuSTAR} exposures.}}
	\label{low}
\end{figure}

\noindent \textbf{Timing properties}: \\
The light curves show the source to vary at significant levels. Daily to yearly variations can be observed in Fig.~\ref{swift_lc}, where the 2-10 keV flux varied by up to a factor of $\sim$5 increasing up to a factor of $\sim$10 when extending the observing time interval to $\sim$1 year. This factor further increases comparing archival observations taken during 2006 with those of 2019. Remarkable flux changes of about $\sim$60\% were observed down to kilo second timescales in the 1-3 keV and 3-10 keV bands, see Fig.~\ref{2992lc}. Interestingly, the larger variations were only observed in the first \textit{XMM-Newton} orbit as, in the second one, smaller amplitude changes were observed. Aside from counts below 1 keV dominated  by a constant ionised component likely emerging from the  Narrow Line Region, the variations in the soft and hard X-rays are of a similar amount. Further evidence of this comes from the consistent shapes of soft and hard SFs in Fig.~\ref{SF}. The lack of a correlation between the hardness ratios and the total flux in Fig.~\ref{swift_ratio} allows us to rule out fast obscuration events to cause the variability which is, instead, intrinsic to the continuum emission dominating the 1-10 keV energy range, see Fig.~\ref{correlation}.\\ 
\indent The two \textit{Swift} monitoring periods also revealed that the source underwent two different variability levels, see Figs.~\ref{SF} and ~\ref{fvar}, where the lower flux state correspond to larger variations. On the other hand, when comparing the excess variance spectra derived from \textit{XMM-Newton}, the opposite trend is observed. In fact, the exposure with larger flux is also the more variable, see Fig.\ref{fvarx}. The former trend has been explained invoking the presence of N randomly flaring sub-units \citep[e.g.][]{Nandra1997,Almaini2000}, and, although it matches the flux changes for the NGC 2992 SFs and in the Fvar spectra (see Fig.~\ref{SF} and Fig.~\ref{fvar}), we notice that log normal flux distribution commonly observed in AGN strongly limits the possible number of uncorrelated active regions \citep[][]{Uttley2005}.\\
\indent Finally, the high S/N  allowed us to fit the excess variance spectra following the prescriptions by \cite{Parker2020}. We successfully tested on our spectra a simple variability model due to the primary continuum with a damping in the Fe K$\alpha$ and in the soft X-ray regions due to the constancy of the reflected and scattered emission. The spectra of both orbits are accounted fairly well by the model. Interestingly, a strong excess around 5 keV is observed in both observations and it is stronger in orbit 2, in agreement with the variable red flaring emission component reported in \cite{Marinucci2020}. From the fit, we found the variance of power-law flux in logarithmic scale to be $\sim$3 times larger in orbit 1 with respect to orbit 2. Moreover, this fit agrees with the nuclear X-rays to dominate the X-ray emission of NGC 2992.\\
\noindent \textbf{Spectral properties}:\\ 
\noindent The main component shaping  the X-ray emission of NGC 2992 is the absorbed primary continuum. The time-average analysis of the two \textit{XMM-Newton} orbits and the \textit{NuSTAR} exposure were consistent with a fairly hard power-law ($\Gamma\sim$1.7) and a line of sight absorbing column density of N$_{\rm H}\sim$8 $\times10^{21}$ cm$^{-2}$, see Table~\ref{xmmaverage} and Fig.\ref{timeaverage}. Our findings are consistent with those based on \textit{XMM-Newton} and \textit{Suzaku} data presented in \cite[][]{Laha2020}.
Moreover, these parameters are fully consistent with those derived on the same data, but via a time resolved analysis, (see Table~\ref{xmmtimeresolved} and Fig.~\ref{xmmtime_resolved}) and also with the values inferred from each of the 123 \textit{Swift} snapshots, Table~\ref{swifttable} and Fig.~\ref{swift_bestfit}. Seeking for the presence of a high energy cut-off we found E$_{\rm cut}$>390 keV. This value is significantly high, especially when compared with the median values for the high energy cut-off of 240 keV and 340 keV as found for a sample of obscured AGN by \cite{Balokovic2021}. The peculiar properties of the NGC 2992 hot corona are further confirmed when testing a Comptonisation continuum where we derive an electron temperature kT>115 keV and an opacity $\tau$<1.2.
The comparison with the coronal properties measured by \cite{Middei2019} reveals the corona in NGC 2992 to be rather extreme, as its properties are compatible only with those of NGC 5506 \cite[][]{Matt2015}. On theoretical grounds, the electron temperature and opacity of the hot corona are responsible for the observed photon index and high energy cut-off, however, despite the large changes in the X-ray flux the spectral properties of the source remain rather constant. The high temperature of the hot plasma in NGC 2992 can possibly explain such a decoupled variability. In accordance with Fig. 6 in \cite{Middei2019} (where iso-$\Gamma$ and iso-E$_{\rm cut-off}$ curves are drawn on the opacity-electron temperature parameter space), to observe a $\Delta\Gamma=0.2$ we need to increase the temperature assuming the coronal opacity not to vary. This increase depends on the actual coronal temperature: if we consider kT=115 keV, consistent with what found in NGC 2992, and kT=50 keV \cite[in agreement with the average temperature of the hot coronae studied in][]{Middei2019}, from $\Gamma=1.7$ to $\Gamma=1.5$ (1.9) we need to increase(decrease) the temperature by a factor of 35\% (25\%) or 20\% (10\%), respectively. Another possible explanation for the decoupled amplitude-spectral variations is discussed in a recent paper \cite[][]{Fernandez2022} where the authors suggest a bulk variation of the Comptonising plasma. For instance, magnetic reconnection events occurring in the close surroundings of the disc may alter the hot plasma \cite[e.g.][]{Poutanen1999,deGouveia2010}. A flare of the hot corona would then boost the number of disc photons affecting the X-ray luminosity but not the spectral shape of the X-ray emission.\\
\indent Distant reflection off an obscuring torus (N$_{\rm H}$=9.6$\pm$2.7) $\times10^{22}$ cm$^{-2}$) accounts for the small spectral curvature in the hard X-rays and its associated narrow Fe K$\alpha$ emission line at 6.4 keV. The reflected spectrum and the Fe K$\alpha$ have a constant behaviour both on hourly and yearly timescales. In fact, no correlation between the \textit{MyTorus} tables and the primary continuum was found and can be ascribed to reflection from cold, distant material, likely the obscuring torus itself. However, the modest broadening of the Fe K$\alpha$ fluorescence line supports the presence of an additional weak (EW$\sim$20 eV) component contributing to the whole flux of this emission feature. This second Fe K$\alpha$ emission component is likely emerging from matter closer with respect to the molecular torus, e.g. the BLR \cite[e.g.][]{Marinucci2018} and may explain the variable emission signature found by \cite{Guolo2021}. We also notice that the K$\alpha$ flux varied by about a factor of 4 compared to the low flux observation in \cite{Marinucci2018}. Moreover it would agree with the findings by \cite{Ghosh2021} that showed the flux of the Fe K$\alpha$ emission line to follow the changes of the primary continuum on yearly timescales. Finally, the 2019 does not require a Fe K$\alpha$ emission line as broad as the one found in the 2003 \textit{XMM-Newton} observation \cite[e.g.][]{Nandra1997,Brenneman2009,Shu2010}.
In the time resolved analysis, the adoption of a single narrow component accounting for this variable fraction of the Fe K$\alpha$ flux is not required by the data, as \textit{MyTorus} already provides a good representation of the Fe K$\alpha$ on 5ks timescales. Finally, transient emission lines have recurrently been observed during the two \textit{XMM-Newton} orbits, see Fig.~\ref{xmmtime_resolved}, and we address the reader to \cite{Marinucci2020} for details.\\
\indent NGC 2992 has been observed at different flux levels (e.g. Fig.~\ref{history}) and in \cite{Marinucci2018} a detailed analysis of XMM-Newton exposures of this object is presented. In order to provide an holistic view of the spectral properties of the central engine in NGC 2992, we tested our 2019 best-fitting model (model 1 presented in Sect. 4.3) on the exposures where the source was found in its lowest and highest states. In particular, we considered the \textit{XMM-Newton} archival observations taken on 2003-05-19 and 2010-11-28 (Obs.IDs. 0147920301\footnote{This exposure, taken in full frame observing mode, is severely affected by pile-up. To mitigate this issue, we used an annular region to extract the source spectrum with r$_{\rm in}$ and r$_{\rm out}$ being 10'' and 40'', respectively.} and 0654910901, respectively). We thus reproduced them directly adopting the best-fit model found for orbit 2 and show it in Fig.~\ref{timeaverage}. We accounted for the different flux states computing the normalisation of the primary emission, its associated reflected component, and the one of the  \textit{apec} and \textit{Cloudy} tables. All the other parameters have been kept fixed to their corresponding best-fit values already quoted in Table~\ref{xmmaverage}. Moreover, to account for the broad emission line found required by the 2003 data, we added a Gaussian component whose width was kept fixed to $\sigma$=400 eV, in accordance with what literature papers \cite[e.g.][]{Nandra1997,Shu2010}. This basic procedure led us to the fits shown in Fig.~\ref{low}, with statistics of $\chi^2$/d.o.f.=218/170 and $\chi^2$/d.o.f.=181/140. In the high flux level, (F$_{\rm 2-10~keV}$=(9.5$\pm$0.1) $\times$10$^{-11}$ erg cm$^{-2}$ s$^{-1}$), the normalisation of the power-law is Norm$_{\rm po}$=(3.00$\pm$0.01)$\times10^{-2}$ ph.\,keV$^{-1}$\, cm$^{2}$ s$^{-1}$, about twice of what found in 2019 while the amount of reflected flux is fully consistent with what found in 2019 as we obtained Norm$_{\rm MyTorus}$=(9.7$\pm$3.7)$\times10^{-2}$ ph.\,keV$^{-1}$\, cm$^{2}$ s$^{-1}$. On the other hand, in the 2010 low flux level exposure (F$_{\rm 2-10~keV}$=(2.9$\pm$0.2) $\times$10$^{-12}$ erg cm$^{-2}$ s$^{-1}$), we found the power-law normalisation to be Norm$_{\rm po}$=(4.7$\pm$0.2)$\times10^{-4}$ ph.\,keV$^{-1}$\, cm$^{2}$ s$^{-1}$ about 20-30 times lower than in 2019. The normalisation of the reflected component modelled using \textit{MyTorus} is Norm$_{\rm MyTorus}$=(9.5$\pm$1.2)$\times10^{-3}$ ph.\,keV$^{-1}$\,$\rm cm^{2}$\,s$^{-1}$, a factor of $\sim$10 less than in 2019.
Therefore, on very long timescales and during a prolonged low state of the source in 2010, the strength of the reflector appears to respond to the continuum. However, the smaller value of the reflected component found in 2010 can be explained by the torus reflecting the primary continuum of NGC 2992 during a low flux state. Observing the light curves in Fig. \ref{swift_lc}, before 2010 NGC 2992 was observed in a very low flux state, even lower than the one in 2021. Such a long term adjustment suggests that reflected spectrum emerges far from the central engine. Finally, in accordance with previous studies, the Fe K$\alpha$ emission line of NGC 2992 has an unresolved component correlating with the primary flux and emerging from the Broad Line Region. However, the current \textit{MyTorus}-based model accounts for whole Fe K$\alpha$ flux (see residuals in Fig.~\ref{low}) and data do not require any additional Gaussian component. Below 1 keV, the non-variable behaviour of the distant scattering off the NLR can be witnessed in the top panels of Fig.~\ref{2992lc} or in the first bin of the excess variance spectra in both Fig.s~\ref{fvar} and ~\ref{fvarx}.\\
\indent In conclusion, the X-ray emission of NGC 2992 is due to a remarkably variable power-law-like continuum, possibly associated with a very hot corona, that is absorbed and reflected by gas whose  N$_{\rm H}$<1.5$\times$10$^{24}$ cm$^{-2}$ (globally Compton-thin). The strong amplitude variations coupled with the very weak spectral changes are somehow suggesting the hot corona in NGC 2992 to be rather peculiar, and additional efforts must be made to clarify the physical properties of this medium. 

\section{Acknowledgements}
We thank the referee for her/his useful comments.
Part of this work is based on archival data, software or online services provided by the Space Science Data Center - ASI. “This work has been partially supported by the ASI-INAF program I/004/11/4. RM acknowledges financial contribution from the agreement ASI-INAF n.2017-14-H.0.  SB acknowledges financial support from ASI under grants ASI-INAF I/037/12/0 and n. 2017-14-H.O. BDM acknowledges support via Ram\'on y Cajal Fellowship RYC2018-025950-I. AL acknowledges support from the HORIZON-2020 grant “Integrated Activities for the High Energy Astrophysics Domain" (AHEAD-2020), G.A. 871158). This  work  is  based  on  observations obtained with: the NuSTAR mission,  a  project  led  by  the  California  Institute  of  Technology,  managed  by  the  Jet  Propulsion  Laboratory  and  funded  by  NASA; XMM-Newton,  an  ESA  science  mission  with  instruments  and  contributions  directly funded  by  ESA  Member  States  and  the  USA  (NASA).
	
	
\section{Data availability}
The \textit{XMM-Newton} data can be downloaded directly from the mission archive at the web page \url{https://www.cosmos.esa.int/web/xmm-newton/xsa}, while \textit{NuSTAR} and \textit{Swift} exposures can be easily downloaded using the \textit{BROWSE} tool at the web pages \url{https://heasarc.gsfc.nasa.gov/cgi-bin/W3Browse/w3browse.pl} or from the official mirror archives hosted by the Space Science Data Center of the Italian Space Agency \url{https://www.ssdc.asi.it/mma.html}.

\bibliographystyle{mnras}

\begin{thebibliography}{}
\makeatletter
\relax
\def\mn@urlcharsother{\let\do\@makeother \do\$\do\&\do\#\do\^\do\_\do\%\do\~}
\def\mn@doi{\begingroup\mn@urlcharsother \@ifnextchar [ {\mn@doi@}
  {\mn@doi@[]}}
\def\mn@doi@[#1]#2{\def\@tempa{#1}\ifx\@tempa\@empty \href
  {http://dx.doi.org/#2} {doi:#2}\else \href {http://dx.doi.org/#2} {#1}\fi
  \endgroup}
\def\mn@eprint#1#2{\mn@eprint@#1:#2::\@nil}
\def\mn@eprint@arXiv#1{\href {http://arxiv.org/abs/#1} {{\tt arXiv:#1}}}
\def\mn@eprint@dblp#1{\href {http://dblp.uni-trier.de/rec/bibtex/#1.xml}
  {dblp:#1}}
\def\mn@eprint@#1:#2:#3:#4\@nil{\def\@tempa {#1}\def\@tempb {#2}\def\@tempc
  {#3}\ifx \@tempc \@empty \let \@tempc \@tempb \let \@tempb \@tempa \fi \ifx
  \@tempb \@empty \def\@tempb {arXiv}\fi \@ifundefined
  {mn@eprint@\@tempb}{\@tempb:\@tempc}{\expandafter \expandafter \csname
  mn@eprint@\@tempb\endcsname \expandafter{\@tempc}}}

\bibitem[\protect\citeauthoryear{{Almaini} et~al.,}{{Almaini}
  et~al.}{2000}]{Almaini2000}
{Almaini} O.,  et~al., 2000, \mn@doi [\mnras]
  {10.1046/j.1365-8711.2000.03385.x}, \href
  {https://ui.adsabs.harvard.edu/abs/2000MNRAS.315..325A} {315, 325}

\bibitem[\protect\citeauthoryear{{Alston} et~al.,}{{Alston}
  et~al.}{2019}]{Alston2019}
{Alston} W.~N.,  et~al., 2019, \mn@doi [\mnras] {10.1093/mnras/sty2527}, \href
  {https://ui.adsabs.harvard.edu/abs/2019MNRAS.482.2088A} {482, 2088}

\bibitem[\protect\citeauthoryear{{Arnaud}}{{Arnaud}}{1996}]{Arnaud1996}
{Arnaud} K.~A.,  1996, {XSPEC: The First Ten Years}.
p.~17

\bibitem[\protect\citeauthoryear{{Awaki}, {Koyama}, {Inoue}  \&
  {Halpern}}{{Awaki} et~al.}{1991}]{Awaki1991}
{Awaki} H.,  {Koyama} K.,  {Inoue} H.,   {Halpern} J.~P.,  1991, \pasj, \href
  {https://ui.adsabs.harvard.edu/abs/1991PASJ...43..195A} {43, 195}

\bibitem[\protect\citeauthoryear{{Awaki}, {Ueno}, {Taniguchi}  \&
  {Weaver}}{{Awaki} et~al.}{2000}]{Awaki2000}
{Awaki} H.,  {Ueno} S.,  {Taniguchi} Y.,   {Weaver} K.~A.,  2000, \mn@doi
  [\apj] {10.1086/309516}, \href
  {https://ui.adsabs.harvard.edu/abs/2000ApJ...542..175A} {542, 175}

\bibitem[\protect\citeauthoryear{{Balokovi{\'c}} et~al.,}{{Balokovi{\'c}}
  et~al.}{2018}]{2Balokovic18}
{Balokovi{\'c}} M.,  et~al., 2018, \mn@doi [\apj] {10.3847/1538-4357/aaa7eb},
  \href {https://ui.adsabs.harvard.edu/abs/2018ApJ...854...42B} {854, 42}

\bibitem[\protect\citeauthoryear{{Balokovi{\'c}}, {Garc{\'\i}a}  \&
  {Cabral}}{{Balokovi{\'c}} et~al.}{2019}]{Balokovic2019}
{Balokovi{\'c}} M.,  {Garc{\'\i}a} J.~A.,   {Cabral} S.~E.,  2019, \mn@doi
  [Research Notes of the American Astronomical Society]
  {10.3847/2515-5172/ab578e}, \href
  {https://ui.adsabs.harvard.edu/abs/2019RNAAS...3..173B} {3, 173}

\bibitem[\protect\citeauthoryear{{Balokovi{\'c}}, {Cabral}, {Brenneman}  \&
  {Urry}}{{Balokovi{\'c}} et~al.}{2021}]{Balokovic2021}
{Balokovi{\'c}} M.,  {Cabral} S.~E.,  {Brenneman} L.,   {Urry} C.~M.,  2021,
  \mn@doi [\apj] {10.3847/1538-4357/abff4d}, \href
  {https://ui.adsabs.harvard.edu/abs/2021ApJ...916...90B} {916, 90}

\bibitem[\protect\citeauthoryear{{Barr} \& {Mushotzky}}{{Barr} \&
  {Mushotzky}}{1986}]{Barr1986}
{Barr} P.,  {Mushotzky} R.~F.,  1986, \mn@doi [\nat] {10.1038/320421a0}, \href
  {https://ui.adsabs.harvard.edu/abs/1986Natur.320..421B} {320, 421}

\bibitem[\protect\citeauthoryear{{Bauer}, {Baltay}, {Coppi}, {Ellman}, {Jerke},
  {Rabinowitz}  \& {Scalzo}}{{Bauer} et~al.}{2009}]{Bauer2009}
{Bauer} A.,  {Baltay} C.,  {Coppi} P.,  {Ellman} N.,  {Jerke} J.,  {Rabinowitz}
  D.,   {Scalzo} R.,  2009, \mn@doi [\apj] {10.1088/0004-637X/696/2/1241},
  \href {https://ui.adsabs.harvard.edu/abs/2009ApJ...696.1241B} {696, 1241}

\bibitem[\protect\citeauthoryear{{Beloborodov}}{{Beloborodov}}{1999}]{Belo99}
{Beloborodov} A.~M.,  1999, in {Poutanen} J.,  {Svensson} R.,  eds,
  Astronomical Society of the Pacific Conference Series Vol. 161, High Energy
  Processes in Accreting Black Holes. p.~295 (\mn@eprint {} {astro-ph/9901108})

\bibitem[\protect\citeauthoryear{{Bianchi}, {Guainazzi}  \&
  {Chiaberge}}{{Bianchi} et~al.}{2006}]{Bianchi2006}
{Bianchi} S.,  {Guainazzi} M.,   {Chiaberge} M.,  2006, \mn@doi [\aap]
  {10.1051/0004-6361:20054091}, \href
  {https://ui.adsabs.harvard.edu/abs/2006A&A...448..499B} {448, 499}

\bibitem[\protect\citeauthoryear{{Bianchi}, {Guainazzi}, {Matt}, {Fonseca
  Bonilla}  \& {Ponti}}{{Bianchi} et~al.}{2009}]{Bian09}
{Bianchi} S.,  {Guainazzi} M.,  {Matt} G.,  {Fonseca Bonilla} N.,   {Ponti} G.,
   2009, \mn@doi [\aap] {10.1051/0004-6361:200810620}, \href
  {http://adsabs.harvard.edu/abs/2009A%26A...495..421B} {495, 421}

\bibitem[\protect\citeauthoryear{{Bianchi}, {Chiaberge}, {Evans}, {Guainazzi},
  {Baldi}, {Matt}  \& {Piconcelli}}{{Bianchi} et~al.}{2010}]{Bianchi2010}
{Bianchi} S.,  {Chiaberge} M.,  {Evans} D.~A.,  {Guainazzi} M.,  {Baldi} R.~D.,
   {Matt} G.,   {Piconcelli} E.,  2010, \mn@doi [\mnras]
  {10.1111/j.1365-2966.2010.16475.x}, \href
  {https://ui.adsabs.harvard.edu/abs/2010MNRAS.405..553B} {405, 553}

\bibitem[\protect\citeauthoryear{{Brenneman} \& {Reynolds}}{{Brenneman} \&
  {Reynolds}}{2009}]{Brenneman2009}
{Brenneman} L.~W.,  {Reynolds} C.~S.,  2009, \mn@doi [\apj]
  {10.1088/0004-637X/702/2/1367}, \href
  {https://ui.adsabs.harvard.edu/abs/2009ApJ...702.1367B} {702, 1367}

\bibitem[\protect\citeauthoryear{{Cappi} et~al.,}{{Cappi}
  et~al.}{1999}]{Cappi1999}
{Cappi} M.,  et~al., 1999, \aap, \href
  {https://ui.adsabs.harvard.edu/abs/1999A&A...344..857C} {344, 857}

\bibitem[\protect\citeauthoryear{{Cash}}{{Cash}}{1979}]{Cash1979}
{Cash} W.,  1979, \mn@doi [\apj] {10.1086/156922}, \href
  {https://ui.adsabs.harvard.edu/abs/1979ApJ...228..939C} {228, 939}

\bibitem[\protect\citeauthoryear{{Colbert}, {Strickland}, {Veilleux}  \&
  {Weaver}}{{Colbert} et~al.}{2005}]{Colbert2005}
{Colbert} E. J.~M.,  {Strickland} D.~K.,  {Veilleux} S.,   {Weaver} K.~A.,
  2005, \mn@doi [\apj] {10.1086/430734}, \href
  {https://ui.adsabs.harvard.edu/abs/2005ApJ...628..113C} {628, 113}

\bibitem[\protect\citeauthoryear{{Dadina}}{{Dadina}}{2007}]{Dadina07}
{Dadina} M.,  2007, \mn@doi [\aap] {10.1051/0004-6361:20065734}, \href
  {http://adsabs.harvard.edu/abs/2007A%26A...461.1209D} {461, 1209}

\bibitem[\protect\citeauthoryear{{De Marco} et~al.,}{{De Marco}
  et~al.}{2020}]{DeMarco2020}
{De Marco} B.,  et~al., 2020, \mn@doi [\aap] {10.1051/0004-6361/201936470},
  \href {https://ui.adsabs.harvard.edu/abs/2020A&A...634A..65D} {634, A65}

\bibitem[\protect\citeauthoryear{{Edelson}, {Turner}, {Pounds}, {Vaughan},
  {Markowitz}, {Marshall}, {Dobbie}  \& {Warwick}}{{Edelson}
  et~al.}{2002}]{Edelson2002}
{Edelson} R.,  {Turner} T.~J.,  {Pounds} K.,  {Vaughan} S.,  {Markowitz} A.,
  {Marshall} H.,  {Dobbie} P.,   {Warwick} R.,  2002, \mn@doi [\apj]
  {10.1086/323779}, \href
  {https://ui.adsabs.harvard.edu/abs/2002ApJ...568..610E} {568, 610}

\bibitem[\protect\citeauthoryear{{Edelson} et~al.,}{{Edelson}
  et~al.}{2015}]{Edelson2015}
{Edelson} R.,  et~al., 2015, \mn@doi [\apj] {10.1088/0004-637X/806/1/129},
  \href {https://ui.adsabs.harvard.edu/abs/2015ApJ...806..129E} {806, 129}

\bibitem[\protect\citeauthoryear{{Fabian}, {Nandra}, {Reynolds}, {Brandt},
  {Otani}, {Tanaka}, {Inoue}  \& {Iwasawa}}{{Fabian} et~al.}{1995}]{Fabian1995}
{Fabian} A.~C.,  {Nandra} K.,  {Reynolds} C.~S.,  {Brandt} W.~N.,  {Otani} C.,
  {Tanaka} Y.,  {Inoue} H.,   {Iwasawa} K.,  1995, \mn@doi [\mnras]
  {10.1093/mnras/277.1.L11}, \href
  {https://ui.adsabs.harvard.edu/abs/1995MNRAS.277L..11F} {277, L11}

\bibitem[\protect\citeauthoryear{{Fabian}, {Lohfink}, {Kara}, {Parker},
  {Vasudevan}  \& {Reynolds}}{{Fabian} et~al.}{2015}]{Fabi15}
{Fabian} A.~C.,  {Lohfink} A.,  {Kara} E.,  {Parker} M.~L.,  {Vasudevan} R.,
  {Reynolds} C.~S.,  2015, \mn@doi [\mnras] {10.1093/mnras/stv1218}, \href
  {http://adsabs.harvard.edu/abs/2015MNRAS.451.4375F} {451, 4375}

\bibitem[\protect\citeauthoryear{{Fabian}, {Lohfink}, {Belmont}, {Malzac}  \&
  {Coppi}}{{Fabian} et~al.}{2017}]{Fabi17}
{Fabian} A.~C.,  {Lohfink} A.,  {Belmont} R.,  {Malzac} J.,   {Coppi} P.,
  2017, \mn@doi [\mnras] {10.1093/mnras/stx221}, \href
  {http://adsabs.harvard.edu/abs/2017MNRAS.467.2566F} {467, 2566}

\bibitem[\protect\citeauthoryear{{Falocco} et~al.,}{{Falocco}
  et~al.}{2017}]{Falocco2017}
{Falocco} S.,  et~al., 2017, \mn@doi [\aap] {10.1051/0004-6361/201731722},
  \href {https://ui.adsabs.harvard.edu/abs/2017A&A...608A..32F} {608, A32}

\bibitem[\protect\citeauthoryear{{Ferland} et~al.,}{{Ferland}
  et~al.}{2017}]{Ferland2017}
{Ferland} G.~J.,  et~al., 2017, \rmxaa, \href
  {https://ui.adsabs.harvard.edu/abs/2017RMxAA..53..385F} {53, 385}

\bibitem[\protect\citeauthoryear{{Fernandez}, {Secrest}, {Johnson}, {Schmitt},
  {Fischer}, {Cigan}  \& {Dorland}}{{Fernandez} et~al.}{2022}]{Fernandez2022}
{Fernandez} L.~C.,  {Secrest} N.~J.,  {Johnson} M.~C.,  {Schmitt} H.~R.,
  {Fischer} T.~C.,  {Cigan} P.~J.,   {Dorland} B.~N.,  2022, arXiv e-prints,
  \href {https://ui.adsabs.harvard.edu/abs/2022arXiv220105152F} {p.
  arXiv:2201.05152}

\bibitem[\protect\citeauthoryear{{Gallo}, {Blue}, {Grupe}, {Komossa}  \&
  {Wilkins}}{{Gallo} et~al.}{2018}]{Gallo2018}
{Gallo} L.~C.,  {Blue} D.~M.,  {Grupe} D.,  {Komossa} S.,   {Wilkins} D.~R.,
  2018, \mn@doi [\mnras] {10.1093/mnras/sty1134}, \href
  {https://ui.adsabs.harvard.edu/abs/2018MNRAS.478.2557G} {478, 2557}

\bibitem[\protect\citeauthoryear{{Ghosh} \& {Pal}}{{Ghosh} \&
  {Pal}}{2021}]{Ghosh2021}
{Ghosh} R.,  {Pal} M.,  2021, \mn@doi [Research Notes of the American
  Astronomical Society] {10.3847/2515-5172/abe863}, \href
  {https://ui.adsabs.harvard.edu/abs/2021RNAAS...5...35G} {5, 35}

\bibitem[\protect\citeauthoryear{{Green}, {McHardy}  \& {Lehto}}{{Green}
  et~al.}{1993}]{Green1993}
{Green} A.~R.,  {McHardy} I.~M.,   {Lehto} H.~J.,  1993, \mn@doi [\mnras]
  {10.1093/mnras/265.3.664}, \href
  {https://ui.adsabs.harvard.edu/abs/1993MNRAS.265..664G} {265, 664}

\bibitem[\protect\citeauthoryear{{Guainazzi} \& {Bianchi}}{{Guainazzi} \&
  {Bianchi}}{2007}]{Guainazzi2007}
{Guainazzi} M.,  {Bianchi} S.,  2007, \mn@doi [\mnras]
  {10.1111/j.1365-2966.2006.11229.x}, \href
  {https://ui.adsabs.harvard.edu/abs/2007MNRAS.374.1290G} {374, 1290}

\bibitem[\protect\citeauthoryear{{Guolo}, {Ruschel-Dutra}, {Grupe}, {Peterson},
  {Storchi-Bergmann}, {Schimoia}, {Nemmen}  \& {Robinson}}{{Guolo}
  et~al.}{2021}]{Guolo2021}
{Guolo} M.,  {Ruschel-Dutra} D.,  {Grupe} D.,  {Peterson} B.~M.,
  {Storchi-Bergmann} T.,  {Schimoia} J.,  {Nemmen} R.,   {Robinson} A.,  2021,
  \mn@doi [\mnras] {10.1093/mnras/stab2550}, \href
  {https://ui.adsabs.harvard.edu/abs/2021MNRAS.508..144G} {508, 144}

\bibitem[\protect\citeauthoryear{{Haardt} \& {Maraschi}}{{Haardt} \&
  {Maraschi}}{1991}]{haar91}
{Haardt} F.,  {Maraschi} L.,  1991, \mn@doi [\apjl] {10.1086/186171}, \href
  {http://adsabs.harvard.edu/abs/1991ApJ...380L..51H} {380, L51}

\bibitem[\protect\citeauthoryear{{Haardt} \& {Maraschi}}{{Haardt} \&
  {Maraschi}}{1993}]{haar93}
{Haardt} F.,  {Maraschi} L.,  1993, \mn@doi [\apj] {10.1086/173020}, \href
  {http://adsabs.harvard.edu/abs/1993ApJ...413..507H} {413, 507}

\bibitem[\protect\citeauthoryear{{Harrison} et~al.,}{{Harrison}
  et~al.}{2013}]{Harr13}
{Harrison} F.~A.,  et~al., 2013, \mn@doi [\apj] {10.1088/0004-637X/770/2/103},
  \href {http://adsabs.harvard.edu/abs/2013ApJ...770..103H} {770, 103}

\bibitem[\protect\citeauthoryear{{Igo} et~al.,}{{Igo} et~al.}{2020}]{Igo2020}
{Igo} Z.,  et~al., 2020, \mn@doi [\mnras] {10.1093/mnras/staa265}, \href
  {https://ui.adsabs.harvard.edu/abs/2020MNRAS.493.1088I} {493, 1088}

\bibitem[\protect\citeauthoryear{{Keel}}{{Keel}}{1996}]{Keel1996}
{Keel} W.~C.,  1996, \mn@doi [\apjs] {10.1086/192326}, \href
  {https://ui.adsabs.harvard.edu/abs/1996ApJS..106...27K} {106, 27}

\bibitem[\protect\citeauthoryear{{Laha} \& {Ghosh}}{{Laha} \&
  {Ghosh}}{2021}]{Laha2021b}
{Laha} S.,  {Ghosh} R.,  2021, \mn@doi [\apj] {10.3847/1538-4357/abfc56}, \href
  {https://ui.adsabs.harvard.edu/abs/2021ApJ...915...93L} {915, 93}

\bibitem[\protect\citeauthoryear{{Laha}, {Markowitz}, {Krumpe}, {Nikutta},
  {Rothschild}  \& {Saha}}{{Laha} et~al.}{2020}]{Laha2020}
{Laha} S.,  {Markowitz} A.~G.,  {Krumpe} M.,  {Nikutta} R.,  {Rothschild} R.,
  {Saha} T.,  2020, \mn@doi [\apj] {10.3847/1538-4357/ab92ab}, \href
  {https://ui.adsabs.harvard.edu/abs/2020ApJ...897...66L} {897, 66}

\bibitem[\protect\citeauthoryear{{Laha} et~al.,}{{Laha}
  et~al.}{2022}]{Laha2022arXiv}
{Laha} S.,  et~al., 2022, arXiv e-prints, \href
  {https://ui.adsabs.harvard.edu/abs/2022arXiv220307446L} {p. arXiv:2203.07446}

\bibitem[\protect\citeauthoryear{{Laurenti}, {Vagnetti}, {Middei}  \&
  {Paolillo}}{{Laurenti} et~al.}{2020}]{Laurenti2020}
{Laurenti} M.,  {Vagnetti} F.,  {Middei} R.,   {Paolillo} M.,  2020, \mn@doi
  [\mnras] {10.1093/mnras/staa3172}, \href
  {https://ui.adsabs.harvard.edu/abs/2020MNRAS.tmp.3249L} {}

\bibitem[\protect\citeauthoryear{{Lawrence} \& {Papadakis}}{{Lawrence} \&
  {Papadakis}}{1993}]{Lawrence1993}
{Lawrence} A.,  {Papadakis} I.,  1993, \mn@doi [\apjl] {10.1086/187002}, \href
  {https://ui.adsabs.harvard.edu/abs/1993ApJ...414L..85L} {414, L85}

\bibitem[\protect\citeauthoryear{{Lusso} \& {Risaliti}}{{Lusso} \&
  {Risaliti}}{2017}]{Lusso2017}
{Lusso} E.,  {Risaliti} G.,  2017, \mn@doi [\aap]
  {10.1051/0004-6361/201630079}, \href
  {https://ui.adsabs.harvard.edu/abs/2017A&A...602A..79L} {602, A79}

\bibitem[\protect\citeauthoryear{{Lusso} et~al.,}{{Lusso}
  et~al.}{2010}]{Lusso2010}
{Lusso} E.,  et~al., 2010, \mn@doi [\aap] {10.1051/0004-6361/200913298}, \href
  {https://ui.adsabs.harvard.edu/abs/2010A&A...512A..34L} {512, A34}

\bibitem[\protect\citeauthoryear{{Madejski} et~al.,}{{Madejski}
  et~al.}{1995}]{Madejski1995}
{Madejski} G.~M.,  et~al., 1995, \mn@doi [\apj] {10.1086/175111}, \href
  {https://ui.adsabs.harvard.edu/abs/1995ApJ...438..672M} {438, 672}

\bibitem[\protect\citeauthoryear{{Malizia}, {Molina}, {Bassani}, {Stephen},
  {Bazzano}, {Ubertini}  \& {Bird}}{{Malizia} et~al.}{2014}]{Mali14}
{Malizia} A.,  {Molina} M.,  {Bassani} L.,  {Stephen} J.~B.,  {Bazzano} A.,
  {Ubertini} P.,   {Bird} A.~J.,  2014, \mn@doi [\apjl]
  {10.1088/2041-8205/782/2/L25}, \href
  {http://adsabs.harvard.edu/abs/2014ApJ...782L..25M} {782, L25}

\bibitem[\protect\citeauthoryear{{Marinucci}, {Bianchi}, {Matt}, {Fabian},
  {Iwasawa}, {Miniutti}  \& {Piconcelli}}{{Marinucci}
  et~al.}{2011}]{Marinucci2011}
{Marinucci} A.,  {Bianchi} S.,  {Matt} G.,  {Fabian} A.~C.,  {Iwasawa} K.,
  {Miniutti} G.,   {Piconcelli} E.,  2011, \mn@doi [\aap]
  {10.1051/0004-6361/201015358}, \href
  {https://ui.adsabs.harvard.edu/abs/2011A&A...526A..36M} {526, A36}

\bibitem[\protect\citeauthoryear{{Marinucci}, {Bianchi}, {Fabbiano}, {Matt},
  {Risaliti}, {Nardini}  \& {Wang}}{{Marinucci} et~al.}{2017}]{Marinucci2017}
{Marinucci} A.,  {Bianchi} S.,  {Fabbiano} G.,  {Matt} G.,  {Risaliti} G.,
  {Nardini} E.,   {Wang} J.,  2017, \mn@doi [\mnras] {10.1093/mnras/stx1551},
  \href {https://ui.adsabs.harvard.edu/abs/2017MNRAS.470.4039M} {470, 4039}

\bibitem[\protect\citeauthoryear{{Marinucci}, {Bianchi}, {Braito}, {Matt},
  {Nardini}  \& {Reeves}}{{Marinucci} et~al.}{2018}]{Marinucci2018}
{Marinucci} A.,  {Bianchi} S.,  {Braito} V.,  {Matt} G.,  {Nardini} E.,
  {Reeves} J.,  2018, \mn@doi [\mnras] {10.1093/mnras/sty1436}, \href
  {https://ui.adsabs.harvard.edu/abs/2018MNRAS.478.5638M} {478, 5638}

\bibitem[\protect\citeauthoryear{{Marinucci}, {Bianchi}, {Braito}, {De Marco},
  {Matt}, {Middei}, {Nardini}  \& {Reeves}}{{Marinucci}
  et~al.}{2020}]{Marinucci2020}
{Marinucci} A.,  {Bianchi} S.,  {Braito} V.,  {De Marco} B.,  {Matt} G.,
  {Middei} R.,  {Nardini} E.,   {Reeves} J.~N.,  2020, \mn@doi [\mnras]
  {10.1093/mnras/staa1683}, \href
  {https://ui.adsabs.harvard.edu/abs/2020MNRAS.496.3412M} {496, 3412}

\bibitem[\protect\citeauthoryear{{Matt}}{{Matt}}{2002}]{Matt2002}
{Matt} G.,  2002, \mn@doi [Philosophical Transactions of the Royal Society of
  London Series A] {10.1098/rsta.2002.1052}, \href
  {https://ui.adsabs.harvard.edu/abs/2002RSPTA.360.2045M} {360, 2045}

\bibitem[\protect\citeauthoryear{{Matt}, {Fabian}  \& {Ross}}{{Matt}
  et~al.}{1993}]{Matt93}
{Matt} G.,  {Fabian} A.~C.,   {Ross} R.~R.,  1993, \mn@doi [\mnras]
  {10.1093/mnras/262.1.179}, \href
  {http://adsabs.harvard.edu/abs/1993MNRAS.262..179M} {262, 179}

\bibitem[\protect\citeauthoryear{{Matt} et~al.,}{{Matt}
  et~al.}{2015}]{Matt2015}
{Matt} G.,  et~al., 2015, \mn@doi [\mnras] {10.1093/mnras/stu2653}, \href
  {https://ui.adsabs.harvard.edu/abs/2015MNRAS.447.3029M} {447, 3029}

\bibitem[\protect\citeauthoryear{{Matzeu}, {Reeves}, {Nardini}, {Braito},
  {Costa}, {Tombesi}  \& {Gofford}}{{Matzeu} et~al.}{2016}]{Matzeu2016mnras}
{Matzeu} G.~A.,  {Reeves} J.~N.,  {Nardini} E.,  {Braito} V.,  {Costa} M.~T.,
  {Tombesi} F.,   {Gofford} J.,  2016, \mn@doi [\mnras] {10.1093/mnras/stw354},
  \href {https://ui.adsabs.harvard.edu/abs/2016MNRAS.458.1311M} {458, 1311}

\bibitem[\protect\citeauthoryear{{Matzeu}, {Reeves}, {Nardini}, {Braito},
  {Turner}  \& {Costa}}{{Matzeu} et~al.}{2017}]{Matzeu2017}
{Matzeu} G.~A.,  {Reeves} J.~N.,  {Nardini} E.,  {Braito} V.,  {Turner} T.~J.,
   {Costa} M.~T.,  2017, \mn@doi [\mnras] {10.1093/mnras/stw2673}, \href
  {https://ui.adsabs.harvard.edu/abs/2017MNRAS.465.2804M} {465, 2804}

\bibitem[\protect\citeauthoryear{{McHardy}, {Koerding}, {Knigge}, {Uttley}  \&
  {Fender}}{{McHardy} et~al.}{2006}]{McHardy2006}
{McHardy} I.~M.,  {Koerding} E.,  {Knigge} C.,  {Uttley} P.,   {Fender} R.~P.,
  2006, \mn@doi [\nat] {10.1038/nature05389}, \href
  {https://ui.adsabs.harvard.edu/abs/2006Natur.444..730M} {444, 730}

\bibitem[\protect\citeauthoryear{{Mehdipour}, {Kaastra}  \&
  {Kallman}}{{Mehdipour} et~al.}{2016}]{Mehdipour2016}
{Mehdipour} M.,  {Kaastra} J.~S.,   {Kallman} T.,  2016, \mn@doi [\aap]
  {10.1051/0004-6361/201628721}, \href
  {https://ui.adsabs.harvard.edu/abs/2016A&A...596A..65M} {596, A65}

\bibitem[\protect\citeauthoryear{{Middei}, {Vagnetti}, {Bianchi}, {La Franca},
  {Paolillo}  \& {Ursini}}{{Middei} et~al.}{2017}]{Middei2017}
{Middei} R.,  {Vagnetti} F.,  {Bianchi} S.,  {La Franca} F.,  {Paolillo} M.,
  {Ursini} F.,  2017, \mn@doi [\aap] {10.1051/0004-6361/201629940}, \href
  {https://ui.adsabs.harvard.edu/abs/2017A&A...599A..82M} {599, A82}

\bibitem[\protect\citeauthoryear{{Middei}, {Bianchi}, {Marinucci}, {Matt},
  {Petrucci}, {Tamborra}  \& {Tortosa}}{{Middei} et~al.}{2019}]{Middei2019}
{Middei} R.,  {Bianchi} S.,  {Marinucci} A.,  {Matt} G.,  {Petrucci} P.~O.,
  {Tamborra} F.,   {Tortosa} A.,  2019, \mn@doi [\aap]
  {10.1051/0004-6361/201935881}, \href
  {https://ui.adsabs.harvard.edu/abs/2019A&A...630A.131M} {630, A131}

\bibitem[\protect\citeauthoryear{{Middei} et~al.,}{{Middei}
  et~al.}{2021}]{Middei2021}
{Middei} R.,  et~al., 2021, \mn@doi [\aap] {10.1051/0004-6361/202039984}, \href
  {https://ui.adsabs.harvard.edu/abs/2021A&A...647A.102M} {647, A102}

\bibitem[\protect\citeauthoryear{{Miller} et~al.,}{{Miller}
  et~al.}{2015}]{Miller2015}
{Miller} J.~M.,  et~al., 2015, \mn@doi [\nat] {10.1038/nature15708}, \href
  {https://ui.adsabs.harvard.edu/abs/2015Natur.526..542M} {526, 542}

\bibitem[\protect\citeauthoryear{{Molendi}, {Bianchi}  \& {Matt}}{{Molendi}
  et~al.}{2003}]{Molendi2003}
{Molendi} S.,  {Bianchi} S.,   {Matt} G.,  2003, \mn@doi [\mnras]
  {10.1046/j.1365-8711.2003.06783.x}, \href
  {https://ui.adsabs.harvard.edu/abs/2003MNRAS.343L...1M} {343, L1}

\bibitem[\protect\citeauthoryear{{Molina} et~al.,}{{Molina}
  et~al.}{2009}]{Molina09}
{Molina} M.,  et~al., 2009, \mn@doi [\mnras]
  {10.1111/j.1365-2966.2009.15257.x}, \href
  {http://adsabs.harvard.edu/abs/2009MNRAS.399.1293M} {399, 1293}

\bibitem[\protect\citeauthoryear{{Molina}, {Bassani}, {Malizia}, {Stephen},
  {Bird}, {Bazzano}  \& {Ubertini}}{{Molina} et~al.}{2013}]{Molina13}
{Molina} M.,  {Bassani} L.,  {Malizia} A.,  {Stephen} J.~B.,  {Bird} A.~J.,
  {Bazzano} A.,   {Ubertini} P.,  2013, \mn@doi [\mnras]
  {10.1093/mnras/stt844}, \href
  {http://adsabs.harvard.edu/abs/2013MNRAS.433.1687M} {433, 1687}

\bibitem[\protect\citeauthoryear{{Murphy} \& {Yaqoob}}{{Murphy} \&
  {Yaqoob}}{2009}]{Murphy2009}
{Murphy} K.~D.,  {Yaqoob} T.,  2009, \mn@doi [\mnras]
  {10.1111/j.1365-2966.2009.15025.x}, \href
  {https://ui.adsabs.harvard.edu/abs/2009MNRAS.397.1549M} {397, 1549}

\bibitem[\protect\citeauthoryear{{Nandra}}{{Nandra}}{2001}]{Nandra2001}
{Nandra} K.,  2001, \mn@doi [Advances in Space Research]
  {10.1016/S0273-1177(01)00409-4}, \href
  {https://ui.adsabs.harvard.edu/abs/2001AdSpR..28..295N} {28, 295}

\bibitem[\protect\citeauthoryear{{Nandra}, {George}, {Mushotzky}, {Turner}  \&
  {Yaqoob}}{{Nandra} et~al.}{1997}]{Nandra1997}
{Nandra} K.,  {George} I.~M.,  {Mushotzky} R.~F.,  {Turner} T.~J.,   {Yaqoob}
  T.,  1997, \mn@doi [\apj] {10.1086/303600}, \href
  {https://ui.adsabs.harvard.edu/abs/1997ApJ...476...70N} {476, 70}

\bibitem[\protect\citeauthoryear{{O'Neill}, {Nandra}, {Papadakis}  \&
  {Turner}}{{O'Neill} et~al.}{2005}]{O'Neill2005}
{O'Neill} P.~M.,  {Nandra} K.,  {Papadakis} I.~E.,   {Turner} T.~J.,  2005,
  \mn@doi [\mnras] {10.1111/j.1365-2966.2005.08860.x}, \href
  {https://ui.adsabs.harvard.edu/abs/2005MNRAS.358.1405O} {358, 1405}

\bibitem[\protect\citeauthoryear{{Padovani} et~al.,}{{Padovani}
  et~al.}{2017}]{Padovani2017}
{Padovani} P.,  et~al., 2017, \mn@doi [\aapr] {10.1007/s00159-017-0102-9},
  \href {https://ui.adsabs.harvard.edu/abs/2017A&ARv..25....2P} {25, 2}

\bibitem[\protect\citeauthoryear{{Paolillo} et~al.,}{{Paolillo}
  et~al.}{2017}]{Paolillo2017}
{Paolillo} M.,  et~al., 2017, \mn@doi [\mnras] {10.1093/mnras/stx1761}, \href
  {https://ui.adsabs.harvard.edu/abs/2017MNRAS.471.4398P} {471, 4398}

\bibitem[\protect\citeauthoryear{{Papadakis}}{{Papadakis}}{2004}]{Papadakis2004}
{Papadakis} I.~E.,  2004, \mn@doi [\mnras] {10.1111/j.1365-2966.2004.07351.x},
  \href {https://ui.adsabs.harvard.edu/abs/2004MNRAS.348..207P} {348, 207}

\bibitem[\protect\citeauthoryear{{Papadakis}, {Chatzopoulos}, {Athanasiadis},
  {Markowitz}  \& {Georgantopoulos}}{{Papadakis} et~al.}{2008}]{Papadakis2008}
{Papadakis} I.~E.,  {Chatzopoulos} E.,  {Athanasiadis} D.,  {Markowitz} A.,
  {Georgantopoulos} I.,  2008, \mn@doi [\aap] {10.1051/0004-6361:200809572},
  \href {https://ui.adsabs.harvard.edu/abs/2008A&A...487..475P} {487, 475}

\bibitem[\protect\citeauthoryear{{Parker}, {Alston}, {Igo}  \&
  {Fabian}}{{Parker} et~al.}{2020}]{Parker2020}
{Parker} M.~L.,  {Alston} W.~N.,  {Igo} Z.,   {Fabian} A.~C.,  2020, \mn@doi
  [\mnras] {10.1093/mnras/stz3470}, \href
  {https://ui.adsabs.harvard.edu/abs/2020MNRAS.492.1363P} {492, 1363}

\bibitem[\protect\citeauthoryear{{Perola}, {Matt}, {Cappi}, {Fiore},
  {Guainazzi}, {Maraschi}, {Petrucci}  \& {Piro}}{{Perola}
  et~al.}{2002}]{Perola2002}
{Perola} G.~C.,  {Matt} G.,  {Cappi} M.,  {Fiore} F.,  {Guainazzi} M.,
  {Maraschi} L.,  {Petrucci} P.~O.,   {Piro} L.,  2002, \mn@doi [\aap]
  {10.1051/0004-6361:20020658}, \href
  {https://ui.adsabs.harvard.edu/abs/2002A&A...389..802P} {389, 802}

\bibitem[\protect\citeauthoryear{{Ponti}, {Cappi}, {Dadina}  \&
  {Malaguti}}{{Ponti} et~al.}{2004}]{Ponti2004}
{Ponti} G.,  {Cappi} M.,  {Dadina} M.,   {Malaguti} G.,  2004, \mn@doi [\aap]
  {10.1051/0004-6361:20031758}, \href
  {https://ui.adsabs.harvard.edu/abs/2004A&A...417..451P} {417, 451}

\bibitem[\protect\citeauthoryear{{Ponti}, {Miniutti}, {Cappi}, {Maraschi},
  {Fabian}  \& {Iwasawa}}{{Ponti} et~al.}{2006}]{Ponti2006}
{Ponti} G.,  {Miniutti} G.,  {Cappi} M.,  {Maraschi} L.,  {Fabian} A.~C.,
  {Iwasawa} K.,  2006, \mn@doi [\mnras] {10.1111/j.1365-2966.2006.10165.x},
  \href {https://ui.adsabs.harvard.edu/abs/2006MNRAS.368..903P} {368, 903}

\bibitem[\protect\citeauthoryear{{Ponti}, {Papadakis}, {Bianchi}, {Guainazzi},
  {Matt}, {Uttley}  \& {Bonilla}}{{Ponti} et~al.}{2012}]{Ponti12}
{Ponti} G.,  {Papadakis} I.,  {Bianchi} S.,  {Guainazzi} M.,  {Matt} G.,
  {Uttley} P.,   {Bonilla} N.~F.,  2012, \mn@doi [\aap]
  {10.1051/0004-6361/201118326}, \href
  {http://adsabs.harvard.edu/abs/2012A%26A...542A..83P} {542, A83}

\bibitem[\protect\citeauthoryear{{Porquet} et~al.,}{{Porquet}
  et~al.}{2018}]{Porquet2018}
{Porquet} D.,  et~al., 2018, \mn@doi [\aap] {10.1051/0004-6361/201731290},
  \href {https://ui.adsabs.harvard.edu/abs/2018A&A...609A..42P} {609, A42}

\bibitem[\protect\citeauthoryear{{Poutanen} \& {Fabian}}{{Poutanen} \&
  {Fabian}}{1999}]{Poutanen1999}
{Poutanen} J.,  {Fabian} A.~C.,  1999, \mn@doi [\mnras]
  {10.1046/j.1365-8711.1999.02735.x}, \href
  {https://ui.adsabs.harvard.edu/abs/1999MNRAS.306L..31P} {306, L31}

\bibitem[\protect\citeauthoryear{{Ricci} et~al.,}{{Ricci}
  et~al.}{2018}]{Ricci18}
{Ricci} C.,  et~al., 2018, \mn@doi [\mnras] {10.1093/mnras/sty1879}, \href
  {http://adsabs.harvard.edu/abs/2018MNRAS.480.1819R} {480, 1819}

\bibitem[\protect\citeauthoryear{{Ricci} et~al.,}{{Ricci}
  et~al.}{2020}]{Ricci2020}
{Ricci} C.,  et~al., 2020, \mn@doi [\apjl] {10.3847/2041-8213/ab91a1}, \href
  {https://ui.adsabs.harvard.edu/abs/2020ApJ...898L...1R} {898, L1}

\bibitem[\protect\citeauthoryear{{Ricci} et~al.,}{{Ricci}
  et~al.}{2021}]{Ricci2021}
{Ricci} C.,  et~al., 2021, \mn@doi [\apjs] {10.3847/1538-4365/abe94b}, \href
  {https://ui.adsabs.harvard.edu/abs/2021ApJS..255....7R} {255, 7}

\bibitem[\protect\citeauthoryear{{Rybicki} \& {Lightman}}{{Rybicki} \&
  {Lightman}}{1979}]{Rybi79}
{Rybicki} G.~B.,  {Lightman} A.~P.,  1979, {Radiative processes in
  astrophysics}

\bibitem[\protect\citeauthoryear{{Schlafly} \& {Finkbeiner}}{{Schlafly} \&
  {Finkbeiner}}{2011}]{Schlafly2011}
{Schlafly} E.~F.,  {Finkbeiner} D.~P.,  2011, \mn@doi [\apj]
  {10.1088/0004-637X/737/2/103}, \href
  {https://ui.adsabs.harvard.edu/abs/2011ApJ...737..103S} {737, 103}

\bibitem[\protect\citeauthoryear{{Serafinelli}, {Vagnetti}  \&
  {Middei}}{{Serafinelli} et~al.}{2017}]{Serafinelli2017}
{Serafinelli} R.,  {Vagnetti} F.,   {Middei} R.,  2017, \mn@doi [\aap]
  {10.1051/0004-6361/201629885}, \href
  {https://ui.adsabs.harvard.edu/abs/2017A&A...600A.101S} {600, A101}

\bibitem[\protect\citeauthoryear{{Shu}, {Yaqoob}, {Murphy}, {Braito}, {Wang}
  \& {Zheng}}{{Shu} et~al.}{2010}]{Shu2010}
{Shu} X.~W.,  {Yaqoob} T.,  {Murphy} K.~D.,  {Braito} V.,  {Wang} J.~X.,
  {Zheng} W.,  2010, \mn@doi [\apj] {10.1088/0004-637X/713/2/1256}, \href
  {https://ui.adsabs.harvard.edu/abs/2010ApJ...713.1256S} {713, 1256}

\bibitem[\protect\citeauthoryear{{Simonetti}, {Cordes}  \&
  {Heeschen}}{{Simonetti} et~al.}{1985}]{Simonetti1985}
{Simonetti} J.~H.,  {Cordes} J.~M.,   {Heeschen} D.~S.,  1985, \mn@doi [\apj]
  {10.1086/163418}, \href
  {https://ui.adsabs.harvard.edu/abs/1985ApJ...296...46S} {296, 46}

\bibitem[\protect\citeauthoryear{{Sobolewska} \& {Papadakis}}{{Sobolewska} \&
  {Papadakis}}{2009}]{Sobolewska2009}
{Sobolewska} M.~A.,  {Papadakis} I.~E.,  2009, \mn@doi [\mnras]
  {10.1111/j.1365-2966.2009.15382.x}, \href
  {https://ui.adsabs.harvard.edu/abs/2009MNRAS.399.1597S} {399, 1597}

\bibitem[\protect\citeauthoryear{{Sunyaev} \& {Titarchuk}}{{Sunyaev} \&
  {Titarchuk}}{1980}]{Sunyaev1980}
{Sunyaev} R.~A.,  {Titarchuk} L.~G.,  1980, \aap, \href
  {https://ui.adsabs.harvard.edu/abs/1980A&A....86..121S} {500, 167}

\bibitem[\protect\citeauthoryear{{Tortosa} et~al.,}{{Tortosa}
  et~al.}{2018}]{Tortosa2018}
{Tortosa} A.,  et~al., 2018, \mn@doi [\mnras] {10.1093/mnras/stx2457}, \href
  {https://ui.adsabs.harvard.edu/abs/2018MNRAS.473.3104T} {473, 3104}

\bibitem[\protect\citeauthoryear{{Trevese}, {Kron}, {Majewski}, {Bershady}  \&
  {Koo}}{{Trevese} et~al.}{1994}]{Trevese1994}
{Trevese} D.,  {Kron} R.~G.,  {Majewski} S.~R.,  {Bershady} M.~A.,   {Koo}
  D.~C.,  1994, \mn@doi [\apj] {10.1086/174661}, \href
  {https://ui.adsabs.harvard.edu/abs/1994ApJ...433..494T} {433, 494}

\bibitem[\protect\citeauthoryear{{Trippe}, {Crenshaw}, {Deo}  \&
  {Dietrich}}{{Trippe} et~al.}{2008}]{Trippe2008}
{Trippe} M.~L.,  {Crenshaw} D.~M.,  {Deo} R.,   {Dietrich} M.,  2008, \mn@doi
  [\aj] {10.1088/0004-6256/135/6/2048}, \href
  {https://ui.adsabs.harvard.edu/abs/2008AJ....135.2048T} {135, 2048}

\bibitem[\protect\citeauthoryear{{Turner}, {George}, {Nandra}  \&
  {Mushotzky}}{{Turner} et~al.}{1997a}]{Turner1997a}
{Turner} T.~J.,  {George} I.~M.,  {Nandra} K.,   {Mushotzky} R.~F.,  1997a,
  \mn@doi [\apjs] {10.1086/313053}, \href
  {https://ui.adsabs.harvard.edu/abs/1997ApJS..113...23T} {113, 23}

\bibitem[\protect\citeauthoryear{{Turner}, {George}, {Nandra}  \&
  {Mushotzky}}{{Turner} et~al.}{1997b}]{Turner1997b}
{Turner} T.~J.,  {George} I.~M.,  {Nandra} K.,   {Mushotzky} R.~F.,  1997b,
  \mn@doi [\apj] {10.1086/304701}, \href
  {https://ui.adsabs.harvard.edu/abs/1997ApJ...488..164T} {488, 164}

\bibitem[\protect\citeauthoryear{{Uttley}, {McHardy}  \& {Papadakis}}{{Uttley}
  et~al.}{2002}]{Uttley2002}
{Uttley} P.,  {McHardy} I.~M.,   {Papadakis} I.~E.,  2002, \mn@doi [\mnras]
  {10.1046/j.1365-8711.2002.05298.x}, \href
  {https://ui.adsabs.harvard.edu/abs/2002MNRAS.332..231U} {332, 231}

\bibitem[\protect\citeauthoryear{{Uttley}, {McHardy}  \& {Vaughan}}{{Uttley}
  et~al.}{2005}]{Uttley2005}
{Uttley} P.,  {McHardy} I.~M.,   {Vaughan} S.,  2005, \mn@doi [\mnras]
  {10.1111/j.1365-2966.2005.08886.x}, \href
  {https://ui.adsabs.harvard.edu/abs/2005MNRAS.359..345U} {359, 345}

\bibitem[\protect\citeauthoryear{{Vagnetti}, {Turriziani}  \&
  {Trevese}}{{Vagnetti} et~al.}{2011}]{Vagnetti2011}
{Vagnetti} F.,  {Turriziani} S.,   {Trevese} D.,  2011, \mn@doi [\aap]
  {10.1051/0004-6361/201118072}, \href
  {https://ui.adsabs.harvard.edu/abs/2011A&A...536A..84V} {536, A84}

\bibitem[\protect\citeauthoryear{{Vagnetti}, {Middei}, {Antonucci}, {Paolillo}
  \& {Serafinelli}}{{Vagnetti} et~al.}{2016}]{Vagnetti2016}
{Vagnetti} F.,  {Middei} R.,  {Antonucci} M.,  {Paolillo} M.,   {Serafinelli}
  R.,  2016, \mn@doi [\aap] {10.1051/0004-6361/201629057}, \href
  {https://ui.adsabs.harvard.edu/abs/2016A&A...593A..55V} {593, A55}

\bibitem[\protect\citeauthoryear{{Vaughan}, {Edelson}, {Warwick}  \&
  {Uttley}}{{Vaughan} et~al.}{2003}]{Vaughan2003}
{Vaughan} S.,  {Edelson} R.,  {Warwick} R.~S.,   {Uttley} P.,  2003, \mn@doi
  [\mnras] {10.1046/j.1365-2966.2003.07042.x}, \href
  {https://ui.adsabs.harvard.edu/abs/2003MNRAS.345.1271V} {345, 1271}

\bibitem[\protect\citeauthoryear{{Vaughan}, {Fabian}, {Ballantyne}, {De Rosa},
  {Piro}  \& {Matt}}{{Vaughan} et~al.}{2004}]{Vaughan2004}
{Vaughan} S.,  {Fabian} A.~C.,  {Ballantyne} D.~R.,  {De Rosa} A.,  {Piro} L.,
   {Matt} G.,  2004, \mn@doi [\mnras] {10.1111/j.1365-2966.2004.07769.x}, \href
  {https://ui.adsabs.harvard.edu/abs/2004MNRAS.351..193V} {351, 193}

\bibitem[\protect\citeauthoryear{{Yaqoob}}{{Yaqoob}}{2012}]{Yaqoob2012}
{Yaqoob} T.,  2012, \mn@doi [\mnras] {10.1111/j.1365-2966.2012.21129.x}, \href
  {https://ui.adsabs.harvard.edu/abs/2012MNRAS.423.3360Y} {423, 3360}

\bibitem[\protect\citeauthoryear{{Yaqoob} et~al.,}{{Yaqoob}
  et~al.}{2007}]{Yaqoob2007}
{Yaqoob} T.,  et~al., 2007, \mn@doi [\pasj] {10.1093/pasj/59.sp1.S283}, \href
  {https://ui.adsabs.harvard.edu/abs/2007PASJ...59S.283Y} {59, 283}

\bibitem[\protect\citeauthoryear{{Zdziarski}, {Johnson}, {Done}, {Smith}  \&
  {McNaron-Brown}}{{Zdziarski} et~al.}{1995}]{Zdziarski1995}
{Zdziarski} A.~A.,  {Johnson} W.~N.,  {Done} C.,  {Smith} D.,   {McNaron-Brown}
  K.,  1995, \mn@doi [\apjl] {10.1086/187716}, \href
  {https://ui.adsabs.harvard.edu/abs/1995ApJ...438L..63Z} {438, L63}

\bibitem[\protect\citeauthoryear{{Zdziarski}, {Johnson}  \&
  {Magdziarz}}{{Zdziarski} et~al.}{1996}]{Zdziarski1996}
{Zdziarski} A.~A.,  {Johnson} W.~N.,   {Magdziarz} P.,  1996, \mn@doi [\mnras]
  {10.1093/mnras/283.1.193}, \href
  {https://ui.adsabs.harvard.edu/abs/1996MNRAS.283..193Z} {283, 193}

\bibitem[\protect\citeauthoryear{{{\.Z}ycki}, {Done}  \& {Smith}}{{{\.Z}ycki}
  et~al.}{1999}]{Zycki1999}
{{\.Z}ycki} P.~T.,  {Done} C.,   {Smith} D.~A.,  1999, \mn@doi [\mnras]
  {10.1046/j.1365-8711.1999.02885.x}, \href
  {https://ui.adsabs.harvard.edu/abs/1999MNRAS.309..561Z} {309, 561}

\bibitem[\protect\citeauthoryear{{de Gouveia Dal Pino}, {Piovezan}  \&
  {Kadowaki}}{{de Gouveia Dal Pino} et~al.}{2010}]{deGouveia2010}
{de Gouveia Dal Pino} E.~M.,  {Piovezan} P.~P.,   {Kadowaki} L.~H.~S.,  2010,
  \mn@doi [\aap] {10.1051/0004-6361/200913462}, \href
  {https://ui.adsabs.harvard.edu/abs/2010A&A...518A...5D} {518, A5}

\bibitem[\protect\citeauthoryear{{de Vries}, {Becker}, {White}  \&
  {Loomis}}{{de Vries} et~al.}{2005}]{DeVries2005}
{de Vries} W.~H.,  {Becker} R.~H.,  {White} R.~L.,   {Loomis} C.,  2005,
  \mn@doi [\aj] {10.1086/427393}, \href
  {https://ui.adsabs.harvard.edu/abs/2005AJ....129..615D} {129, 615}

\bibitem[\protect\citeauthoryear{{di Clemente}, {Giallongo}, {Natali},
  {Trevese}  \& {Vagnetti}}{{di Clemente} et~al.}{1996}]{diClemente1996}
{di Clemente} A.,  {Giallongo} E.,  {Natali} G.,  {Trevese} D.,   {Vagnetti}
  F.,  1996, \mn@doi [\apj] {10.1086/177261}, \href
  {https://ui.adsabs.harvard.edu/abs/1996ApJ...463..466D} {463, 466}

\makeatother
\end{thebibliography}
\input{NGC2992REVISEDBIS.bbl}

\clearpage
\onecolumn
\appendix
	
\section{Some extra material}

\begin{longtable}{r r r r r r r}
	\caption{\label{swifttable} Best-fit parameters derived form the analysis of XRT exposures. Fluxes in the 0.5-2 and 2-10 keV are reported in units of $\times10^{-12}$ and $\times10^{-11}$ erg cm$^{-2}$ s$^{-1}$, respectively. The column density accounts for 10$^{22}$ cm$^{-2}.$ and normalisations are listed in units of photons keV$^{-1}$ cm$^{-2}$ s$^{-1}$.}
	\setlength{\tabcolsep}{.1 pt}\\
	\hline\hline
Date& Obs. ID. & F$_{\rm 0.5-2~keV}$& F$_{\rm 2-10~keV}$ & N$_{\rm H}$ & $\Gamma$  & Norm \\	\hline
	\endfirsthead
	\caption{continued.}\\
	\hline\hline
Date& Obs. ID. & F$_{\rm 0.5-2~keV}$& F$_{\rm 2-10~keV}$ & N$_{\rm H}$ & $\Gamma$  & Norm \\	\hline
	\endhead
	\hline
	\endfoot
	\hline
2006-06-14&00035344002&1.16$\pm$0.23&2.30$\pm$0.23&0.23$\pm$0.22&0.90$\pm$0.20&0.0008$\pm$0.0002\\
2006-06-25&00035344003&0.68$\pm$0.09&0.98$\pm$0.09&0.19$\pm$0.14&1.05$\pm$0.17&0.0005$\pm$0.0001\\
2006-07-06&00035344004&0.67$\pm$0.09&0.97$\pm$0.09&0.16$\pm$0.13&0.90$\pm$0.14&0.0004$\pm$0.0001\\
2006-07-07&00035344005&0.85$\pm$0.06&0.59$\pm$0.06&0.22$\pm$0.07&1.16$\pm$0.08&0.0006$\pm$0.0001\\
2006-07-11&00035344006&0.78$\pm$0.16&1.37$\pm$0.16&0.18$\pm$0.16&1.19$\pm$0.24&0.0005$\pm$0.0002\\
2006-07-12&00035344007&0.80$\pm$0.21&1.63$\pm$0.21&0.34$\pm$0.28&1.29$\pm$0.28&0.0007$\pm$0.0003\\
2015-12-02&00081055001&6.60$\pm$0.27&3.14$\pm$0.27&0.89$\pm$0.10&1.52$\pm$0.06&0.0123$\pm$0.0011\\
2017-06-02&00035344008&4.19$\pm$0.29&2.86$\pm$0.29&0.58$\pm$0.12&1.31$\pm$0.09&0.0052$\pm$0.0006\\
2017-06-04&00035344009&4.65$\pm$0.45&4.77$\pm$0.45&0.80$\pm$0.22&1.45$\pm$0.13&0.0077$\pm$0.0015\\
2019-03-26&00035344010&11.80$\pm$0.71&6.87$\pm$0.71&0.81$\pm$0.13&1.63$\pm$0.09&0.0211$\pm$0.0026\\
2019-03-30&00035344011&14.24$\pm$0.91&7.82$\pm$0.91&0.79$\pm$0.14&1.70$\pm$0.09&0.0258$\pm$0.0033\\
2019-04-03&00035344012&8.72$\pm$0.62&6.18$\pm$0.62&0.84$\pm$0.18&1.54$\pm$0.11&0.0155$\pm$0.0024\\
2019-04-07&00035344013&10.01$\pm$0.83&7.48$\pm$0.83&0.79$\pm$0.19&1.57$\pm$0.12&0.017$\pm$0.003\\
2019-04-11&00035344014&6.33$\pm$0.58&5.32$\pm$0.58&0.44$\pm$0.15&1.24$\pm$0.11&0.0064$\pm$0.0010\\
2019-04-14&00035344015&10.83$\pm$0.77&7.95$\pm$0.77&0.74$\pm$0.15&1.56$\pm$0.10&0.0174$\pm$0.0025\\
2019-04-19&00035344016&7.75$\pm$0.55&6.03$\pm$0.55&0.57$\pm$0.14&1.29$\pm$0.10&0.0094$\pm$0.0013\\
2019-04-23&00035344017&8.78$\pm$0.60&6.32$\pm$0.60&0.63$\pm$0.13&1.39$\pm$0.10&0.0118$\pm$0.0016\\
2019-05-01&00035344019&10.21$\pm$0.82&7.67$\pm$0.82&0.90$\pm$0.17&1.61$\pm$0.11&0.0198$\pm$0.0031\\
2019-05-04&00035344020&7.28$\pm$0.63&6.76$\pm$0.63&0.75$\pm$0.20&1.42$\pm$0.13&0.0113$\pm$0.002\\
2019-05-06&00035344021&9.73$\pm$0.68&6.24$\pm$0.68&0.81$\pm$0.15&1.58$\pm$0.1&0.0171$\pm$0.0024\\
2019-05-08&00035344022&9.55$\pm$0.62&6.47$\pm$0.62&0.81$\pm$0.15&1.48$\pm$0.09&0.0161$\pm$0.0021\\
2019-05-10&00035344023&9.46$\pm$1.33&11.36$\pm$1.33&0.70$\pm$0.24&1.50$\pm$0.17&0.0143$\pm$0.0034\\
2019-05-12&00035344024&11.91$\pm$5.47&9.60$\pm$1.47&0.76$\pm$0.57&1.53$\pm$0.39&0.0194$\pm$0.0109\\
2019-05-14&00035344025&12.80$\pm$0.68&7.36$\pm$0.68&0.80$\pm$0.13&1.57$\pm$0.08&0.0221$\pm$0.0026\\
2019-05-16&00035344026&15.48$\pm$0.92&6.97$\pm$0.92&0.85$\pm$0.13&1.74$\pm$0.09&0.0302$\pm$0.0036\\
2019-05-18&00035344027&10.39$\pm$0.68&6.92$\pm$0.68&0.82$\pm$0.15&1.54$\pm$0.10&0.0181$\pm$0.0025\\
2019-05-20&00035344028&11.53$\pm$0.71&6.33$\pm$0.71&0.93$\pm$0.15&1.69$\pm$0.09&0.0238$\pm$0.0031\\
2019-05-22&00035344029&12.44$\pm$0.74&7.10$\pm$0.74&0.88$\pm$0.15&1.67$\pm$0.10&0.0243$\pm$0.0032\\
2019-05-24&00035344030&11.33$\pm$0.74&6.80$\pm$0.74&0.57$\pm$0.11&1.47$\pm$0.09&0.0145$\pm$0.0017\\
2019-05-26&00035344031&10.34$\pm$0.68&6.20$\pm$0.68&0.66$\pm$0.12&1.51$\pm$0.09&0.0151$\pm$0.0019\\
2019-05-28&00035344032&8.42$\pm$0.64&6.09$\pm$0.64&0.87$\pm$0.18&1.54$\pm$0.11&0.0155$\pm$0.0023\\
2019-05-30&00035344033&9.43$\pm$0.67&6.56$\pm$0.67&0.87$\pm$0.17&1.56$\pm$0.11&0.0175$\pm$0.0026\\
2019-06-01&00035344034&10.19$\pm$0.74&7.04$\pm$0.74&0.62$\pm$0.15&1.46$\pm$0.10&0.0138$\pm$0.002\\
2019-06-03&00035344035&7.67$\pm$0.59&6.19$\pm$0.59&0.52$\pm$0.13&1.26$\pm$0.10&0.0086$\pm$0.0012\\
2019-06-05&00035344036&15.33$\pm$0.92&8.89$\pm$0.92&0.83$\pm$0.15&1.64$\pm$0.10&0.0282$\pm$0.0037\\
2019-06-07&00035344037&8.15$\pm$0.61&6.60$\pm$0.61&0.62$\pm$0.13&1.33$\pm$0.09&0.0106$\pm$0.0014\\
2019-06-09&00035344038&5.87$\pm$0.49&5.00$\pm$0.49&0.80$\pm$0.20&1.53$\pm$0.13&0.01$\pm$0.0018\\
2019-06-11&00035344039&10.12$\pm$0.66&6.27$\pm$0.66&0.88$\pm$0.14&1.59$\pm$0.09&0.0192$\pm$0.0025\\
2019-06-16&00035344040&7.40$\pm$0.58&5.25$\pm$0.58&0.60$\pm$0.14&1.38$\pm$0.10&0.0095$\pm$0.0013\\
2019-06-20&00035344041&10.63$\pm$0.66&6.17$\pm$0.66&0.83$\pm$0.14&1.61$\pm$0.09&0.0192$\pm$0.0025\\
2019-06-24&00035344042&6.99$\pm$0.62&5.94$\pm$0.62&0.74$\pm$0.19&1.51$\pm$0.13&0.011$\pm$0.002\\
2019-06-27&00035344043&4.82$\pm$1.37&5.29$\pm$1.37&0.13$\pm$0.23&0.62$\pm$0.27&0.0026$\pm$0.001\\
2019-10-09&00035344044&11.99$\pm$0.80&7.50$\pm$0.80&0.77$\pm$0.13&1.57$\pm$0.09&0.0201$\pm$0.0025\\
2019-10-13&00035344045&4.22$\pm$0.40&4.00$\pm$0.40&0.39$\pm$0.13&1.21$\pm$0.11&0.004$\pm$0.0006\\
2019-10-17&00035344046&5.57$\pm$0.49&4.71$\pm$0.49&0.68$\pm$0.18&1.39$\pm$0.12&0.0079$\pm$0.0013\\
2019-10-21&00035344047&5.71$\pm$0.67&6.88$\pm$0.67&0.52$\pm$0.22&1.31$\pm$0.16&0.0065$\pm$0.0015\\
2019-10-25&00035344048&6.21$\pm$0.64&5.65$\pm$0.64&0.80$\pm$0.23&1.57$\pm$0.15&0.0107$\pm$0.0022\\
2019-10-29&00035344049&5.86$\pm$0.55&6.52$\pm$0.55&0.49$\pm$0.20&1.14$\pm$0.13&0.0061$\pm$0.0012\\
2019-11-02&00035344050&4.05$\pm$0.45&4.56$\pm$0.45&0.43$\pm$0.18&1.19$\pm$0.14&0.004$\pm$0.0008\\
2019-11-06&00035344051&4.57$\pm$0.61&5.98$\pm$0.61&0.64$\pm$0.24&1.42$\pm$0.17&0.0063$\pm$0.0015\\
2019-11-08&00035344052&4.99$\pm$0.42&3.87$\pm$0.42&0.60$\pm$0.15&1.43$\pm$0.11&0.0066$\pm$0.001\\
2019-11-10&00035344053&7.22$\pm$0.61&5.93$\pm$0.61&0.77$\pm$0.17&1.47$\pm$0.11&0.0115$\pm$0.0018\\
2019-11-12&00035344054&7.29$\pm$0.58&5.55$\pm$0.58&0.88$\pm$0.17&1.63$\pm$0.11&0.0141$\pm$0.0022\\
2019-11-14&00035344055&9.14$\pm$0.71&7.40$\pm$0.71&0.54$\pm$0.13&1.27$\pm$0.10&0.0106$\pm$0.0014\\
2019-11-16&00035344056&7.68$\pm$0.58&6.07$\pm$0.58&0.82$\pm$0.17&1.48$\pm$0.11&0.013$\pm$0.0019\\
2019-11-17&00035344057&10.62$\pm$0.69&6.58$\pm$0.69&0.85$\pm$0.13&1.57$\pm$0.09&0.0193$\pm$0.0024\\
2019-11-20&00035344058&8.13$\pm$0.65&6.70$\pm$0.65&0.76$\pm$0.17&1.41$\pm$0.11&0.0126$\pm$0.0019\\
2019-11-22&00035344059&7.80$\pm$0.98&9.19$\pm$0.98&0.52$\pm$0.2&1.31$\pm$0.15&0.0089$\pm$0.0018\\
2019-11-26&00035344061&4.92$\pm$0.49&4.27$\pm$0.49&0.63$\pm$0.18&1.42$\pm$0.13&0.0067$\pm$0.0012\\
2019-11-28&00035344062&10.67$\pm$0.83&7.15$\pm$0.83&0.66$\pm$0.15&1.58$\pm$0.11&0.0159$\pm$0.0024\\
2019-11-30&00035344063&7.84$\pm$0.73&6.63$\pm$0.73&0.78$\pm$0.18&1.53$\pm$0.12&0.013$\pm$0.0022\\
2019-12-02&00035344064&7.08$\pm$0.63&4.54$\pm$0.63&0.72$\pm$0.17&1.77$\pm$0.12&0.0122$\pm$0.002\\
2019-12-04&00035344065&10.82$\pm$0.75&7.92$\pm$0.75&0.57$\pm$0.12&1.37$\pm$0.09&0.0135$\pm$0.0017\\
2019-12-06&00035344066&14.79$\pm$1.07&8.81$\pm$1.07&0.83$\pm$0.15&1.72$\pm$0.10&0.028$\pm$0.004\\
2019-12-08&00035344067&11.25$\pm$0.78&7.60$\pm$0.78&0.65$\pm$0.15&1.47$\pm$0.10&0.016$\pm$0.0023\\
2019-12-10&00035344068&7.88$\pm$1.00&8.39$\pm$1.00&0.69$\pm$0.26&1.54$\pm$0.18&0.012$\pm$0.0029\\
2019-12-12&00035344069&9.70$\pm$0.67&6.20$\pm$0.67&0.70$\pm$0.13&1.55$\pm$0.09&0.015$\pm$0.0019\\
2019-12-14&00035344070&9.36$\pm$0.74&6.57$\pm$0.74&0.89$\pm$0.19&1.66$\pm$0.12&0.0185$\pm$0.0031\\
2019-12-31&00035344071&2.96$\pm$0.32&2.91$\pm$0.32&0.53$\pm$0.18&1.51$\pm$0.15&0.0037$\pm$0.0007\\
2020-01-28&00035344072&11.64$\pm$0.97&8.21$\pm$0.97&0.71$\pm$0.15&1.63$\pm$0.12&0.0187$\pm$0.0029\\
2020-03-25&00035344073&12.08$\pm$0.79&7.11$\pm$0.79&0.74$\pm$0.14&1.55$\pm$0.09&0.0193$\pm$0.0025\\
2020-05-19&00035344074&6.97$\pm$0.58&5.79$\pm$0.58&0.83$\pm$0.19&1.58$\pm$0.12&0.0125$\pm$0.0021\\
2021-01-24&00035344075&3.31$\pm$0.29&3.11$\pm$0.29&0.49$\pm$0.14&1.29$\pm$0.11&0.0037$\pm$0.0006\\
2021-01-29&00035344076&2.95$\pm$0.41&3.98$\pm$0.41&0.65$\pm$0.25&1.41$\pm$0.18&0.0041$\pm$0.001\\
2021-02-03&00035344077&1.54$\pm$0.84&9.94$\pm$0.84&0.58$\pm$0.68&1.11$\pm$0.47&0.0017$\pm$0.0012\\
2021-02-08&00035344078&1.60$\pm$0.30&2.09$\pm$0.30&0.83$\pm$0.36&1.82$\pm$0.26&0.0032$\pm$0.0011\\
2021-02-13&00035344079&2.98$\pm$0.32&3.07$\pm$0.32&0.77$\pm$0.23&1.52$\pm$0.15&0.0049$\pm$0.001\\
2021-02-18&00035344080&2.22$\pm$0.42&3.81$\pm$0.42&0.58$\pm$0.33&1.34$\pm$0.23&0.0028$\pm$0.0009\\
2021-02-22&00035344081&5.27$\pm$0.49&4.36$\pm$0.49&0.84$\pm$0.24&1.71$\pm$0.15&0.0101$\pm$0.0021\\
2021-02-28&00035344082&1.65$\pm$0.22&2.27$\pm$0.22&0.32$\pm$0.17&1.17$\pm$0.17&0.0014$\pm$0.0003\\
2021-03-05&00035344083&1.50$\pm$0.19&2.32$\pm$0.19&0.28$\pm$0.19&0.94$\pm$0.16&0.0011$\pm$0.0002\\
2021-03-10&00035344084&1.04$\pm$0.19&1.46$\pm$0.19&0.12$\pm$0.14&1.14$\pm$0.20&0.0006$\pm$0.0002\\
2021-03-15&00035344085&1.37$\pm$0.22&1.71$\pm$0.22&0.42$\pm$0.20&1.50$\pm$0.20&0.0015$\pm$0.0004\\
2021-03-25&00035344088&2.23$\pm$0.25&2.78$\pm$0.25&0.38$\pm$0.16&1.18$\pm$0.14&0.0021$\pm$0.0004\\
2021-03-30&00035344089&2.00$\pm$0.30&2.95$\pm$0.30&0.35$\pm$0.21&1.09$\pm$0.17&0.0017$\pm$0.0004\\
2021-04-04&00035344090&2.00$\pm$0.26&2.19$\pm$0.26&0.41$\pm$0.17&1.40$\pm$0.16&0.0021$\pm$0.0004\\
2021-04-09&00035344091&3.14$\pm$0.34&3.27$\pm$0.34&0.84$\pm$0.27&1.61$\pm$0.17&0.0058$\pm$0.0014\\
2021-04-14&00035344092&1.52$\pm$0.22&1.65$\pm$0.22&0.50$\pm$0.22&1.58$\pm$0.20&0.0019$\pm$0.0005\\
2021-04-19&00035344093&2.31$\pm$0.23&2.68$\pm$0.23&0.28$\pm$0.14&0.99$\pm$0.13&0.0017$\pm$0.0003\\
2021-04-24&00035344094&1.85$\pm$0.27&2.57$\pm$0.27&0.61$\pm$0.25&1.36$\pm$0.18&0.0024$\pm$0.0006\\
2021-04-30&00035344096&2.76$\pm$0.30&3.11$\pm$0.30&0.62$\pm$0.21&1.38$\pm$0.15&0.0036$\pm$0.0008\\
2021-05-04&00035344097&1.76$\pm$0.23&2.70$\pm$0.23&0.15$\pm$0.12&0.93$\pm$0.15&0.0011$\pm$0.0002\\
2021-05-09&00035344098&1.01$\pm$0.29&2.40$\pm$0.29&0.43$\pm$0.34&1.37$\pm$0.32&0.0011$\pm$0.0004\\
2021-05-12&00035344099&1.34$\pm$0.26&2.47$\pm$0.26&0.24$\pm$0.20&1.08$\pm$0.21&0.0010$\pm$0.0003\\
2021-05-14&00035344100&2.08$\pm$0.27&2.80$\pm$0.27&0.46$\pm$0.17&1.26$\pm$0.15&0.0022$\pm$0.0004\\
2021-05-19&00035344101&2.12$\pm$0.29&2.85$\pm$0.29&0.58$\pm$0.28&1.38$\pm$0.20&0.0027$\pm$0.0007\\
2021-05-24&00035344102&2.32$\pm$0.25&3.08$\pm$0.25&0.40$\pm$0.17&0.96$\pm$0.13&0.0020$\pm$0.0004\\
2021-05-29&00035344103&1.38$\pm$0.30&3.46$\pm$0.30&0.20$\pm$0.16&0.82$\pm$0.23&0.0008$\pm$0.0002\\
2021-06-03&00035344104&0.90$\pm$0.19&1.74$\pm$0.19&0.30$\pm$0.22&1.11$\pm$0.21&0.0007$\pm$0.0002\\
2021-06-08&00035344105&1.88$\pm$0.41&3.37$\pm$0.41&0.72$\pm$0.39&1.68$\pm$0.29&0.0031$\pm$0.0012\\
2021-06-13&00035344106&2.27$\pm$0.33&2.91$\pm$0.33&0.70$\pm$0.31&1.54$\pm$0.21&0.0035$\pm$0.001\\
2021-06-18&00035344107&1.47$\pm$0.18&2.28$\pm$0.18&0.25$\pm$0.15&0.95$\pm$0.15&0.0011$\pm$0.0002\\
2021-06-23&00035344108&3.00$\pm$0.28&2.78$\pm$0.28&0.66$\pm$0.19&1.42$\pm$0.13&0.0042$\pm$0.0008\\
2021-06-27&00035344109&2.43$\pm$0.27&2.57$\pm$0.27&0.51$\pm$0.19&1.35$\pm$0.15&0.0028$\pm$0.0006\\
2021-07-03&00035344110&2.29$\pm$0.36&3.52$\pm$0.36&1.10$\pm$0.36&1.70$\pm$0.21&0.0056$\pm$0.0017\\
2021-07-08&00035344111&2.60$\pm$0.29&3.03$\pm$0.29&0.46$\pm$0.20&1.19$\pm$0.15&0.0026$\pm$0.0006\\
2021-07-13&00035344112&4.03$\pm$0.39&3.66$\pm$0.39&0.64$\pm$0.19&1.46$\pm$0.14&0.0056$\pm$0.0011\\
2021-10-08&00035344113&3.58$\pm$0.37&3.42$\pm$0.37&0.55$\pm$0.17&1.43$\pm$0.13&0.0044$\pm$0.0008\\
2021-10-13&00035344114&3.61$\pm$0.30&3.21$\pm$0.30&0.51$\pm$0.13&1.27$\pm$0.11&0.0040$\pm$0.0006\\
2021-10-18&00035344115&3.55$\pm$0.45&4.91$\pm$0.45&0.80$\pm$0.27&1.43$\pm$0.17&0.0058$\pm$0.0014\\
2021-10-19&00035344116&4.61$\pm$0.85&8.54$\pm$0.85&0.77$\pm$0.38&1.43$\pm$0.25&0.0073$\pm$0.0025\\
2021-10-21&00035344117&2.99$\pm$0.52&5.37$\pm$0.52&0.70$\pm$0.30&1.38$\pm$0.22&0.0043$\pm$0.0013\\
2021-10-23&00035344118&2.59$\pm$0.37&3.20$\pm$0.37&0.74$\pm$0.26&1.55$\pm$0.19&0.0041$\pm$0.0011\\
2021-10-28&00035344119&4.01$\pm$0.36&3.38$\pm$0.36&0.65$\pm$0.18&1.49$\pm$0.13&0.0057$\pm$0.001\\
2021-11-02&00035344120&5.66$\pm$0.57&6.94$\pm$0.57&0.66$\pm$0.24&1.25$\pm$0.15&0.0075$\pm$0.0016\\
2021-11-07&00035344121&3.57$\pm$0.32&3.31$\pm$0.32&0.55$\pm$0.17&1.31$\pm$0.12&0.0042$\pm$0.0007\\
2021-11-12&00035344122&5.43$\pm$0.51&5.01$\pm$0.51&0.62$\pm$0.17&1.40$\pm$0.13&0.0073$\pm$0.0013\\
2021-11-17&00035344123&4.04$\pm$0.44&3.65$\pm$0.44&0.91$\pm$0.25&1.71$\pm$0.16&0.0083$\pm$0.0018\\
2021-11-22&00035344124&6.69$\pm$0.62&6.20$\pm$0.62&0.62$\pm$0.18&1.31$\pm$0.12&0.0085$\pm$0.0015\\
2021-12-02&00035344125&6.71$\pm$0.61&6.75$\pm$0.61&0.42$\pm$0.14&1.20$\pm$0.12&0.0065$\pm$0.001\\
2021-12-12&00035344126&6.67$\pm$0.76&7.98$\pm$0.76&0.55$\pm$0.20&1.26$\pm$0.14&0.0078$\pm$0.0015\\
2021-12-24&00035344128&3.50$\pm$0.35&3.85$\pm$0.35&0.35$\pm$0.15&1.10$\pm$0.12&0.0030$\pm$0.0005\\
2021-12-28&00035344129&4.30$\pm$0.30&3.90$\pm$0.30&0.65$\pm$0.20&1.35$\pm$0.20&0.0057$\pm$0.0003\\
\end{longtable}

\begin{table*}
		\caption{\small{Best  fit  parameters  of  the  time-resolved  \textit{XMM-Newton}-\textit{NuSTAR}  analysis. For each spectral slice the corresponding ks are quoted. Fluxes are in units of erg cm$^{-2}$ s$^{-1} \times10^{-11}$ while Gaussian normalisation accounts for $10^{-5}$ ph. cm$^{-2}$ s$^{-1}$. The quoted normalisation for power law and \textit{MyTorus} are in units of ph. keV$^{-1}$ cm$^{-2}$ s$^{-1}$. Finally, N$_{\rm H}$ and N$_{\rm H}^{\rm MyT}$ are in units of 10$^{22}$ and 10$^{24}$ cm$^{-2}$, respectively.}}
		\label{xmmtimeresolved}
		\setlength{\tabcolsep}{4.2pt}
		\begin{tabular}{c c c c c c c c c c c c c}
Slice & F$_{\rm 0.3-2~keV}$& F$_{\rm 2-10~keV}$& N$_{\rm H}$ &$\Gamma$& Norm$_{\rm po}$  &N$_{\rm H}^{\rm MyT}$ &Norm$^{\rm MyT}$  &Fe$_{\rm He\alpha}$ &   Fe$_{\rm Ly\alpha}$&Red flare& $\chi^2$ & d.o.f \\
5&1.51$\pm$0.03&9.59$\pm$0.42&0.77$\pm$0.02&1.76$\pm$0.03&0.025$\pm$0.001&<0.77&0.26$\pm$0.31&<2.62&3.22$\pm$2.16&<3.3&160&153\\
10&1.57$\pm$0.03&10.19$\pm$0.25&0.78$\pm$0.02&1.74$\pm$0.03&0.027$\pm$0.001&<0.78&0.13$\pm$0.23&<3.78&<4.13&6.43$\pm$2.9&164&156\\
15&1.47$\pm$0.02&9.88$\pm$0.26&0.77$\pm$0.02&1.71$\pm$0.03&0.025$\pm$0.001&0.77$\pm$0.06&0.12$0\pm$0.11&<3.38&<1.44&<5.0&167&154\\
20&1.35$\pm$0.03&9.14$\pm$0.41&0.77$\pm$0.02&1.69$\pm$0.04&0.022$\pm$0.001&<0.77&0.24$\pm$0.13&2.96$\pm$2.05&<4.12&<1.26&164&152\\
25&1.24$\pm$0.01&8.47$\pm$0.26&0.78$\pm$0.02&1.70$\pm$0.03&0.021$\pm$0.001&0.78$\pm$0.17&0.07$\pm$0.24&<2.62&<3.89&<3.03&136&155\\
30&1.13$\pm$0.01&7.69$\pm$0.22&0.79$\pm$0.02&1.70$\pm$0.03&0.019$\pm$0.001&0.79$\pm$0.17&0.05$\pm$0.06&2.76$\pm$1.91&3.35$\pm$1.97&<3.39&210&151\\
35&1.07$\pm$0.02&7.48$\pm$0.26&0.75$\pm$0.02&1.66$\pm$0.04&0.017$\pm$0.001&<0.75&0.10$\pm$0.23&2.57$\pm$1.89&<2.04&<1.93&132&150\\
40&1.03$\pm$0.01&7.01$\pm$0.22&0.78$\pm$0.03&1.71$\pm$0.03&0.017$\pm$0.001&<0.78&0.08$\pm$0.24&<2.62&4.01$\pm$1.96&<1.24&143&150\\
45&1.02$\pm$0.02&7.13$\pm$0.33&0.74$\pm$0.02&1.65$\pm$0.04&0.016$\pm$0.001&<0.74&0.22$\pm$0.26&<2.14&2.32$\pm$1.89&<0.88&169&151\\
50&1.10$\pm$0.02&7.45$\pm$0.61&0.75$\pm$0.02&1.64$\pm$0.04&0.016$\pm$0.001&<0.75&0.33$\pm$0.17&<1.66&<1.05&<2.42&163&151\\
55&1.18$\pm$0.01&8.13$\pm$0.26&0.76$\pm$0.02&1.68$\pm$0.03&0.019$\pm$0.001&0.76$\pm$0.13&0.07$\pm$0.05&<1.72&<2.39&<1.85&173&151\\
60&1.30$\pm$0.02&8.75$\pm$0.55&0.76$\pm$0.03&1.68$\pm$0.03&0.021$\pm$0.001&<0.76&0.3$\pm$0.23&<2.62&2.75$\pm$2.05&<3.86&155&154\\
65&1.34$\pm$0.03&9.13$\pm$0.38&0.79$\pm$0.02&1.70$\pm$0.03&0.022$\pm$0.001&<0.79&0.2$\pm$0.26&<1.86&<4.03&<4.56&179&154\\
70&1.39$\pm$0.01&9.73$\pm$0.23&0.78$\pm$0.02&1.66$\pm$0.03&0.023$\pm$0.001&0.78$\pm$0.75&0.05$\pm$0.09&<1.63&<3.23&<2.2&123&153\\
75&1.40$\pm$0.02&9.62$\pm$0.32&0.79$\pm$0.02&1.71$\pm$0.03&0.024$\pm$0.001&<0.79&0.15$\pm$0.27&2.13$\pm$2.08&<2.17&<2.54&152&151\\
80&1.36$\pm$0.01&9.47$\pm$0.30&0.78$\pm$0.02&1.69$\pm$0.03&0.023$\pm$0.001&0.78$\pm$0.27&0.05$\pm$0.07&<1.95&<2.15&<3.86&158&155\\
85&1.34$\pm$0.01&9.15$\pm$0.48&0.78$\pm$0.02&1.70$\pm$0.03&0.022$\pm$0.001&<0.78&0.03$\pm$0.01&<3.98&<2.07&<1.8&200&154\\
90&1.36$\pm$0.02&9.43$\pm$0.61&0.80$\pm$0.03&1.67$\pm$0.02&0.022$\pm$0.001&<0.8&0.54$\pm$0.17&2.33$\pm$2.02&<3.08&<1.3&181&153\\
95&1.33$\pm$0.03&9.12$\pm$0.47&0.79$\pm$0.02&1.68$\pm$0.03&0.022$\pm$0.001&<0.79&0.21$\pm$0.19&<2.29&<3.94&<2.94&130&152\\
100&1.29$\pm$0.02&8.99$\pm$0.27&0.78$\pm$0.02&1.69$\pm$0.03&0.022$\pm$0.001&0.78$\pm$0.06&0.1$\pm$0.11&<3.94&<1.06&2.78$\pm$2.64&161&155\\
105&1.23$\pm$0.02&8.67$\pm$0.30&0.77$\pm$0.02&1.67$\pm$0.04&0.020$\pm$0.001&0.77$\pm$0.12&0.1$\pm$0.21&<3.88&<1.66&<1.77&189&154\\
110&1.15$\pm$0.01&7.98$\pm$0.30&0.80$\pm$0.02&1.71$\pm$0.03&0.020$\pm$0.001&0.80$\pm$0.38&0.06$\pm$0.08&<3.63&<2.6&<3.22&133&152\\
115&1.09$\pm$0.03&7.86$\pm$0.43&0.75$\pm$0.03&1.62$\pm$0.04&0.016$\pm$0.001&<0.75&0.21$\pm$0.24&<2.25&<3.68&<1.79&161&152\\
120&1.07$\pm$0.02&7.72$\pm$0.23&0.76$\pm$0.03&1.63$\pm$0.04&0.017$\pm$0.001&<0.76&0.12$\pm$0.21&3.48$\pm$1.94&3.29$\pm$1.97&<0.77&158&151\\
125&1.08$\pm$0.01&7.76$\pm$0.46&0.75$\pm$0.02&1.62$\pm$0.03&0.017$\pm$0.001&<0.75&0.02$\pm$0.01&<1.99&3.36$\pm$2.01&<3.08&178&151\\
130&1.14$\pm$0.02&8.01$\pm$0.24&0.79$\pm$0.02&1.67$\pm$0.03&0.019$\pm$0.001&0.79$\pm$0.13&0.06$\pm$0.14&<3.27&2.61$\pm$1.99&<4.24&162&152\\
177&0.95$\pm$0.02&6.78$\pm$0.32&0.76$\pm$0.02&1.64$\pm$0.03&0.015$\pm$0.001&0.76$\pm$0.04&0.17$\pm$0.16&<1.2&<4.7&<2.12&135&147\\
182&0.99$\pm$0.02&7.07$\pm$0.21&0.73$\pm$0.01&1.60$\pm$0.02&0.015$\pm$0.001&0.73$\pm$0.03&0.07$\pm$0.05&1.55$\pm$1.12&<1.68&<1.75&165&151\\
187&1.02$\pm$0.01&7.39$\pm$0.30&0.75$\pm$0.02&1.65$\pm$0.02&0.016$\pm$0.001&0.75$\pm$0.04&0.12$\pm$0.05&3.65$\pm$1.18&2.01$\pm$1.19&<2.05&189&151\\
191&1.03$\pm$0.01&7.29$\pm$0.25&0.78$\pm$0.02&1.68$\pm$0.02&0.017$\pm$0.001&0.78$\pm$0.04&0.09$\pm$0.05&2.38$\pm$1.63&2.14$\pm$1.62&<3.92&447&393\\
196&0.99$\pm$0.01&7.15$\pm$0.28&0.79$\pm$0.02&1.68$\pm$0.02&0.016$\pm$0.001&0.79$\pm$0.04&0.09$\pm$0.05&2.49$\pm$1.61&<2.77&<2.25&424&390\\
202&0.99$\pm$0.01&7.02$\pm$0.25&0.79$\pm$0.02&1.68$\pm$0.02&0.016$\pm$0.001&0.79$\pm$0.04&0.08$\pm$0.04&<1.48&<2.8&<1.54&469&401\\
208&1.00$\pm$0.01&7.16$\pm$0.25&0.80$\pm$0.02&1.69$\pm$0.02&0.017$\pm$0.001&0.80$\pm$0.04&0.08$\pm$0.04&<1.02&<3.61&1.97$\pm$1.76&440&420\\
214&1.01$\pm$0.01&7.16$\pm$0.26&0.78$\pm$0.02&1.68$\pm$0.02&0.017$\pm$0.001&0.78$\pm$0.04&0.09$\pm$0.06&<2.37&2.26$\pm$1.53&<1.09&447&429\\
220&0.97$\pm$0.01&6.99$\pm$0.28&0.77$\pm$0.02&1.67$\pm$0.02&0.016$\pm$0.001&0.77$\pm$0.04&0.10$\pm$0.05&2.37$\pm$1.58&<3.91&3.93$\pm$1.98&438&415\\
225&0.97$\pm$0.01&6.92$\pm$0.27&0.78$\pm$0.02&1.67$\pm$0.02&0.016$\pm$0.001&0.78$\pm$0.03&0.09$\pm$0.05&<2.09&1.6$\pm$1.23&2.96$\pm$1.93&430&420\\
231&0.97$\pm$0.01&7.06$\pm$0.25&0.81$\pm$0.02&1.67$\pm$0.02&0.016$\pm$0.001&0.81$\pm$0.07&0.07$\pm$0.06&2.07$\pm$1.54&<2.67&<3.12&456&427\\
237&0.93$\pm$0.01&6.74$\pm$0.27&0.81$\pm$0.02&1.69$\pm$0.02&0.015$\pm$0.001&0.81$\pm$0.03&0.13$\pm$0.07&2.03$\pm$1.5&<2.52&<2.53&485&426\\
243&0.97$\pm$0.01&6.85$\pm$0.28&0.77$\pm$0.02&1.67$\pm$0.02&0.016$\pm$0.001&0.77$\pm$0.04&0.10$\pm$0.07&1.6$\pm$1.48&<2.62&<2.92&490&428\\
249&1.04$\pm$0.01&7.50$\pm$0.26&0.78$\pm$0.02&1.66$\pm$0.02&0.017$\pm$0.001&0.78$\pm$0.06&0.07$\pm$0.08&1.73$\pm$1.62&<3.14&<2.43&433&430\\
254&1.10$\pm$0.01&7.67$\pm$0.27&0.79$\pm$0.02&1.70$\pm$0.02&0.018$\pm$0.001&0.79$\pm$0.04&0.10$\pm$0.05&<2.03&1.88$\pm$1.58&<2.72&438&426\\
260&1.08$\pm$0.01&7.81$\pm$0.29&0.82$\pm$0.02&1.71$\pm$0.02&0.019$\pm$0.001&0.82$\pm$0.04&0.10$\pm$0.06&<3.1&<2.42$\pm$1.65&<4.1&482&435\\
266&1.03$\pm$0.01&7.42$\pm$0.28&0.81$\pm$0.02&1.68$\pm$0.02&0.017$\pm$0.001&0.81$\pm$0.04&0.10$\pm$0.07&<2.9&<2.02&2.62$\pm$2.05&427&404\\
272&0.99$\pm$0.01&7.23$\pm$0.31&0.78$\pm$0.02&1.67$\pm$0.02&0.016$\pm$0.001&0.78$\pm$0.04&0.11$\pm$0.06&<2.32&4.18$\pm$1.66&3.38$\pm$2.03&401&414\\
278&1.02$\pm$0.01&7.54$\pm$0.28&0.77$\pm$0.02&1.64$\pm$0.02&0.016$\pm$0.001&0.77$\pm$0.04&0.09$\pm$0.06&<1.11&<2.81&3.14$\pm$2.08&410&403\\
283&1.06$\pm$0.01&7.59$\pm$0.29&0.80$\pm$0.02&1.69$\pm$0.02&0.018$\pm$0.001&0.80$\pm$0.03&0.12$\pm$0.07&<1.17&<1.65&<1.76&391&397\\
289&1.05$\pm$0.01&7.63$\pm$0.28&0.79$\pm$0.02&1.67$\pm$0.02&0.017$\pm$0.001&0.79$\pm$0.04&0.14$\pm$0.1&<2.86&2.25$\pm$1.61&<3.89&439&413\\
295&1.04$\pm$0.01&7.48$\pm$0.26&0.78$\pm$0.02&1.66$\pm$0.02&0.017$\pm$0.001&0.78$\pm$0.03&0.13$\pm$0.08&<2.82&2.48$\pm$1.59&2.14$\pm$2.01&480&423\\
301&1.09$\pm$0.01&7.89$\pm$0.26&0.80$\pm$0.02&1.68$\pm$0.02&0.018$\pm$0.001&0.80$\pm$0.03&0.10$\pm$0.05&1.77$\pm$1.62&1.74$\pm$1.61&<2.29&501&439\\
304&1.06$\pm$0.03&8.0$\pm$0.5&0.77$\pm$0.03&1.62$\pm$0.05&0.017$\pm$0.001&<0.77&0.22$\pm$0.2&3.77$\pm$2.53&<3.36&4.41$\pm$3.27&133&144\\
\hline
\end{tabular}
\end{table*}
%
%
\begin{figure*}
\centering 
  \subfigure{\includegraphics[width=0.19\linewidth]{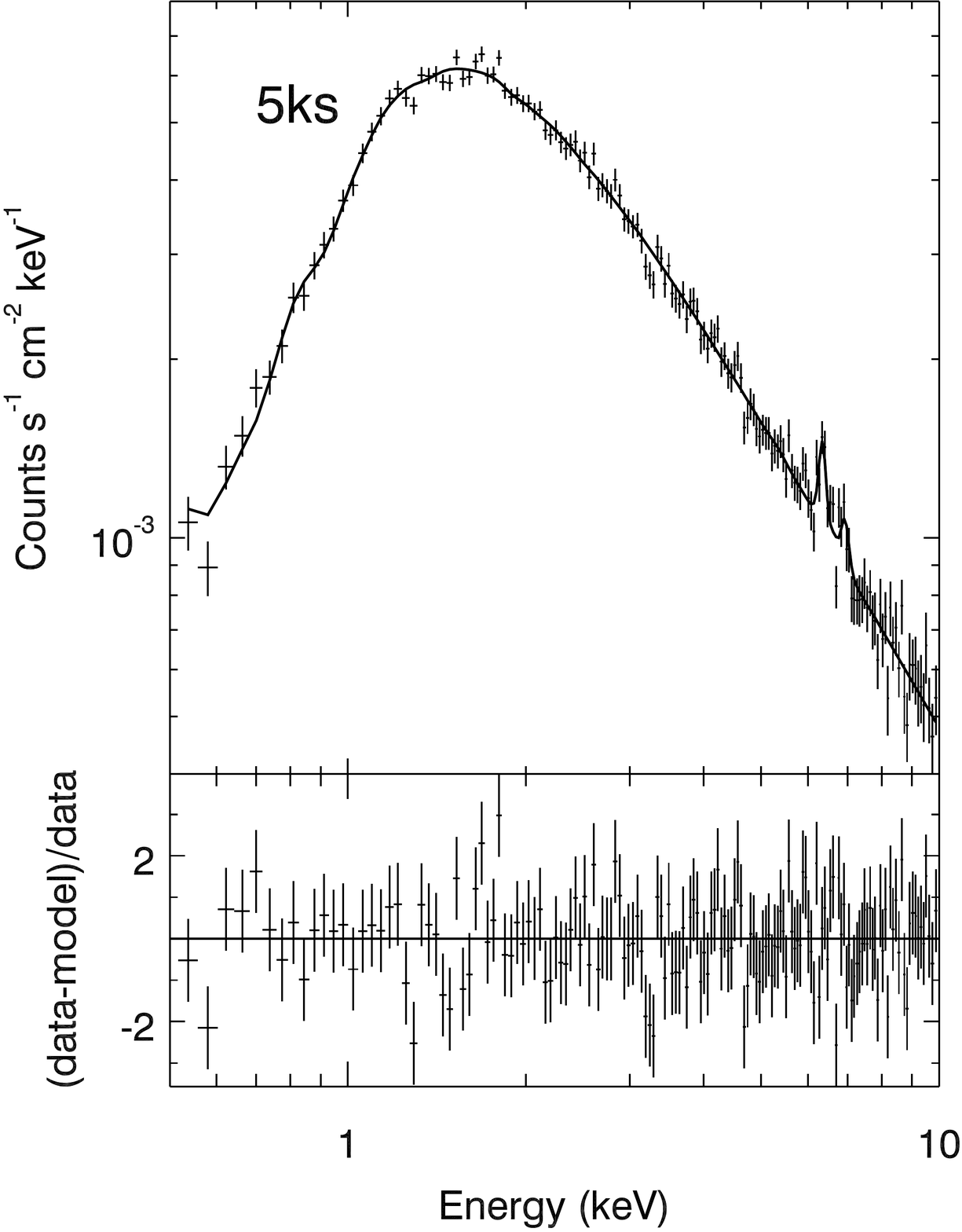}}
  \subfigure{\includegraphics[width=0.19\linewidth]{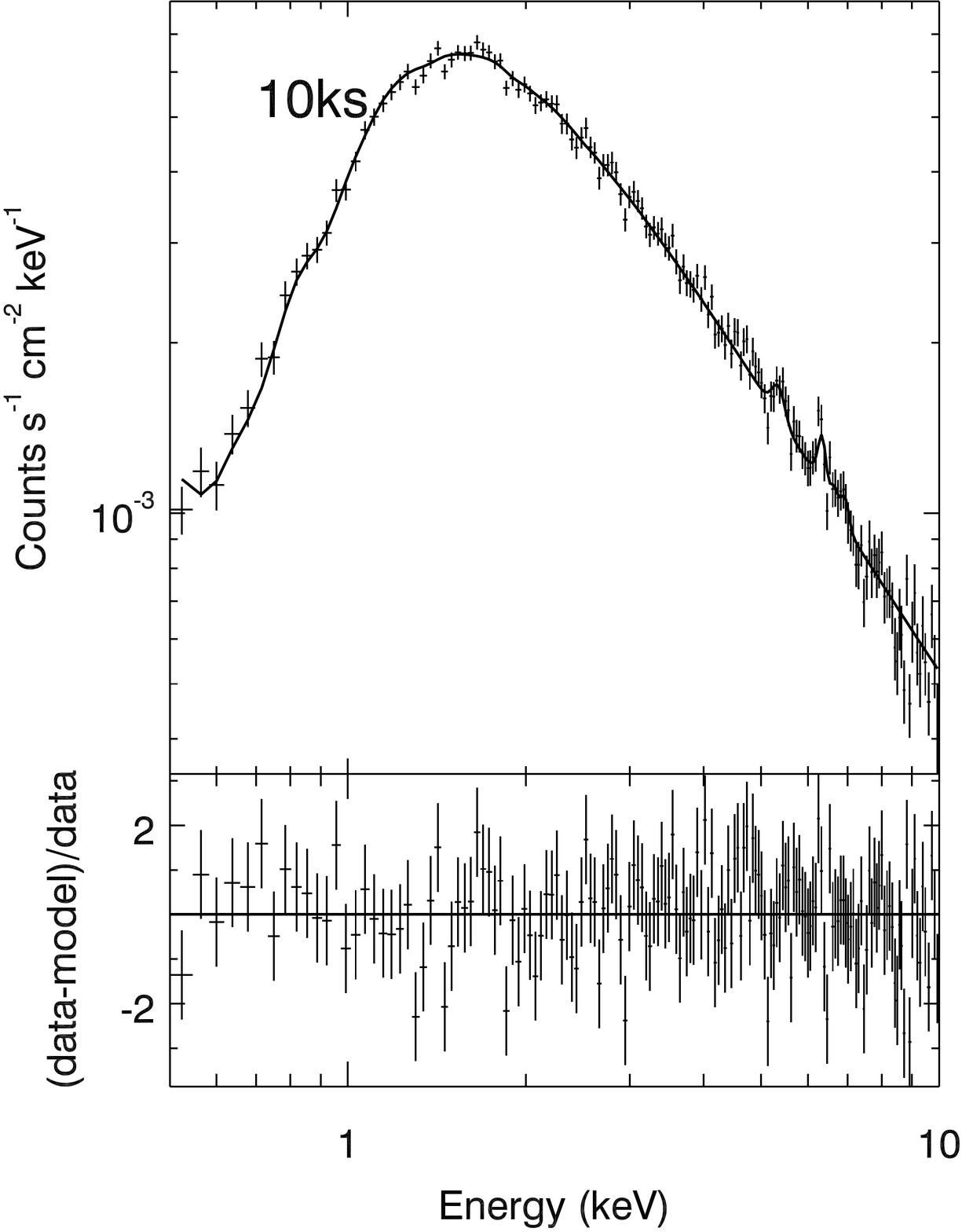}}
  \subfigure{\includegraphics[width=0.19\linewidth]{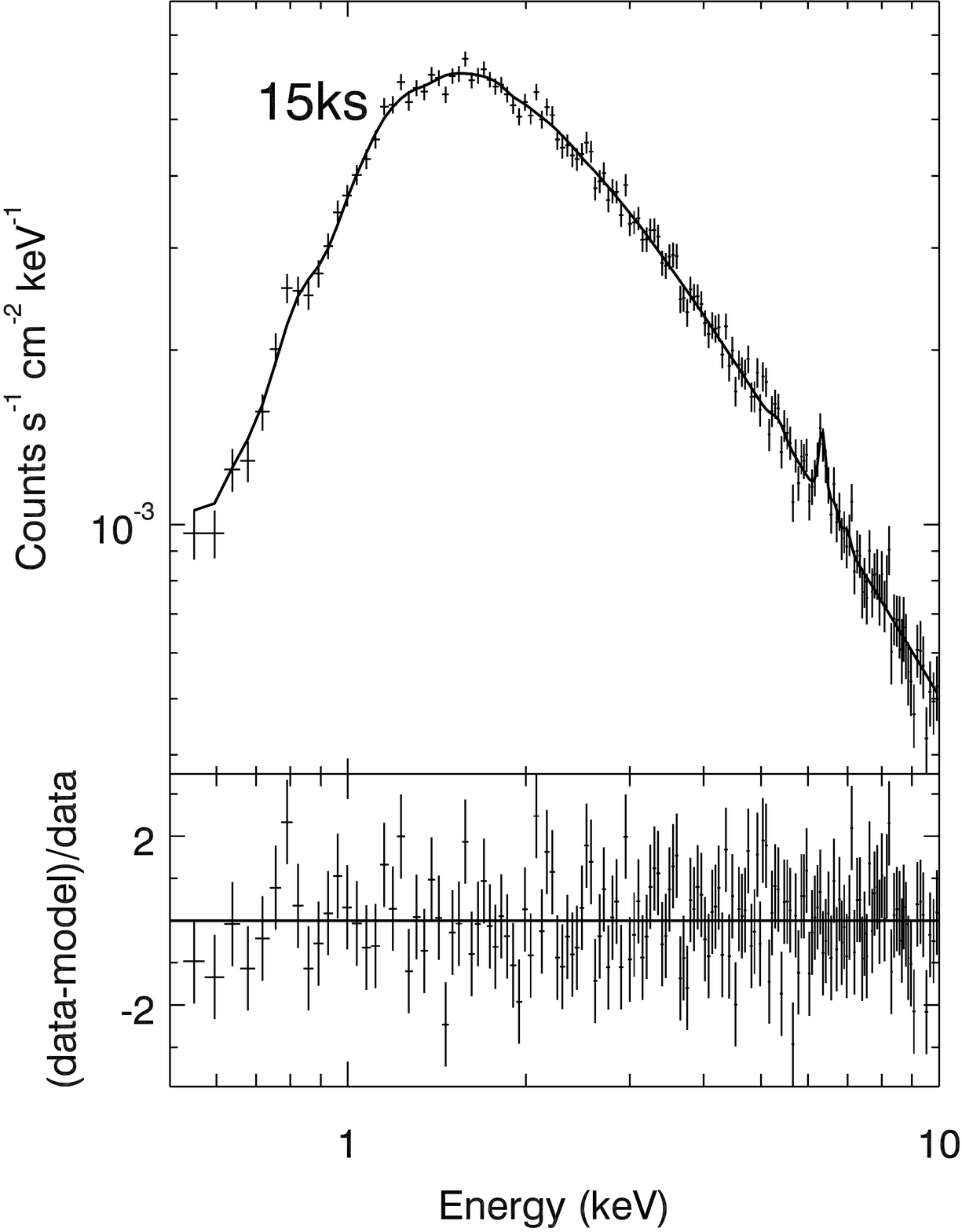}}
  \subfigure{\includegraphics[width=0.19\linewidth]{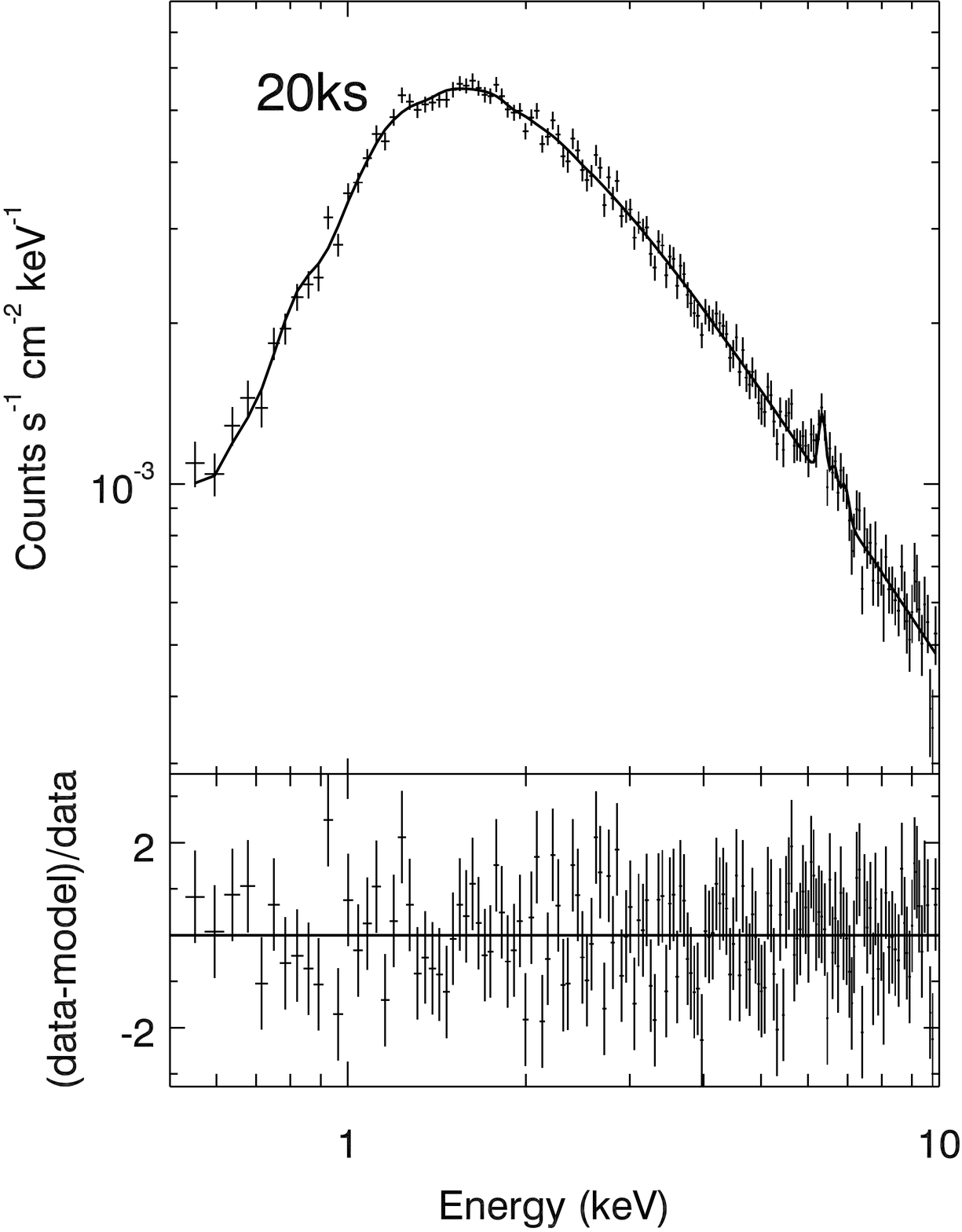}}
  \subfigure{\includegraphics[width=0.19\linewidth]{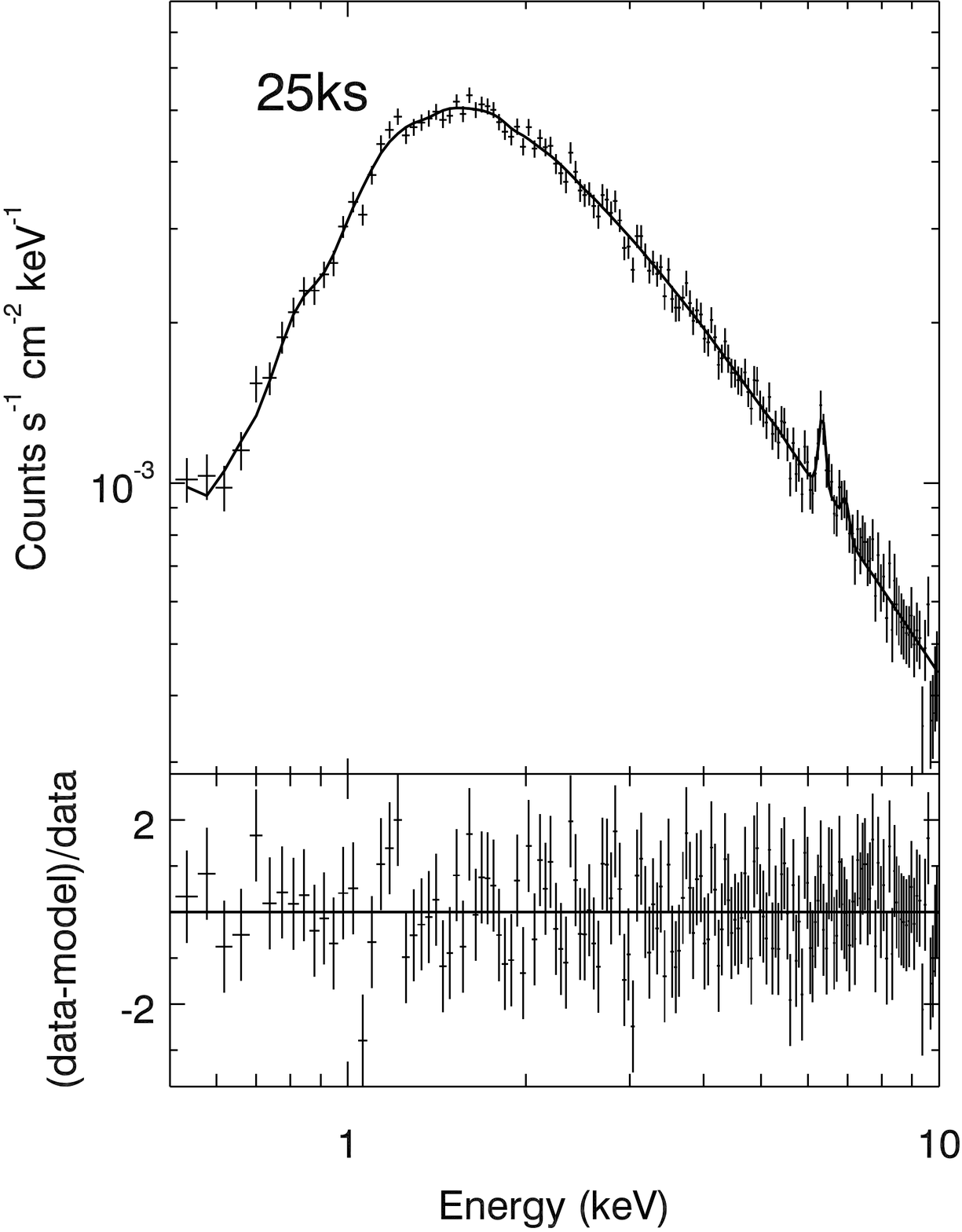}}

  \subfigure{\includegraphics[width=0.19\textwidth]{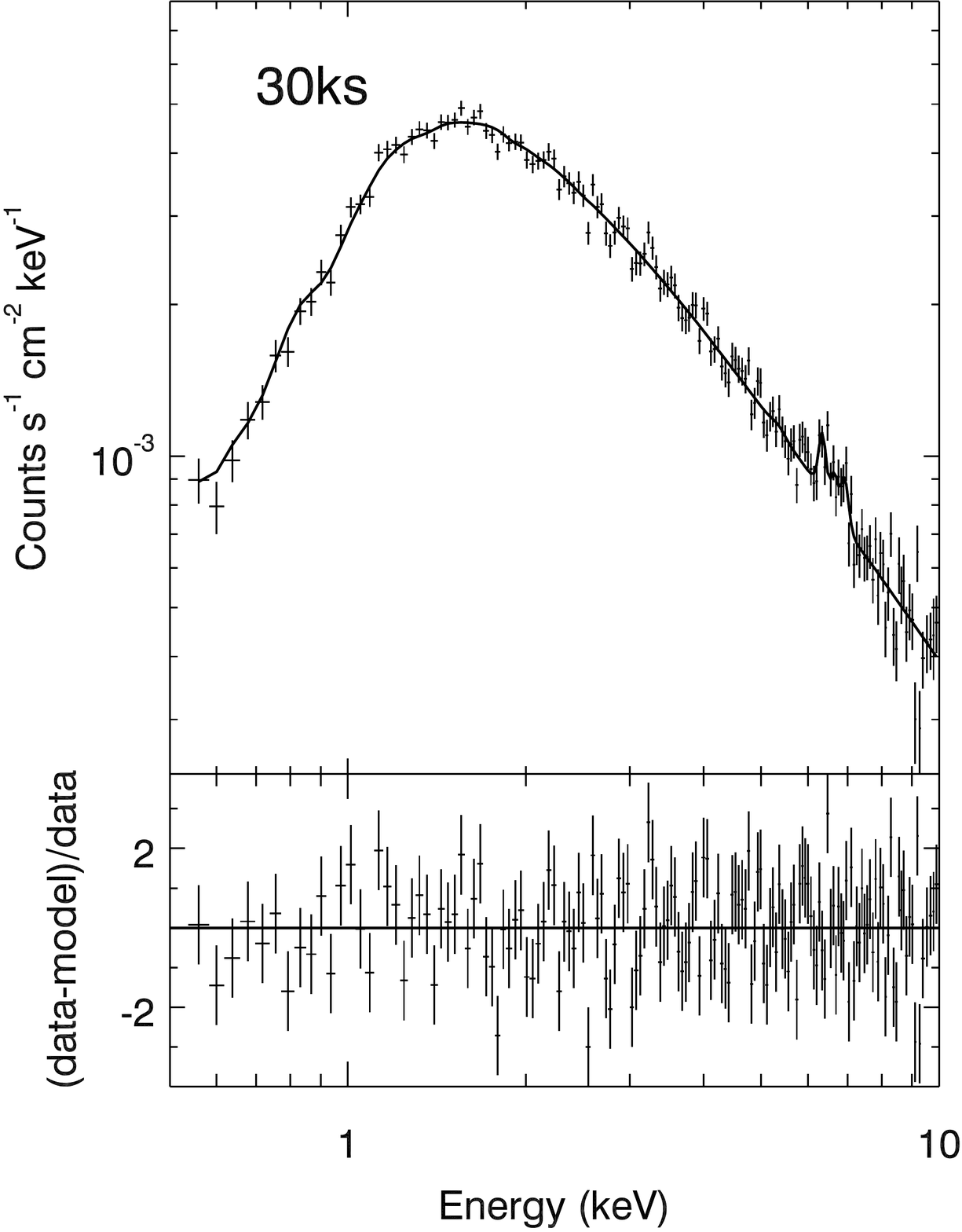}}
  \subfigure{\includegraphics[width=0.19\textwidth]{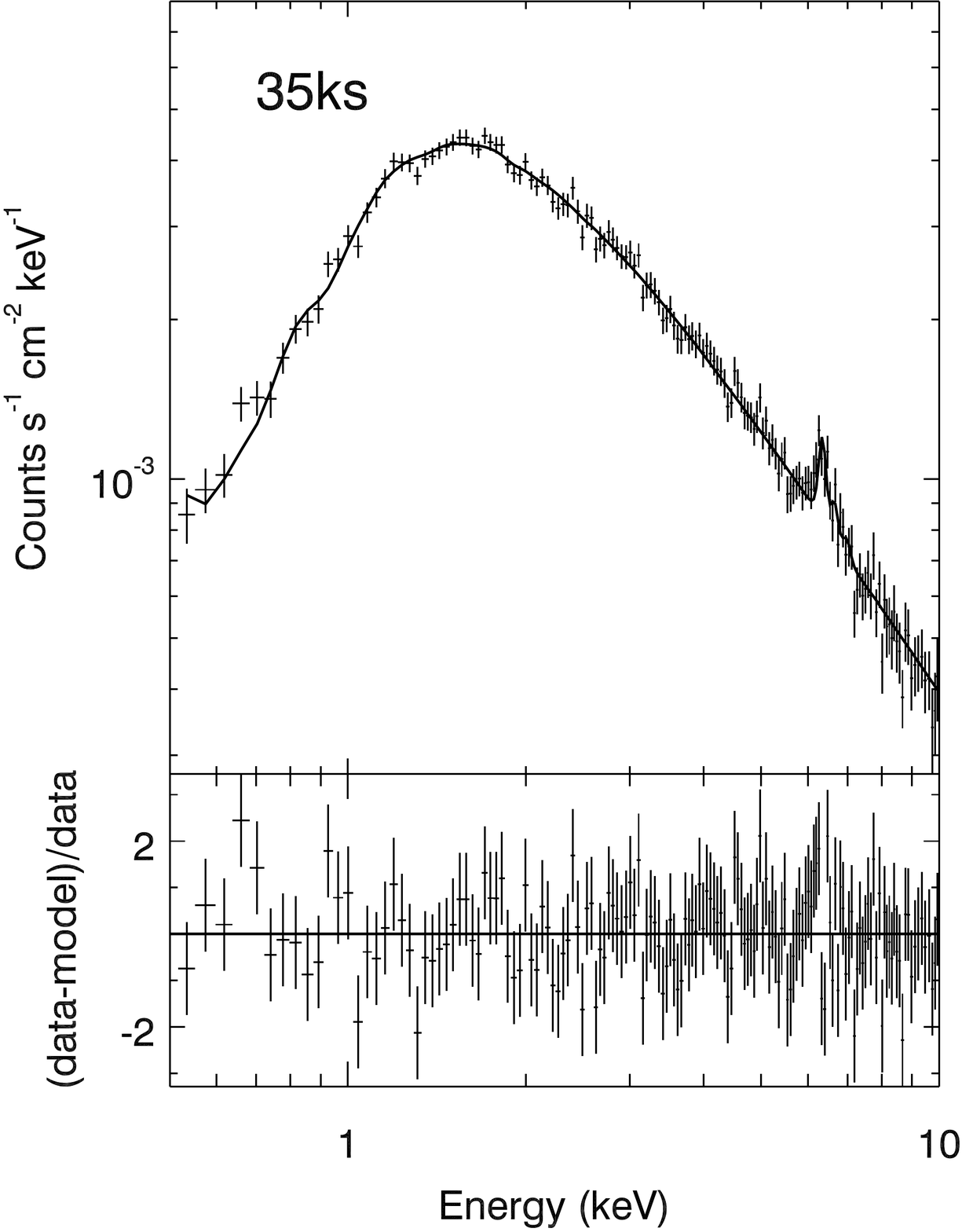}}
  \subfigure{\includegraphics[width=0.19\textwidth]{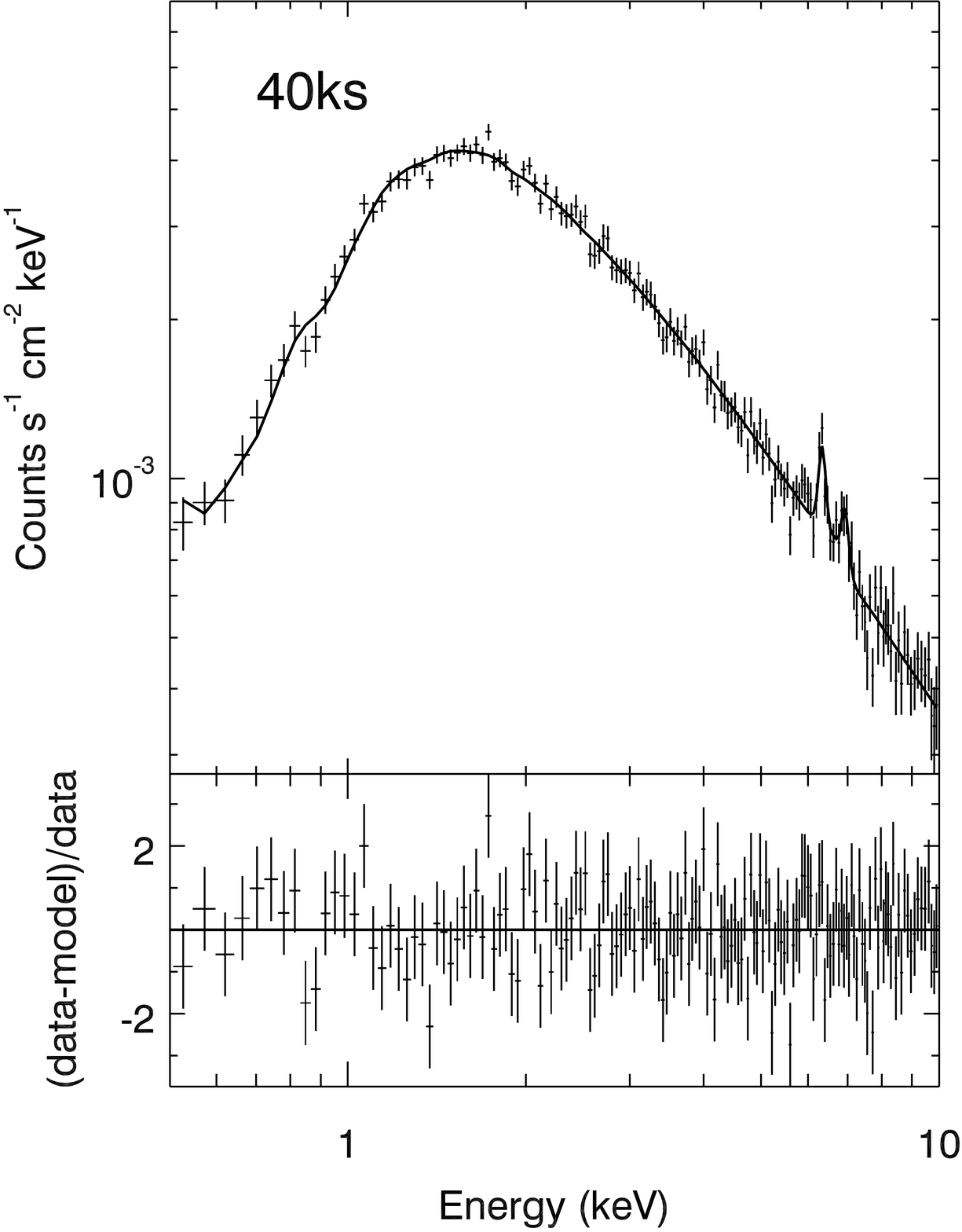}}
  \subfigure{\includegraphics[width=0.19\textwidth]{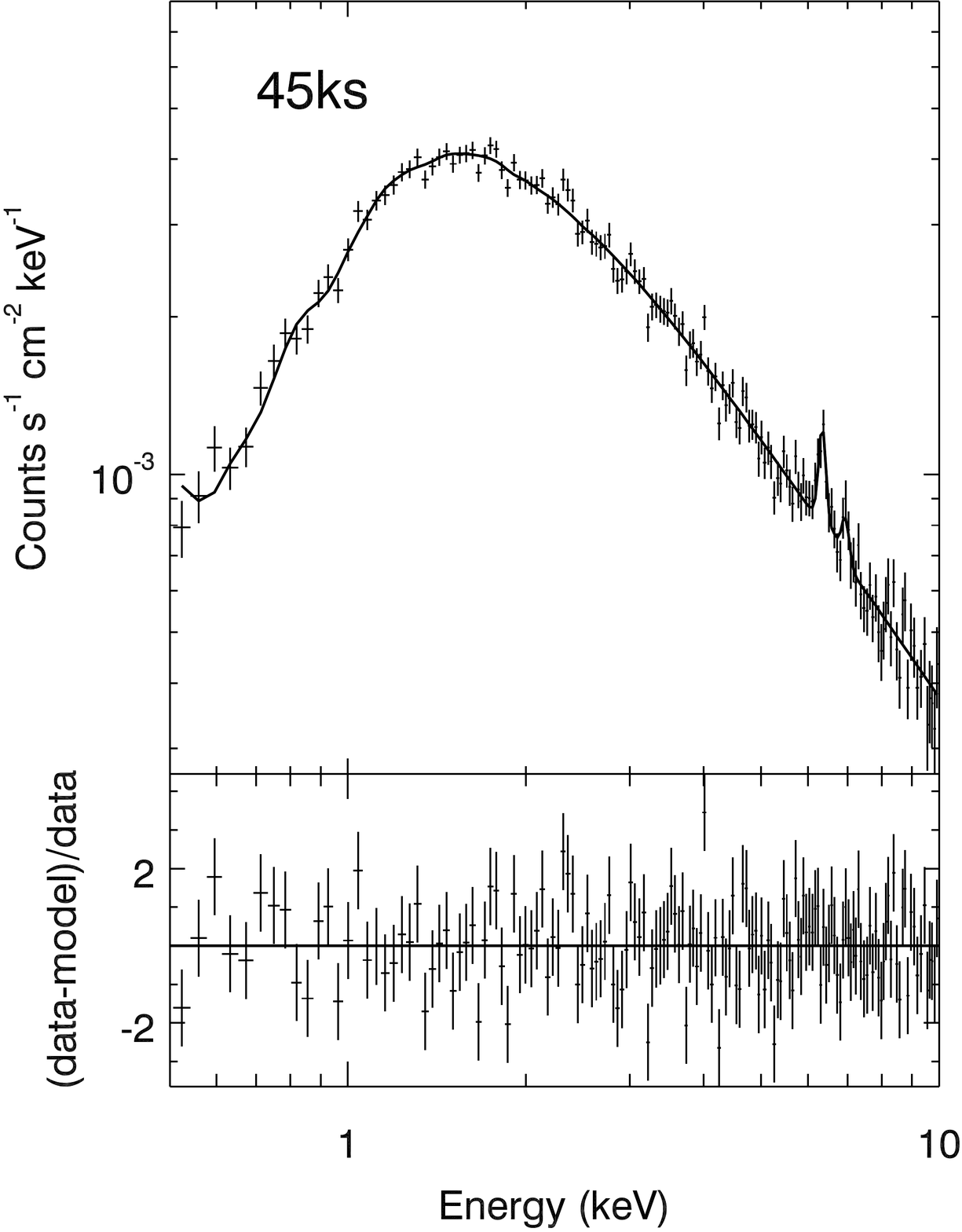}}
  \subfigure{\includegraphics[width=0.19\textwidth]{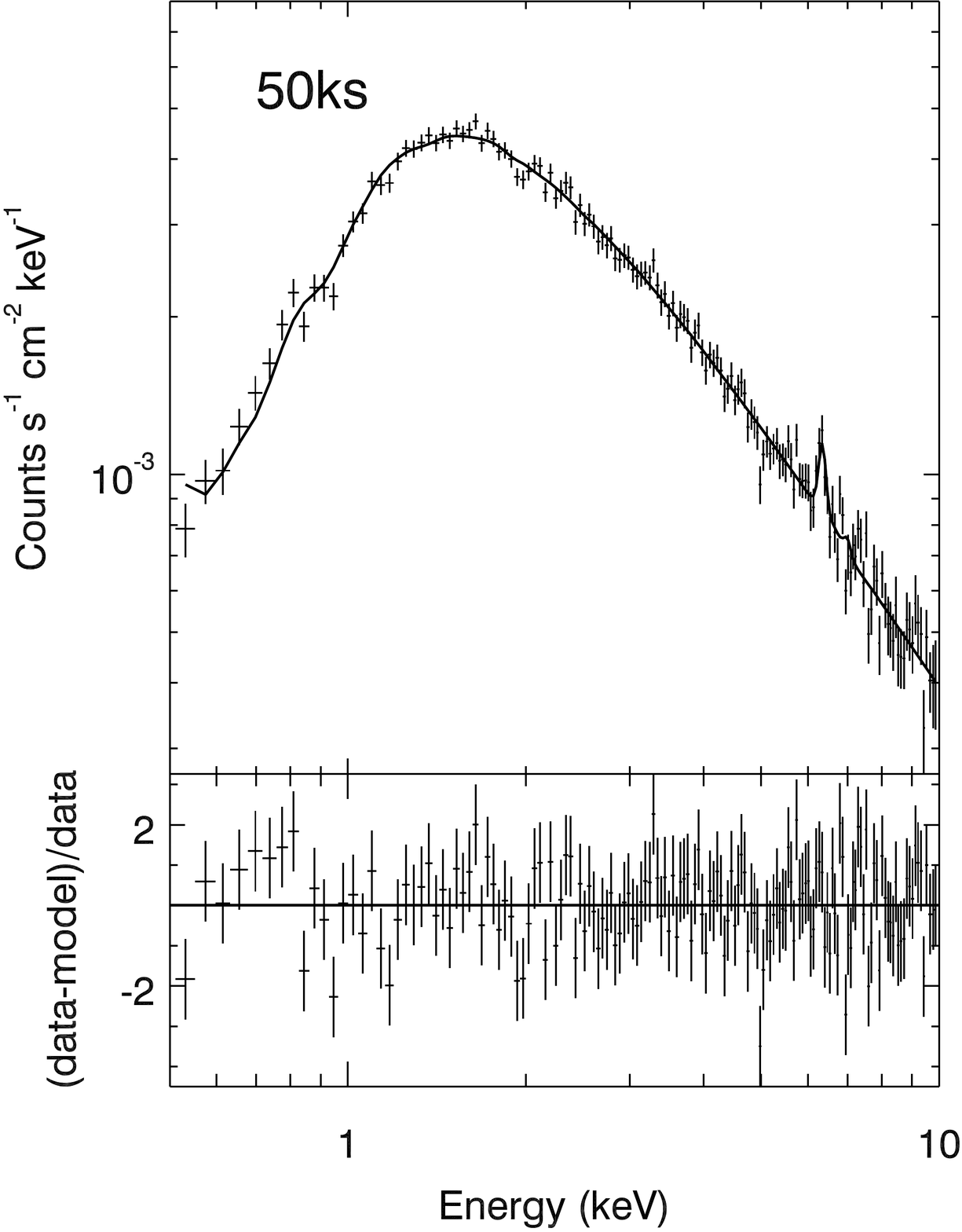}}

  \subfigure{\includegraphics[width=0.19\textwidth]{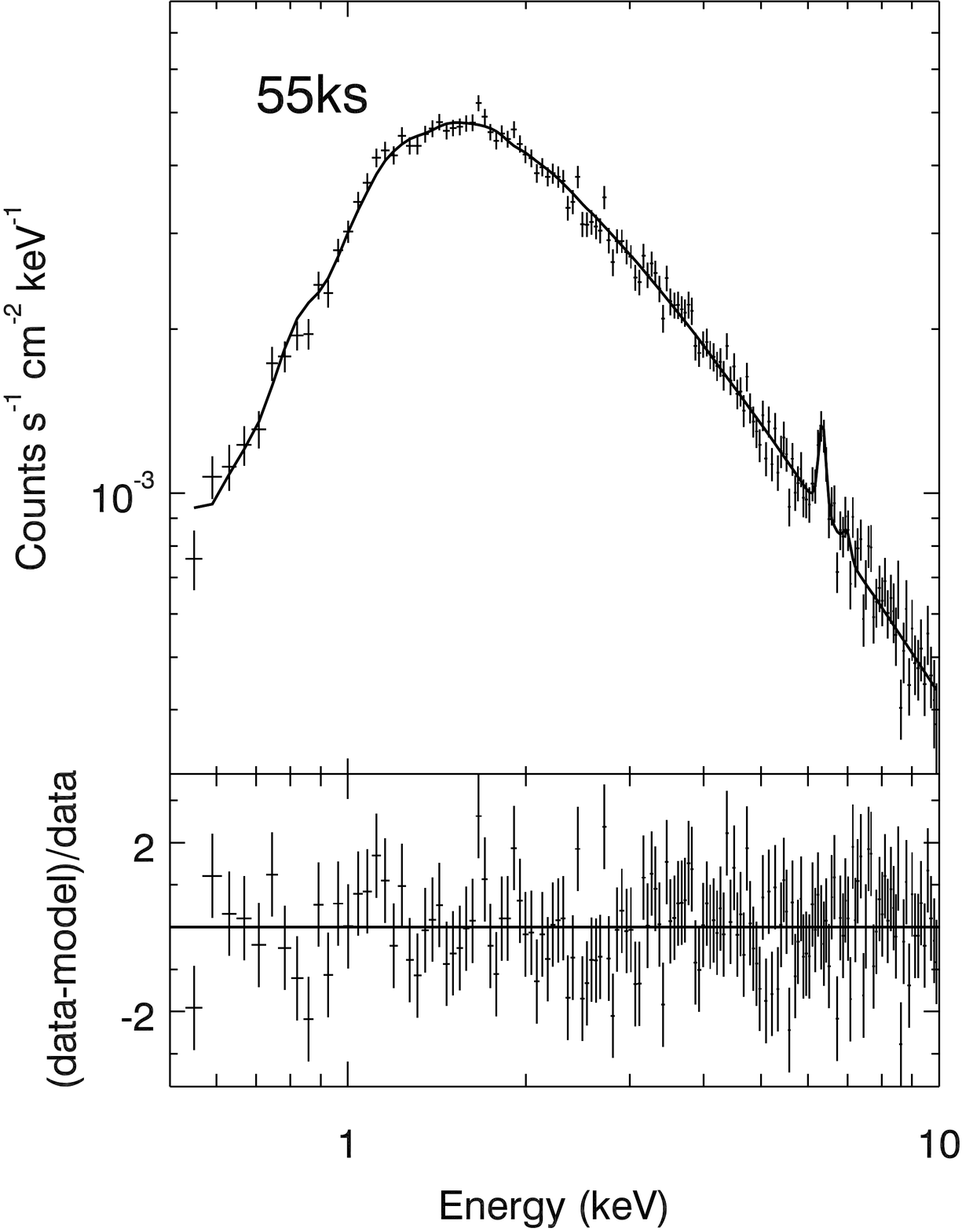}}
  \subfigure{\includegraphics[width=0.19\textwidth]{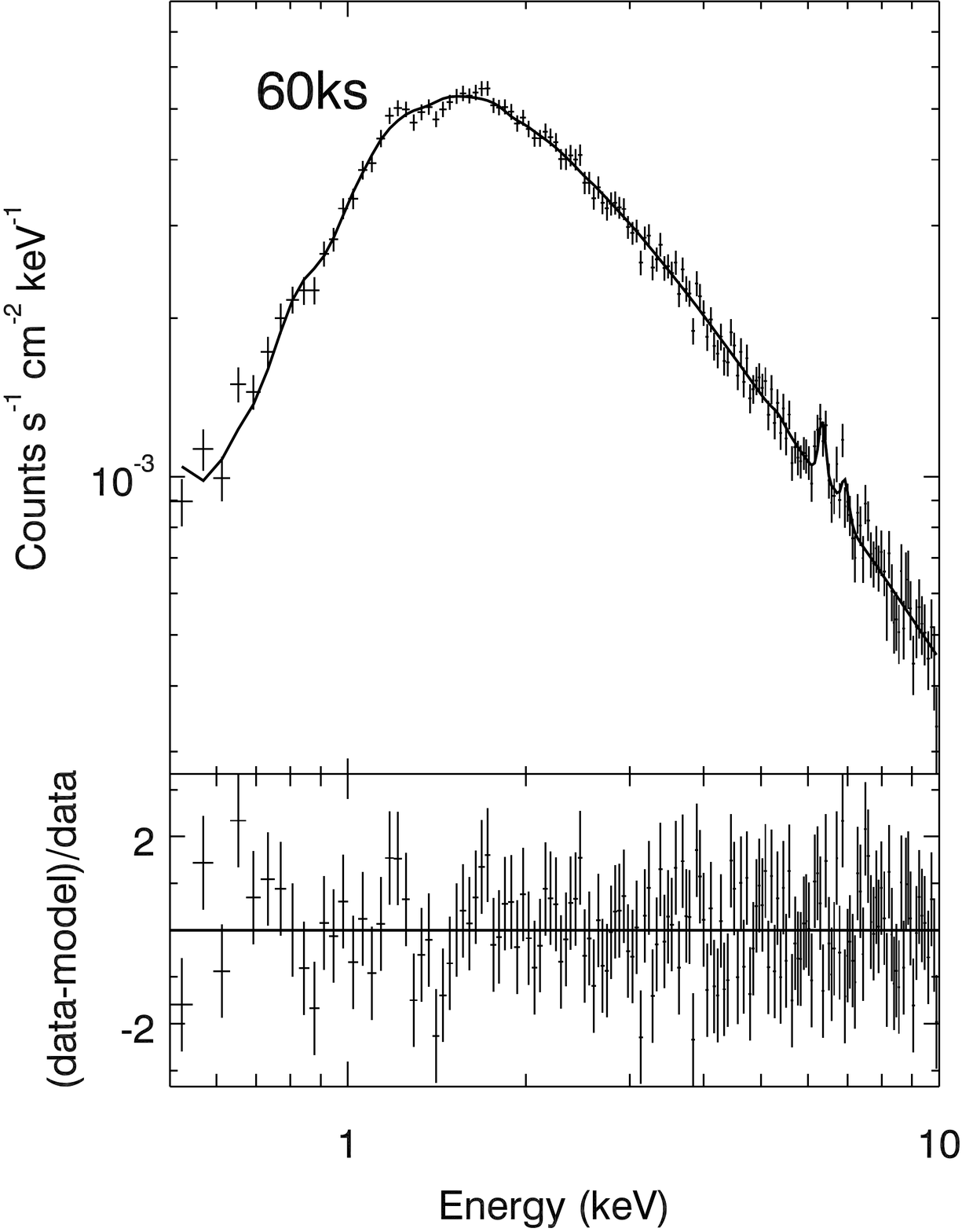}}
  \subfigure{\includegraphics[width=0.19\textwidth]{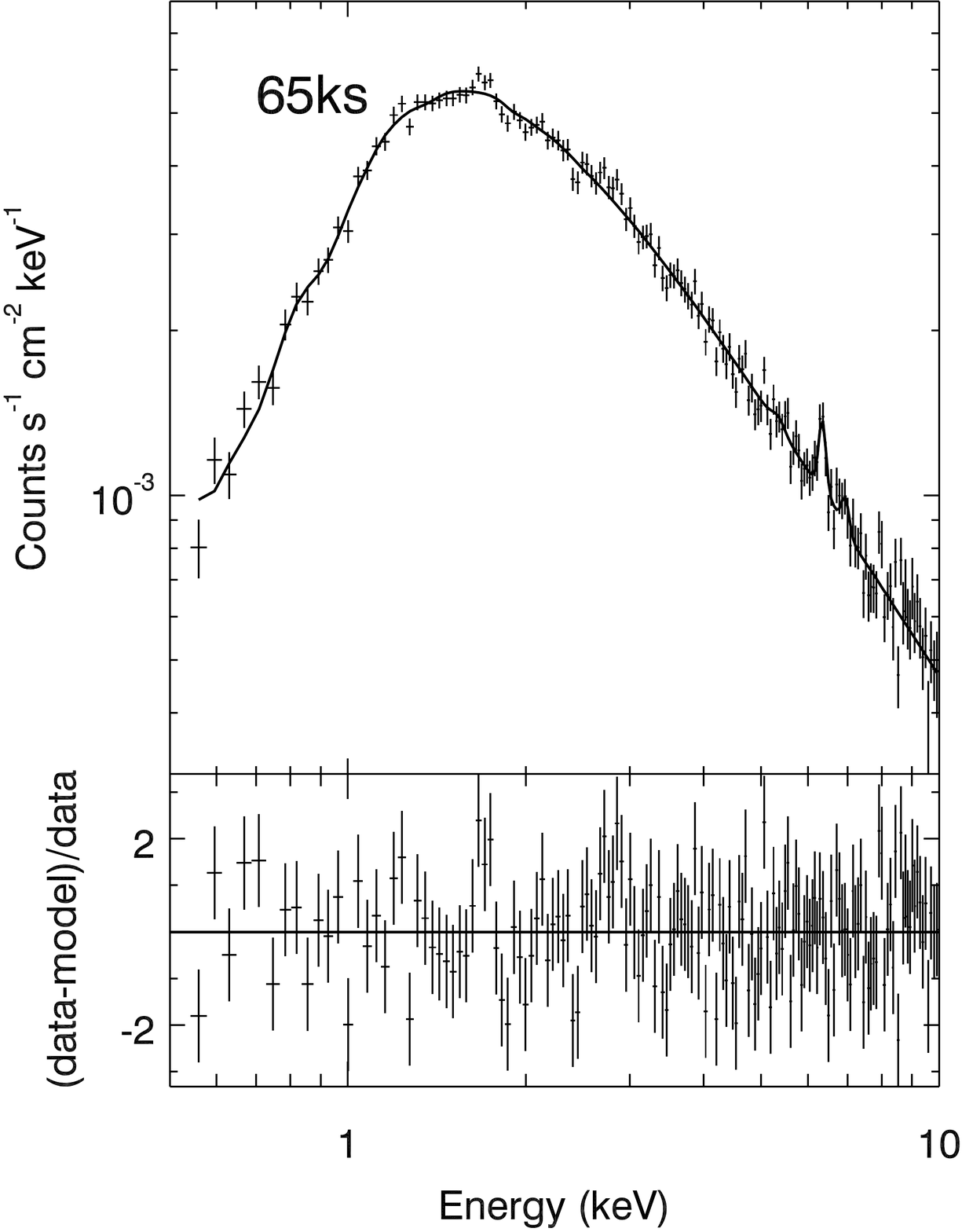}}
  \subfigure{\includegraphics[width=0.19\textwidth]{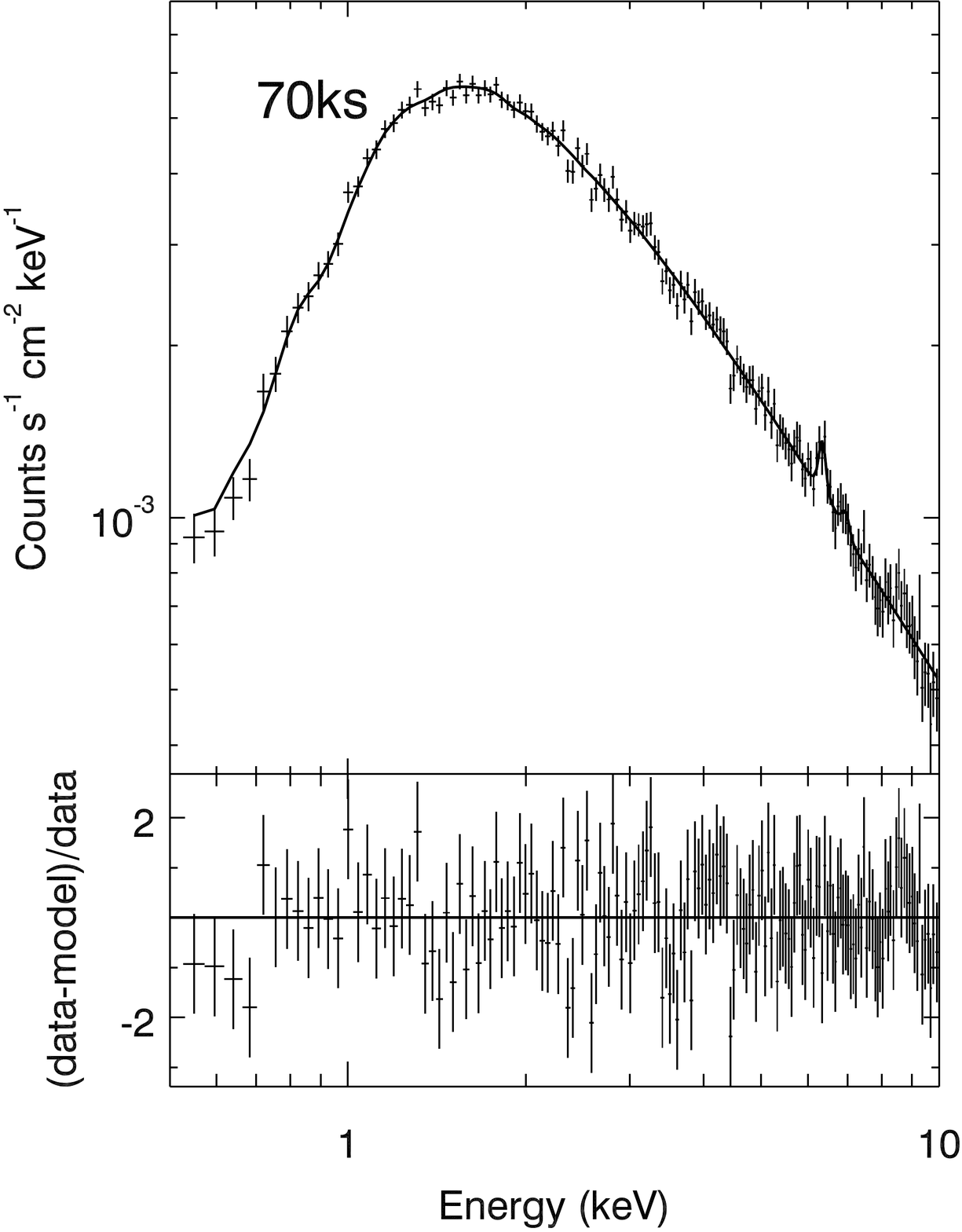}}
  \subfigure{\includegraphics[width=0.19\textwidth]{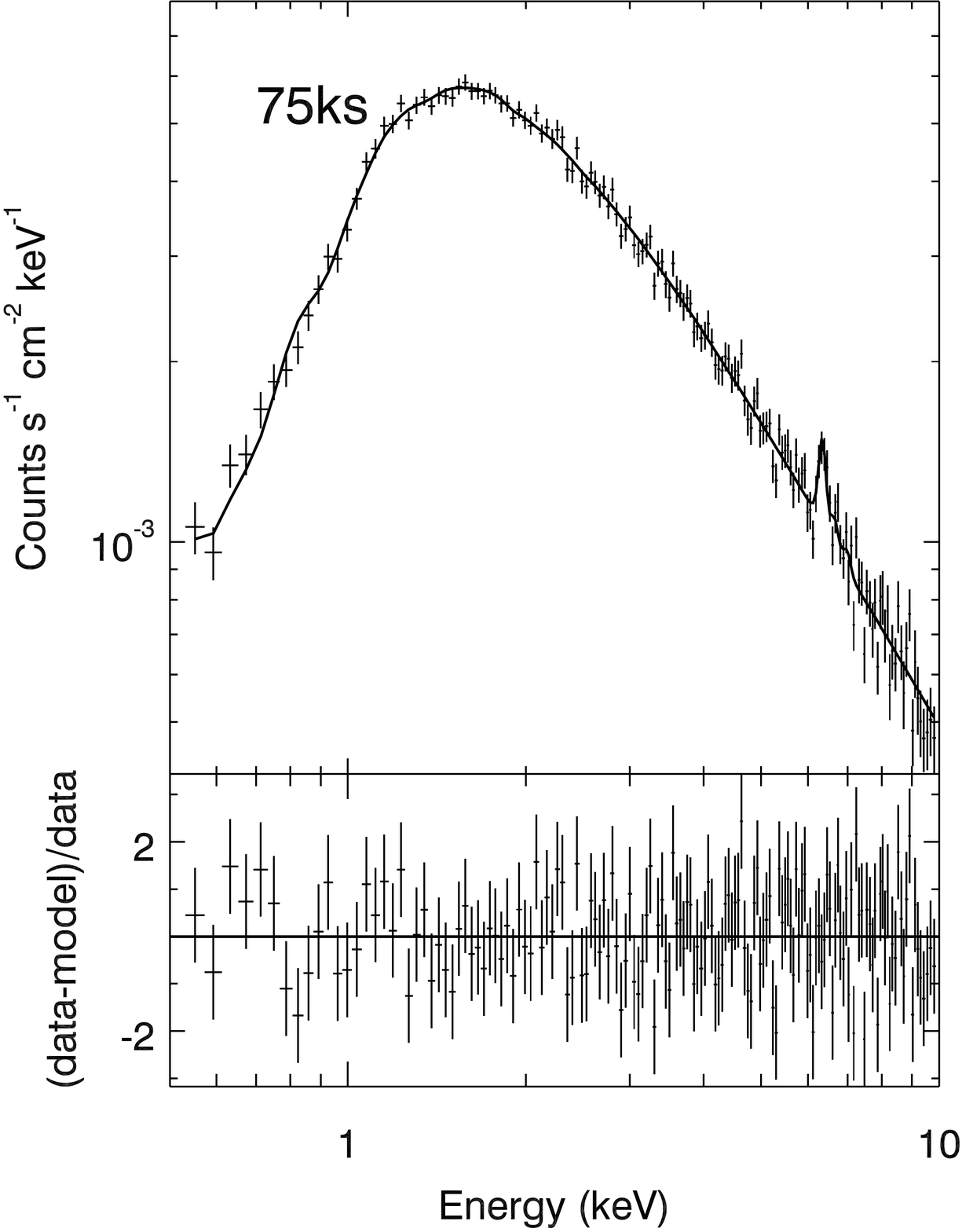}}
  
  \subfigure{\includegraphics[width=0.19\textwidth]{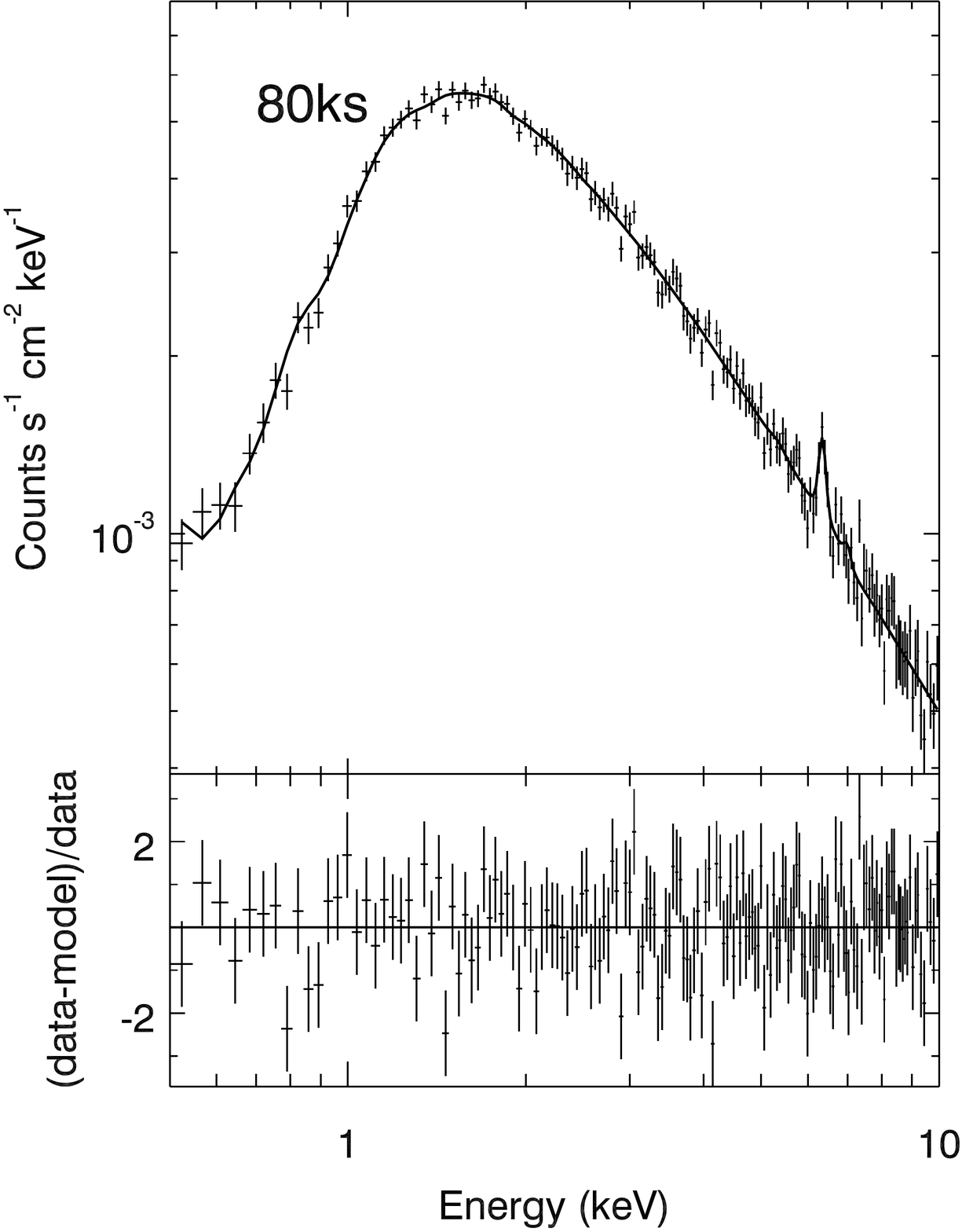}}
  \subfigure{\includegraphics[width=0.19\textwidth]{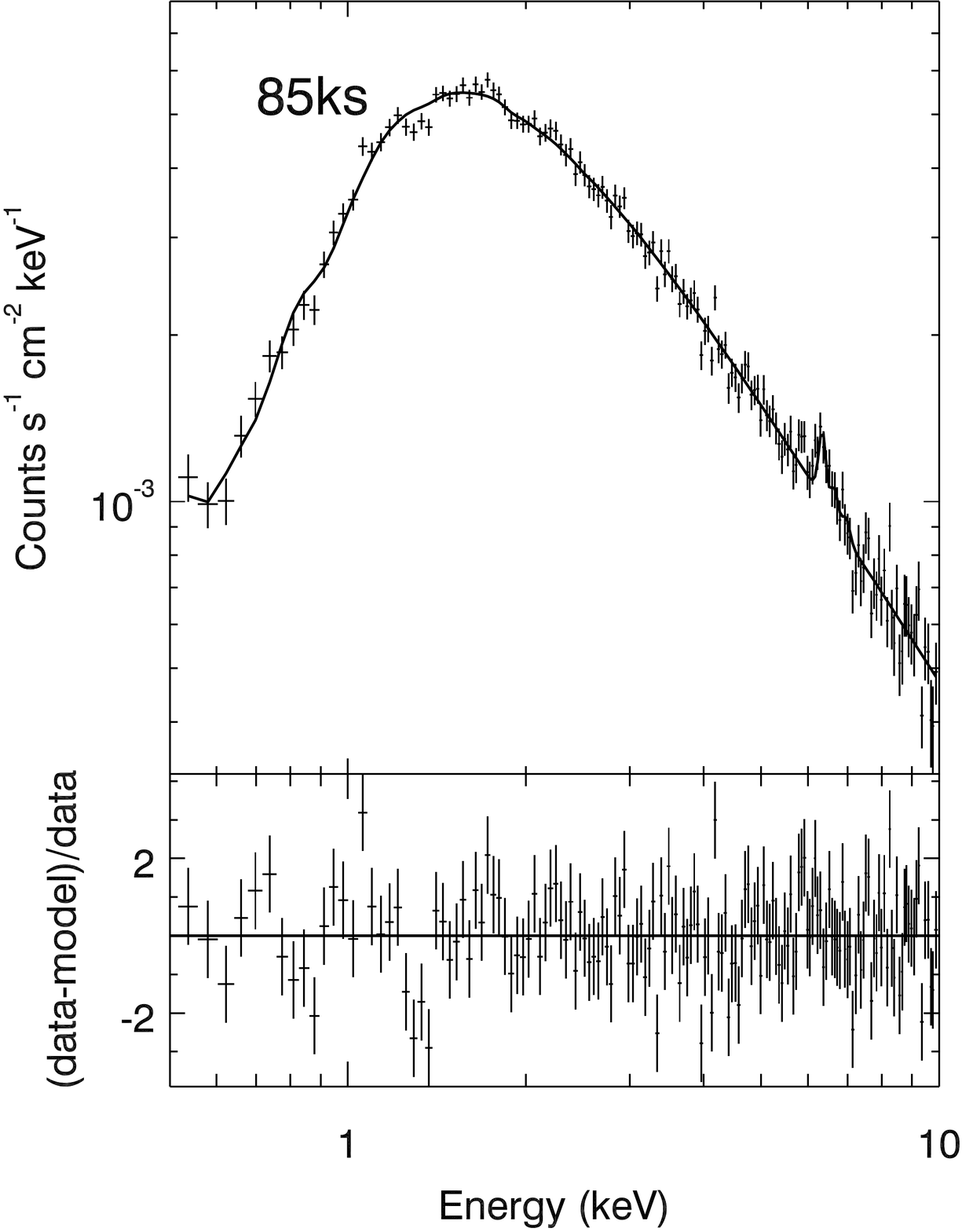}}
  \subfigure{\includegraphics[width=0.19\textwidth]{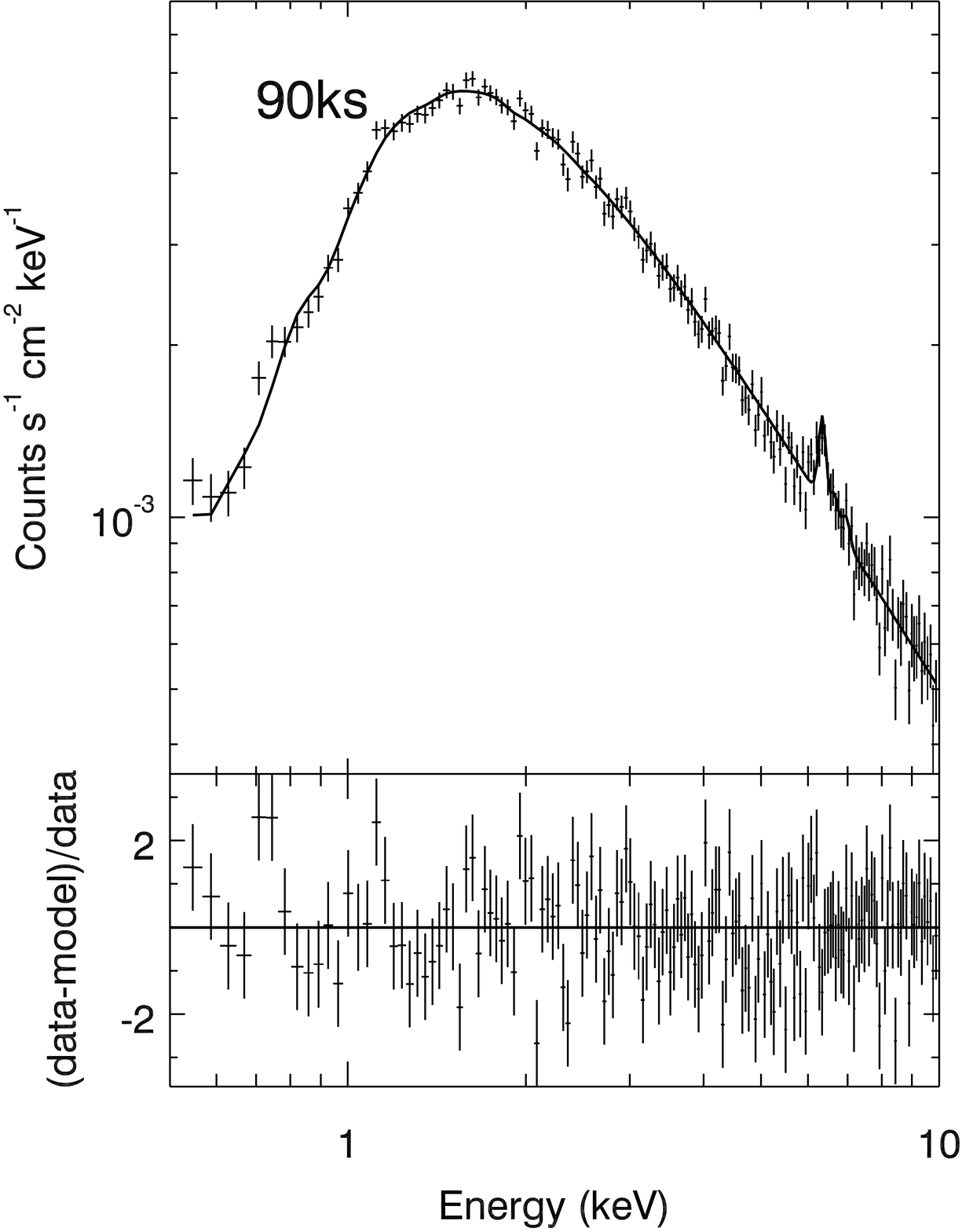}}
  \subfigure{\includegraphics[width=0.19\textwidth]{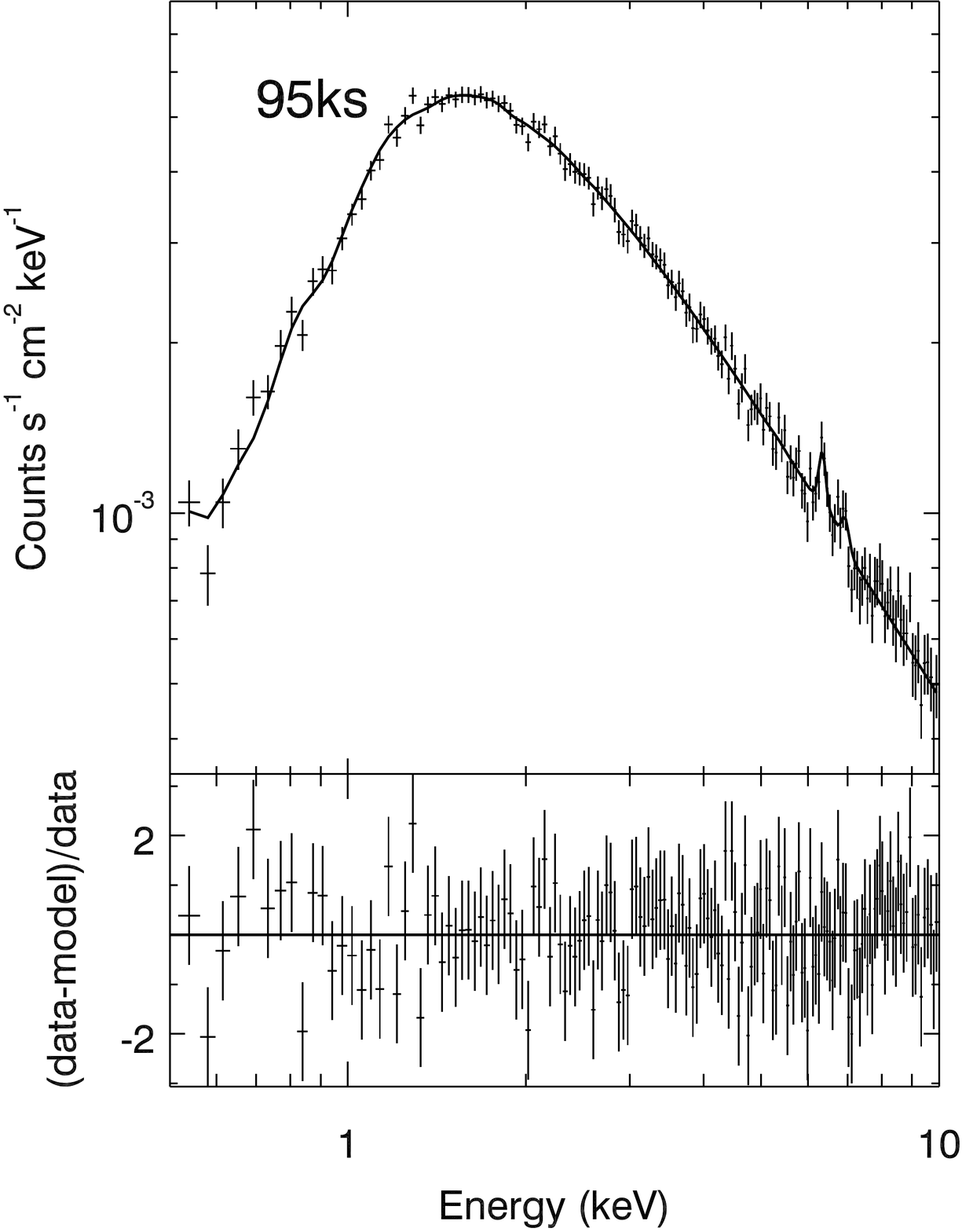}}
  \subfigure{\includegraphics[width=0.19\textwidth]{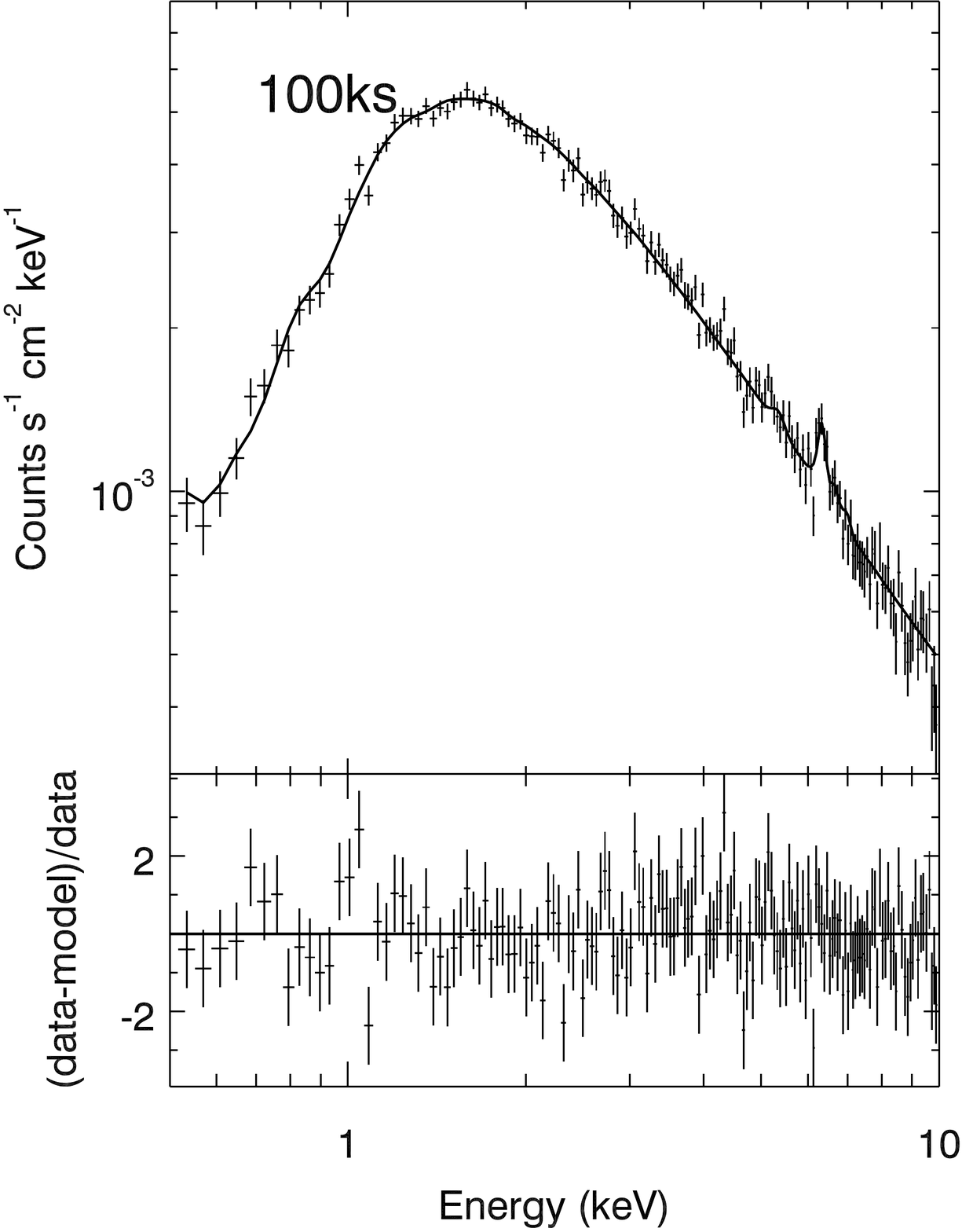}}
  
  \subfigure{\includegraphics[width=0.19\textwidth]{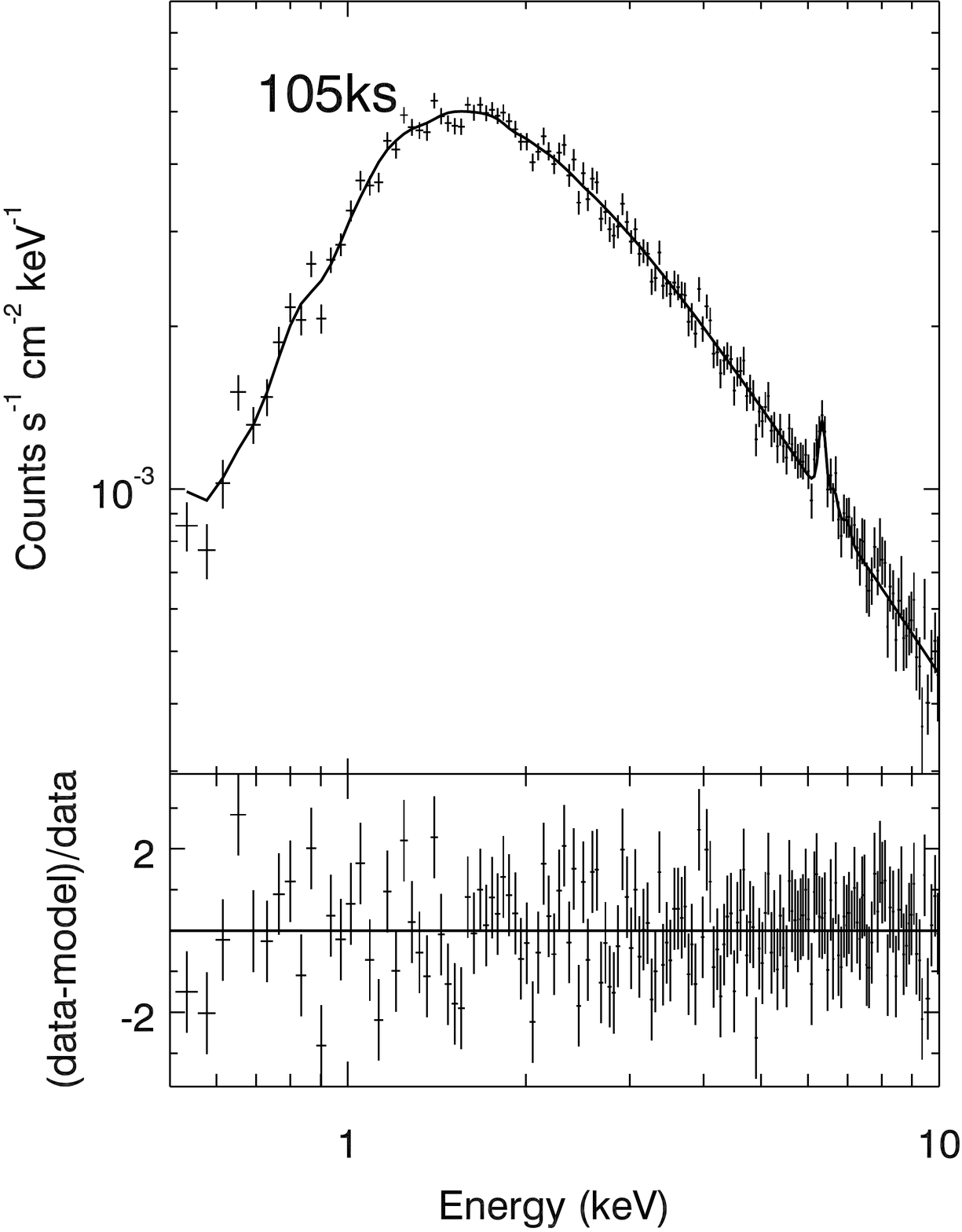}}
  \subfigure{\includegraphics[width=0.19\textwidth]{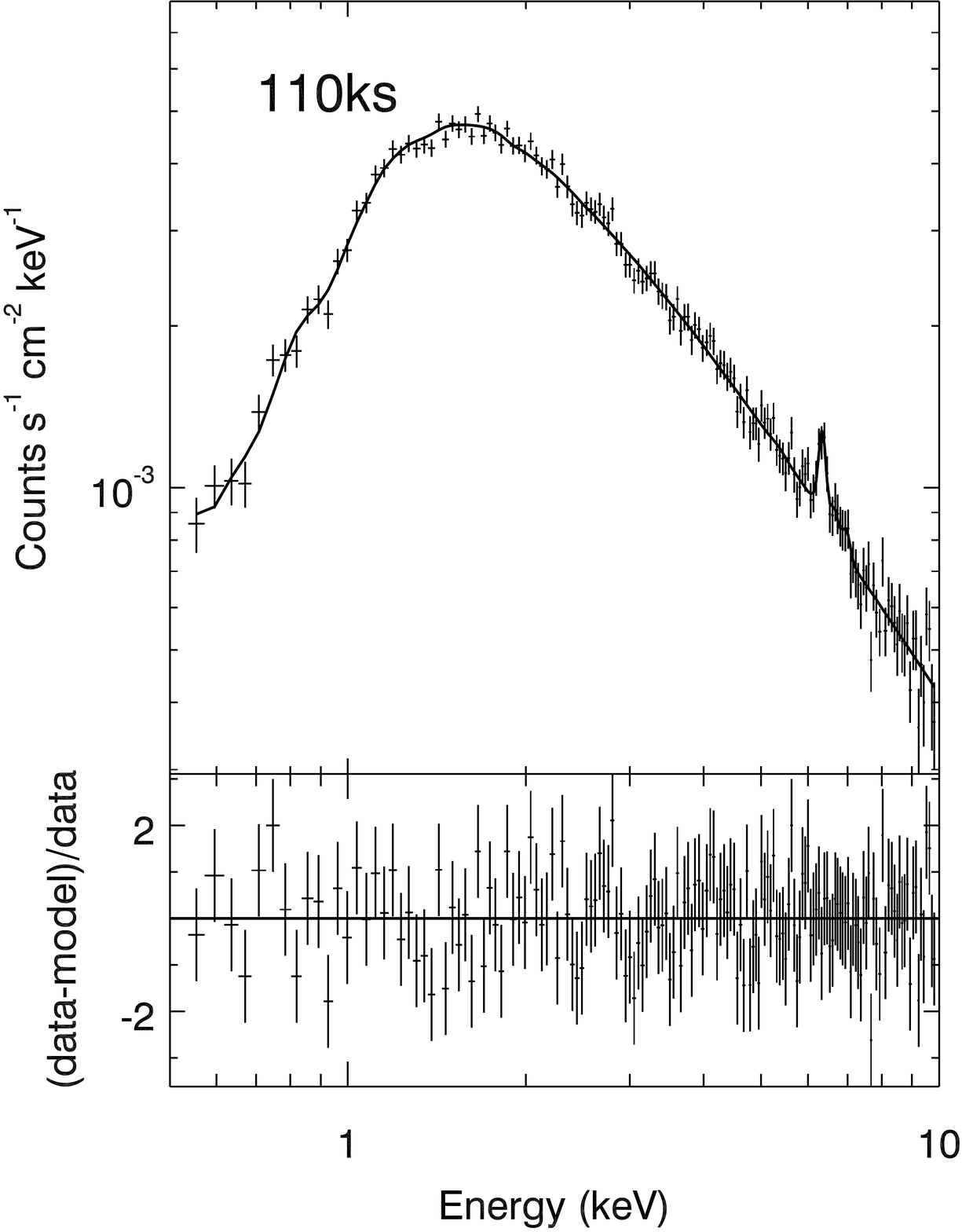}}
  \subfigure{\includegraphics[width=0.19\textwidth]{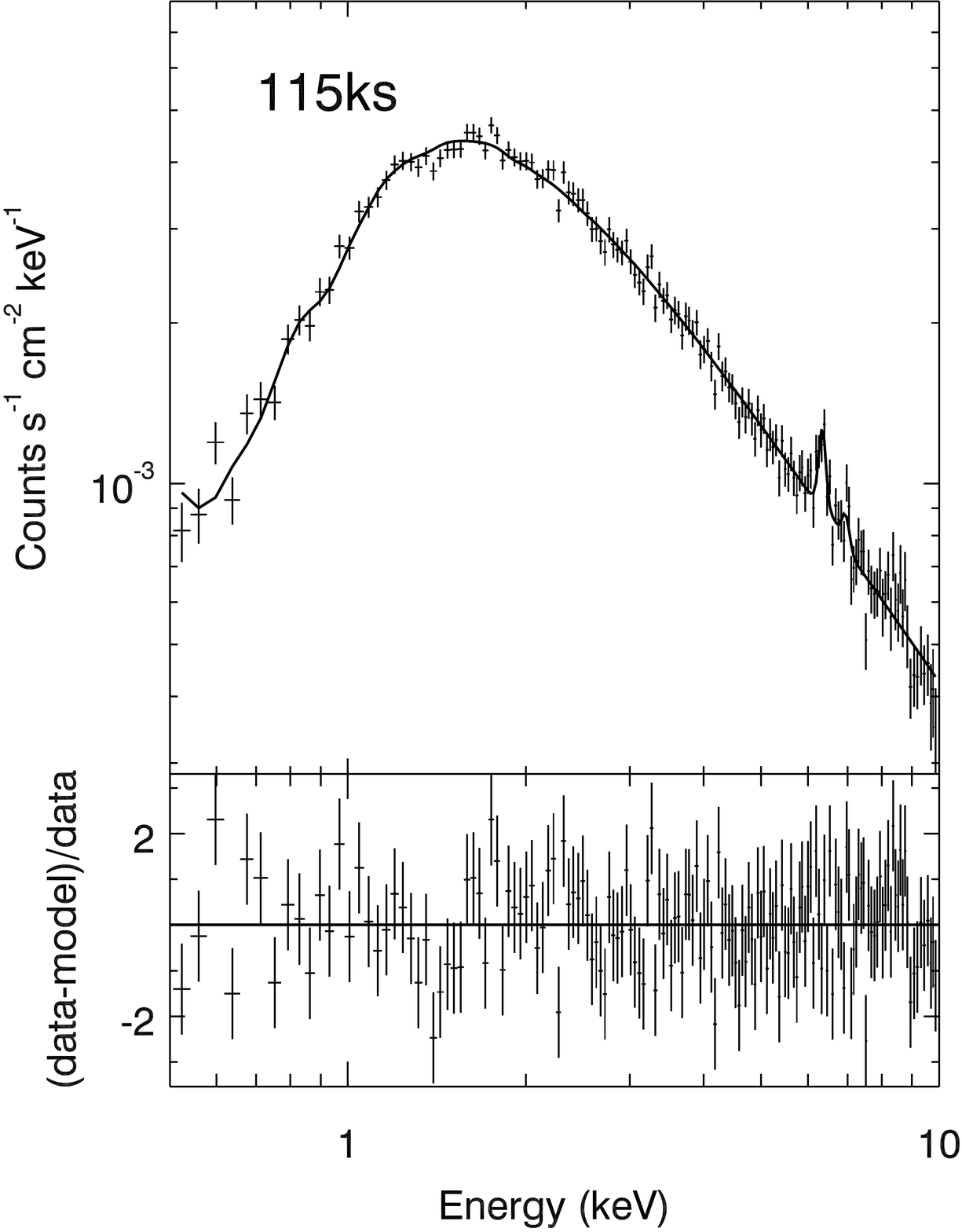}}
  \subfigure{\includegraphics[width=0.19\textwidth]{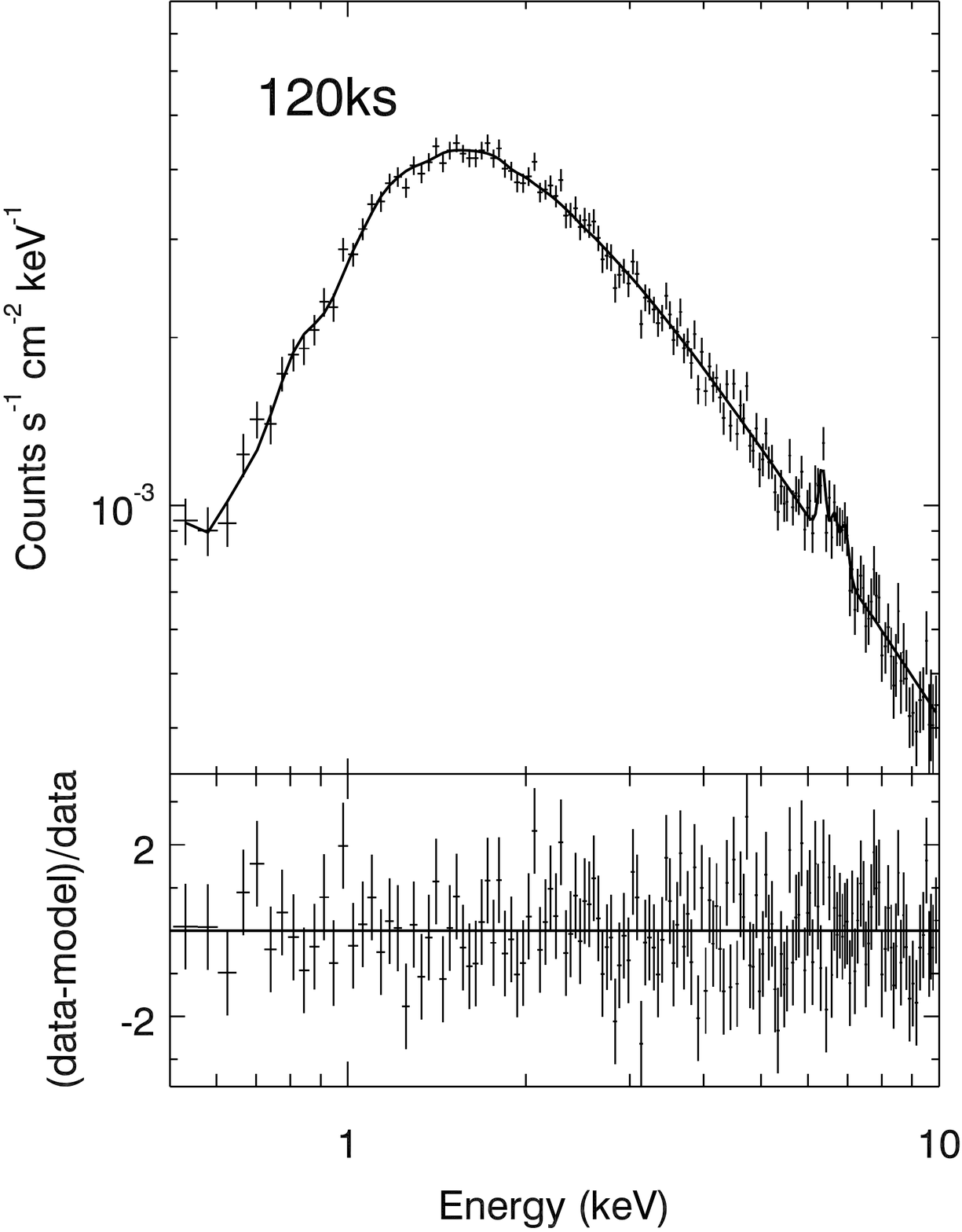}}
  \subfigure{\includegraphics[width=0.19\textwidth]{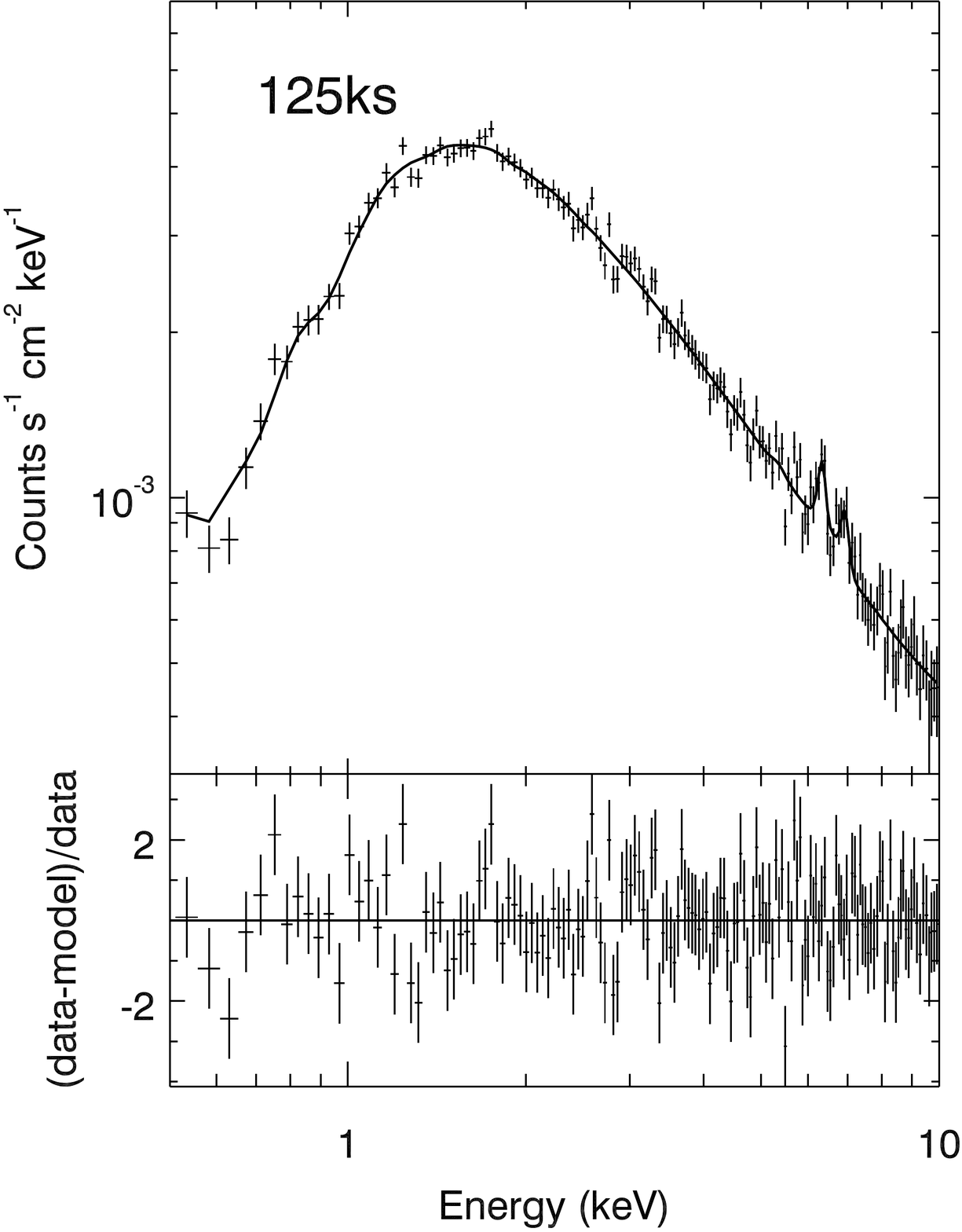}}
\caption{Spectral slices $\sim$5 ks long best-fitted in accordance with Sect. 5.4}
\label{slices1}
\end{figure*}

\begin{figure*}
\centering 

  \subfigure{\includegraphics[width=0.19\textwidth]{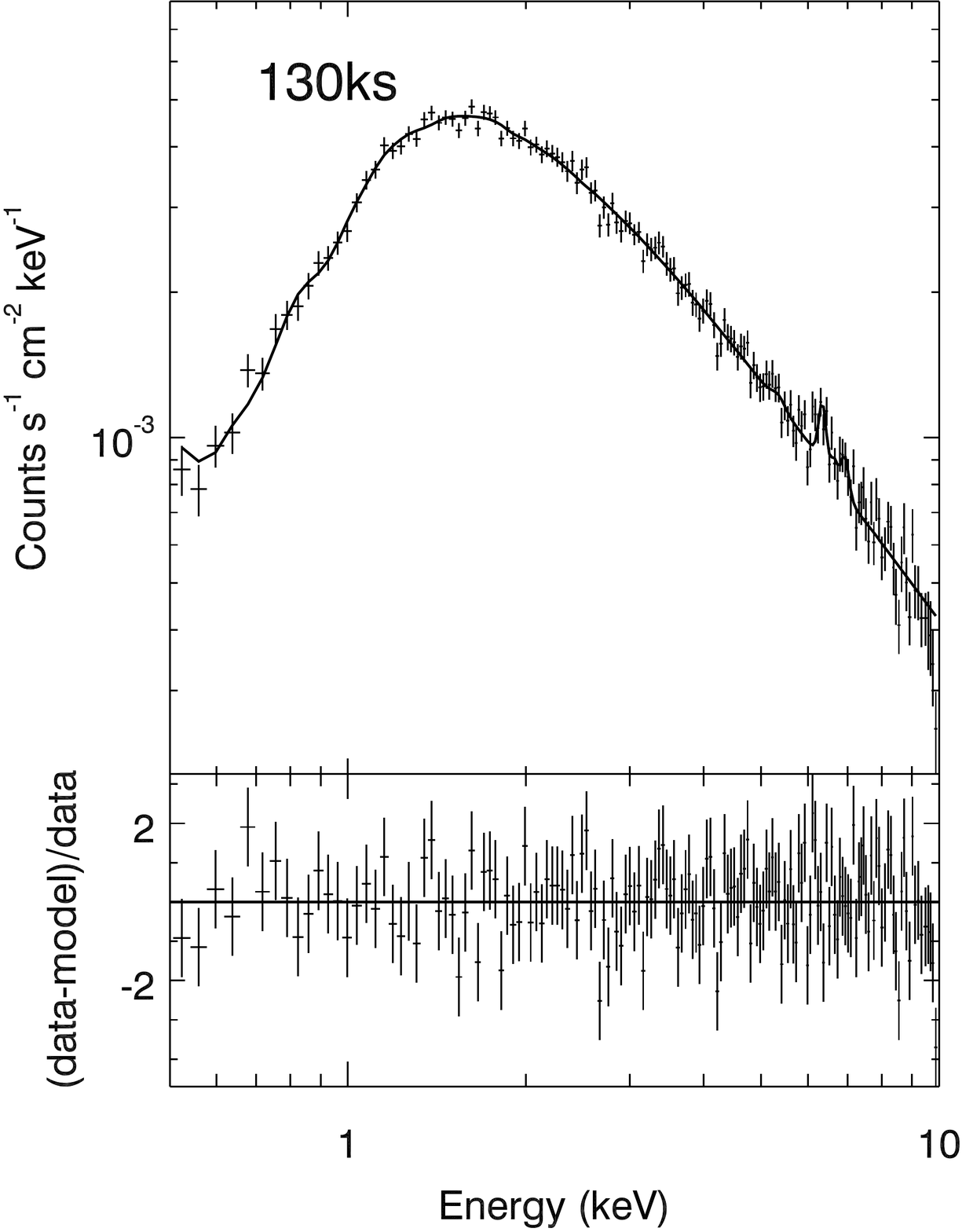}}
  \subfigure{\includegraphics[width=0.19\textwidth]{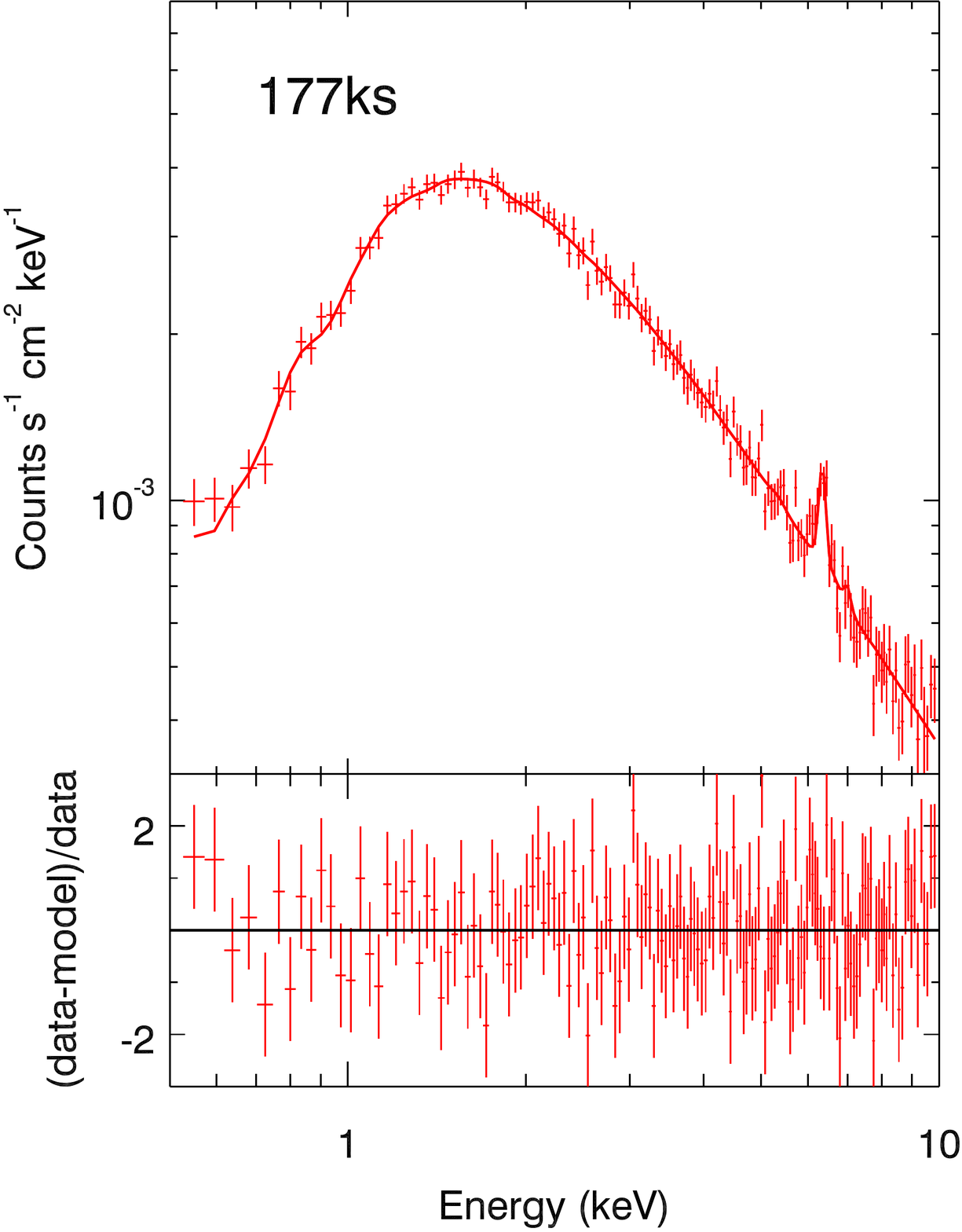}}
  \subfigure{\includegraphics[width=0.19\textwidth]{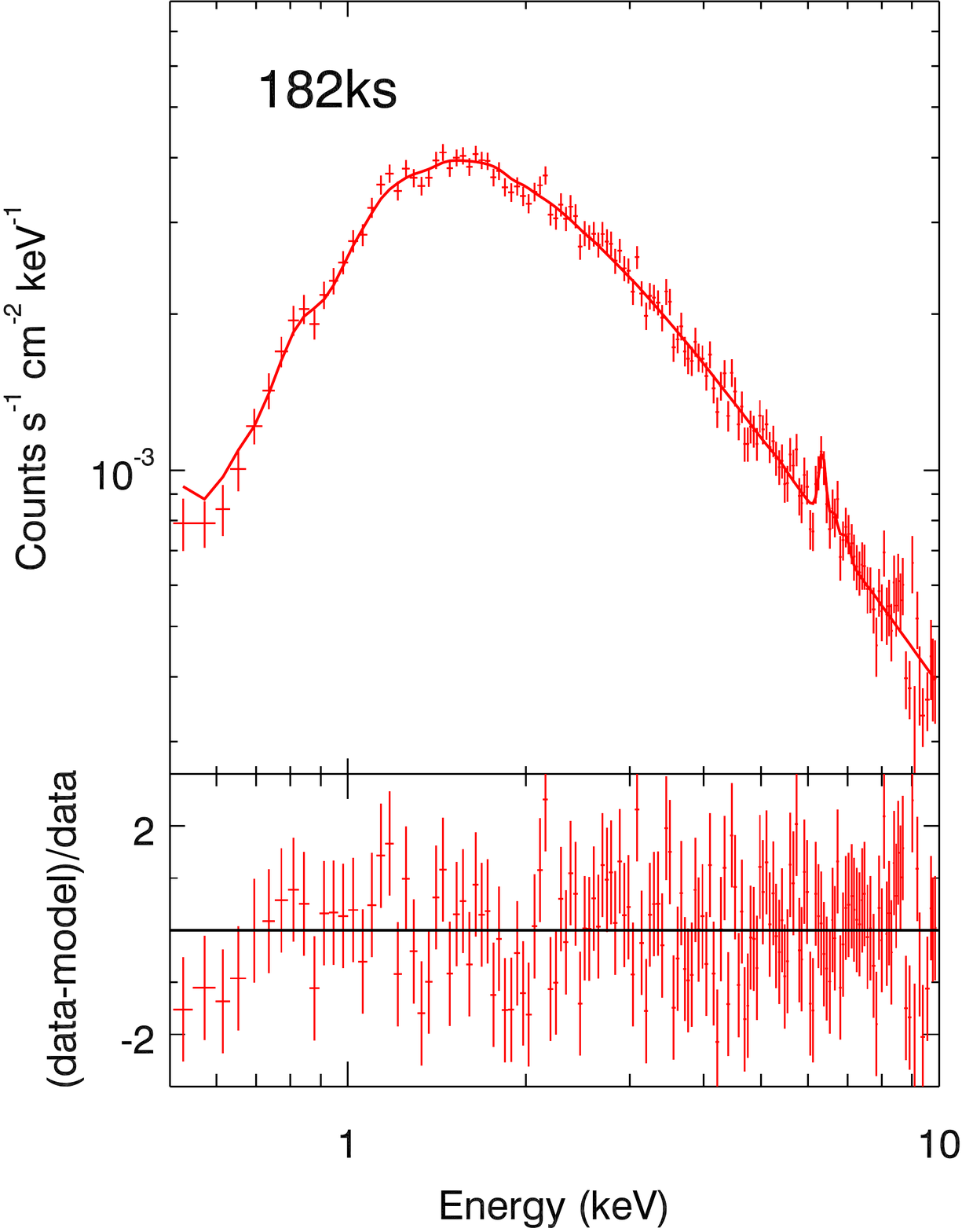}}
  \subfigure{\includegraphics[width=0.19\textwidth]{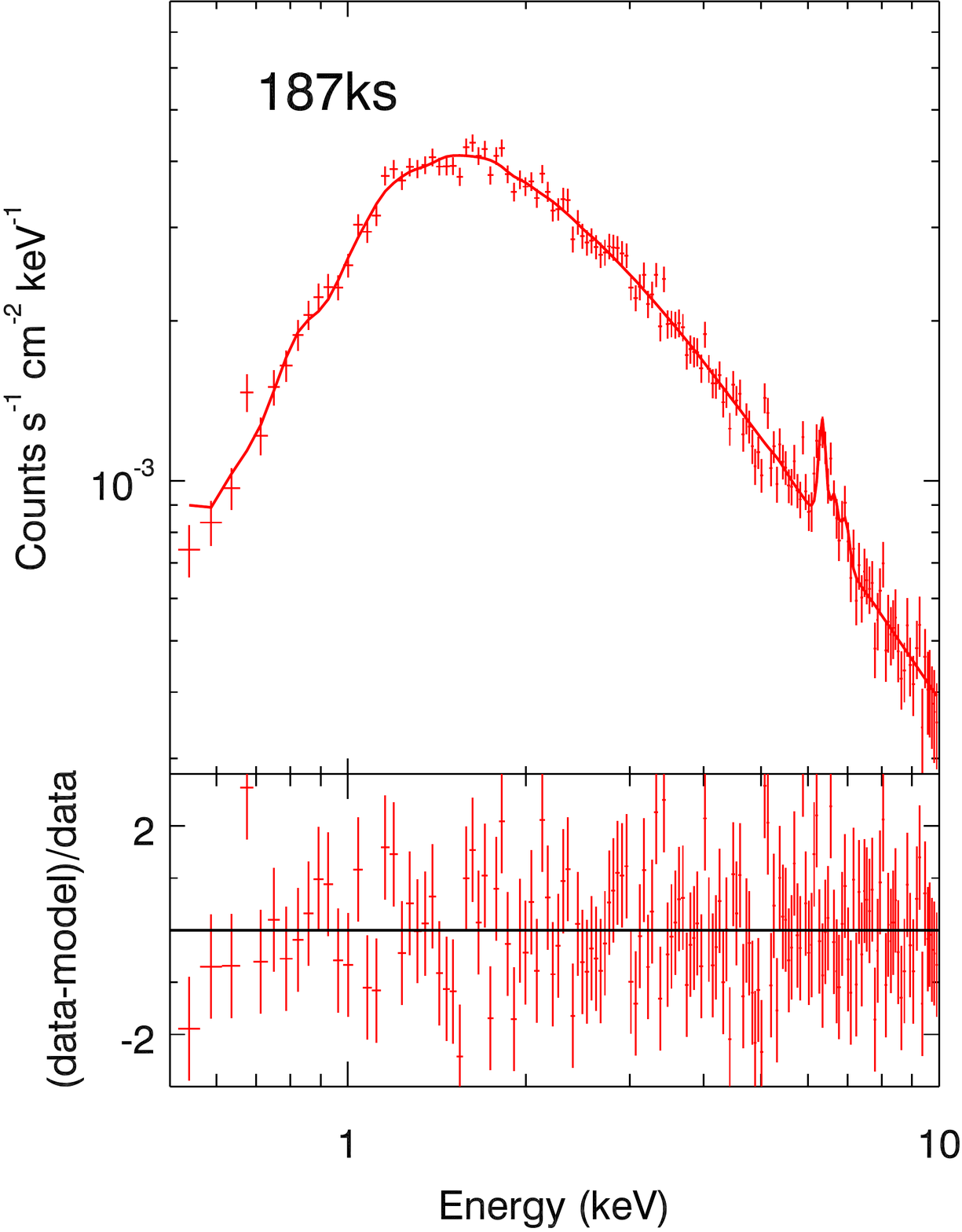}}
  \subfigure{\includegraphics[width=0.19\textwidth]{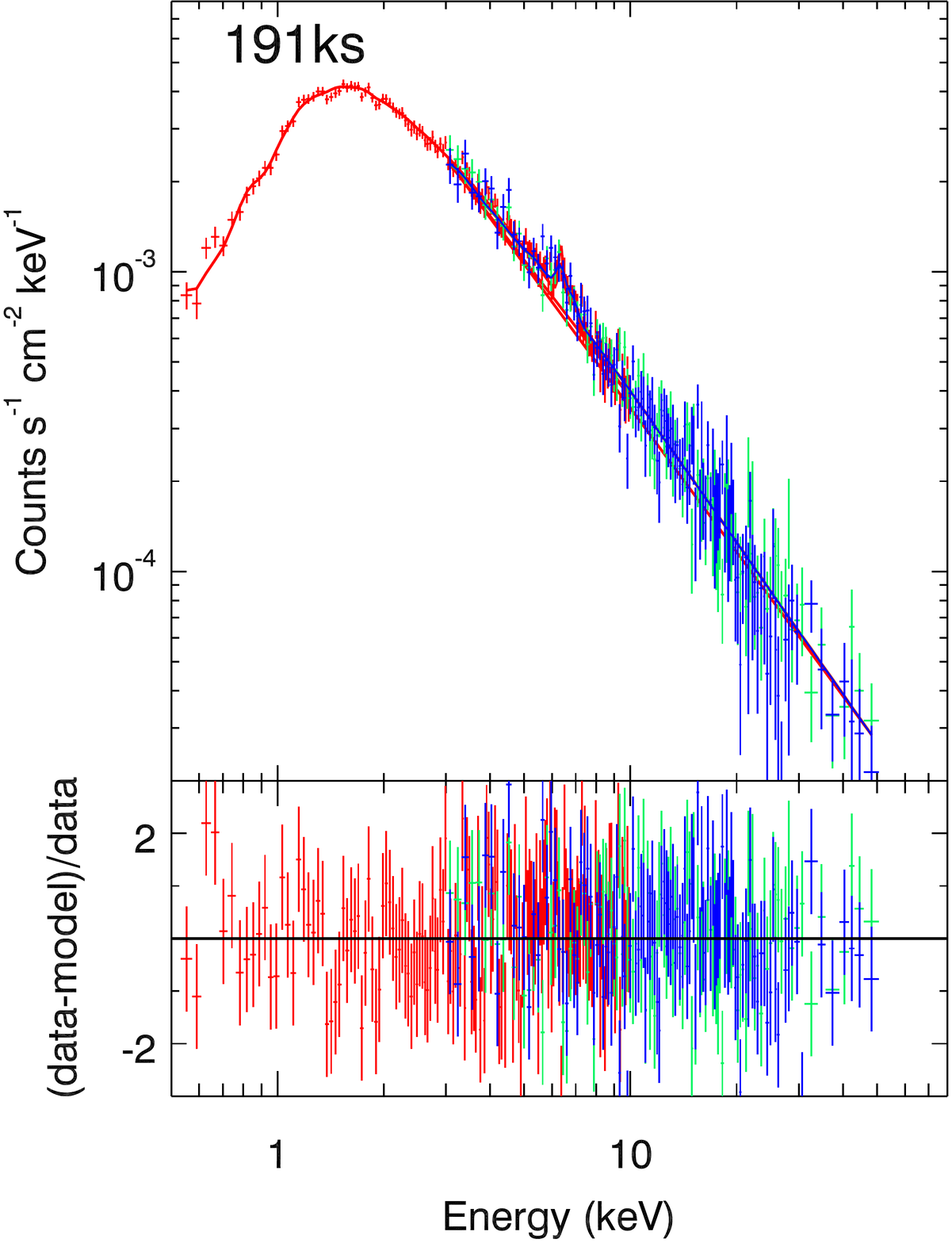}}
  
  \subfigure{\includegraphics[width=0.19\textwidth]{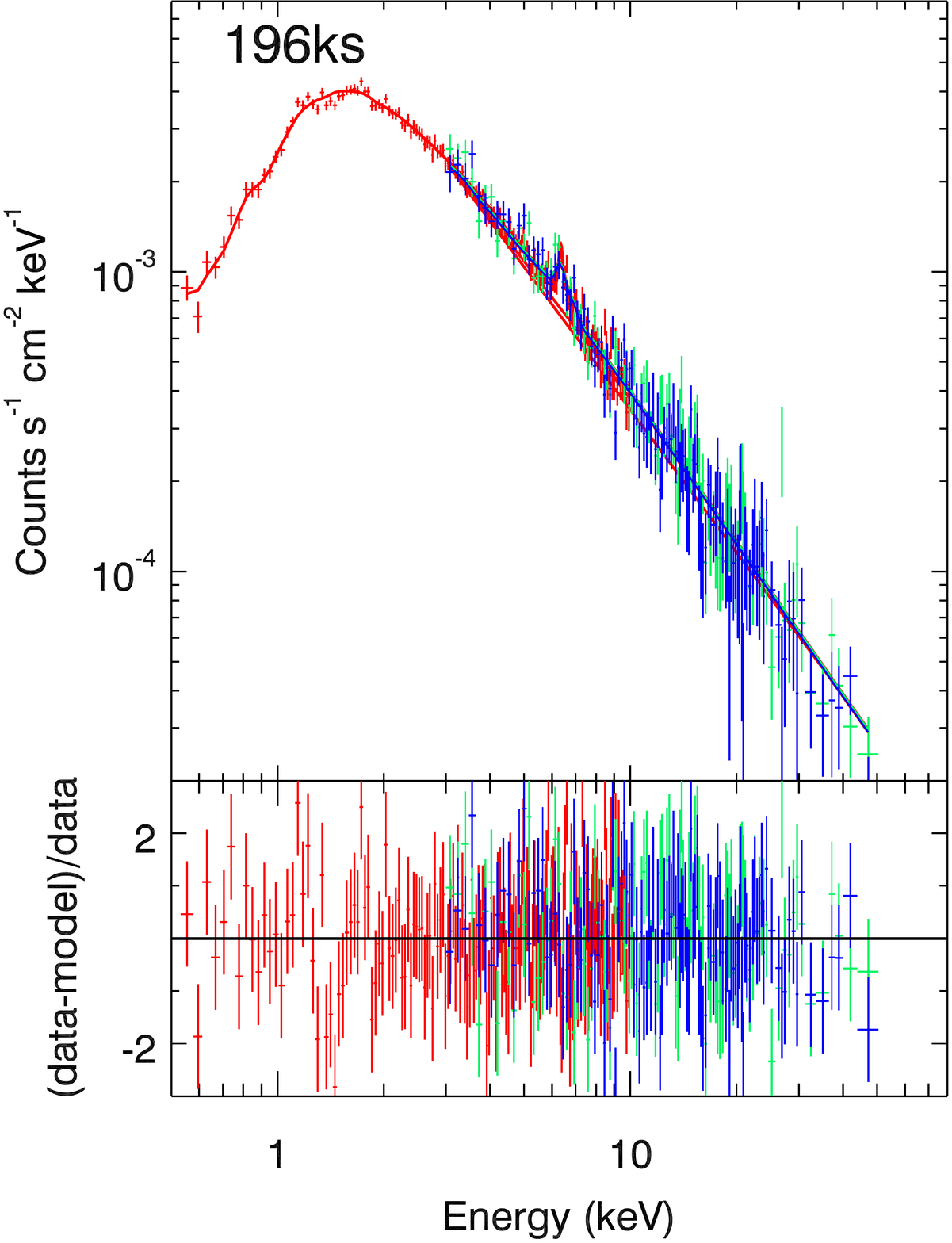}}
  \subfigure{\includegraphics[width=0.19\textwidth]{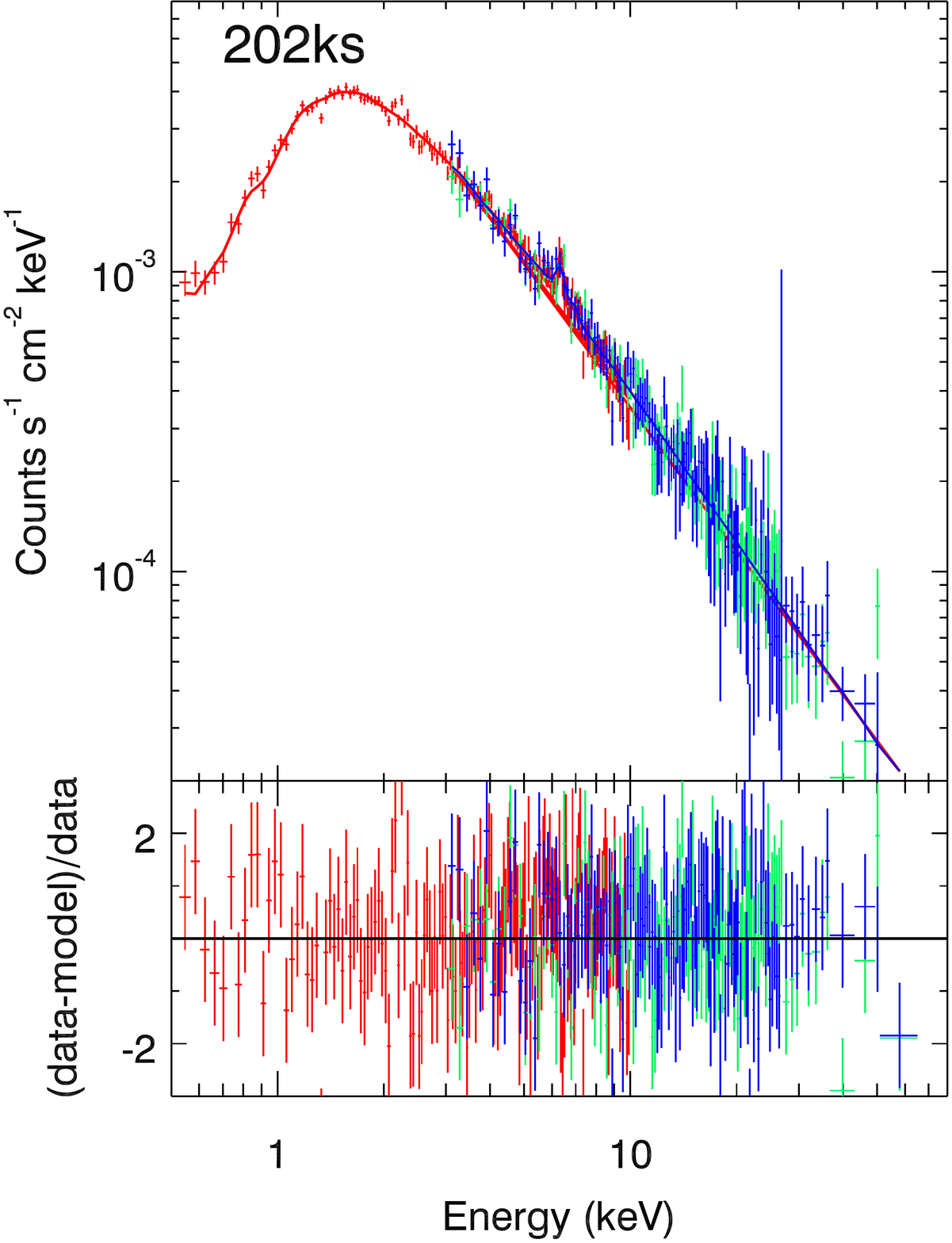}}
  \subfigure{\includegraphics[width=0.19\textwidth]{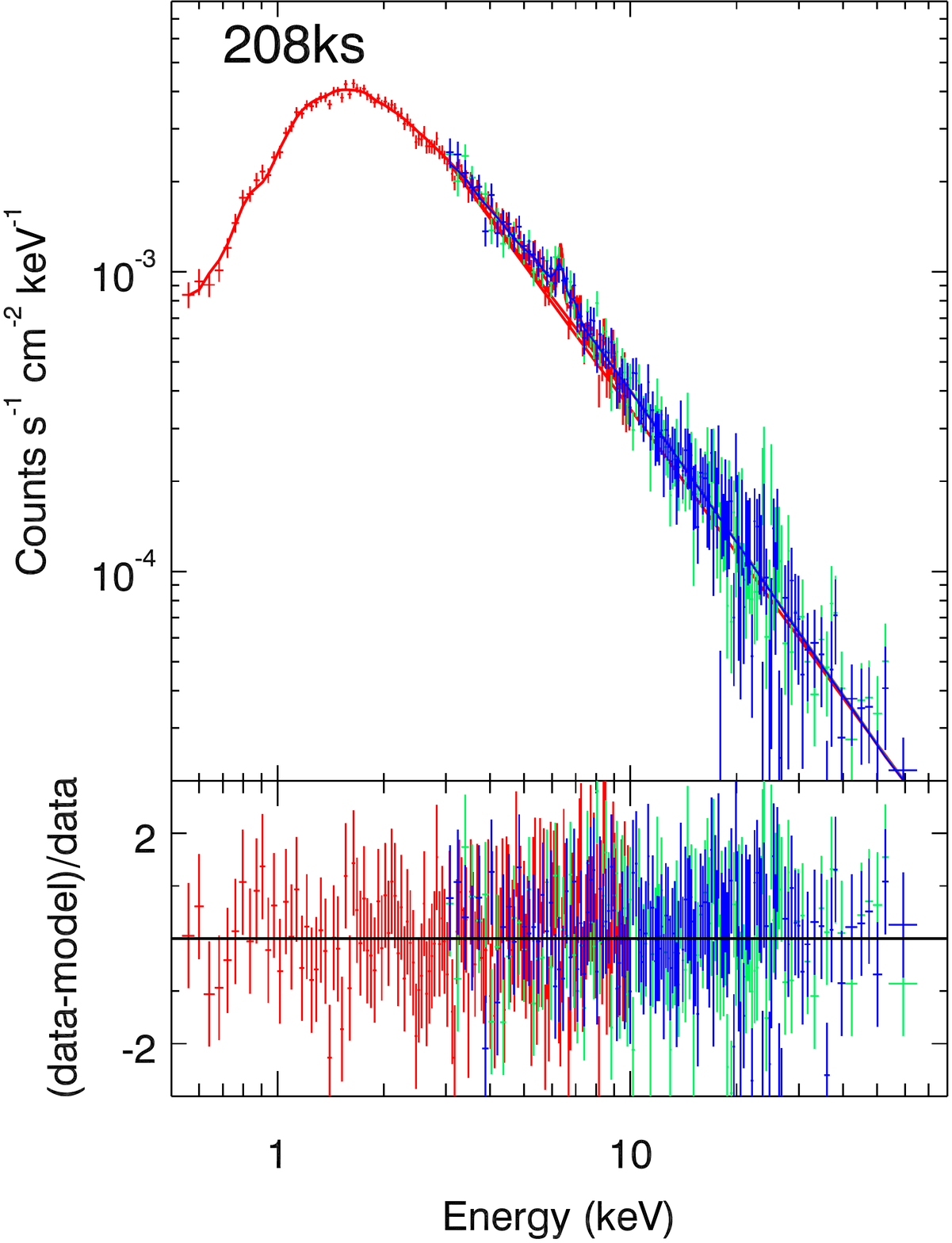}}
  \subfigure{\includegraphics[width=0.19\textwidth]{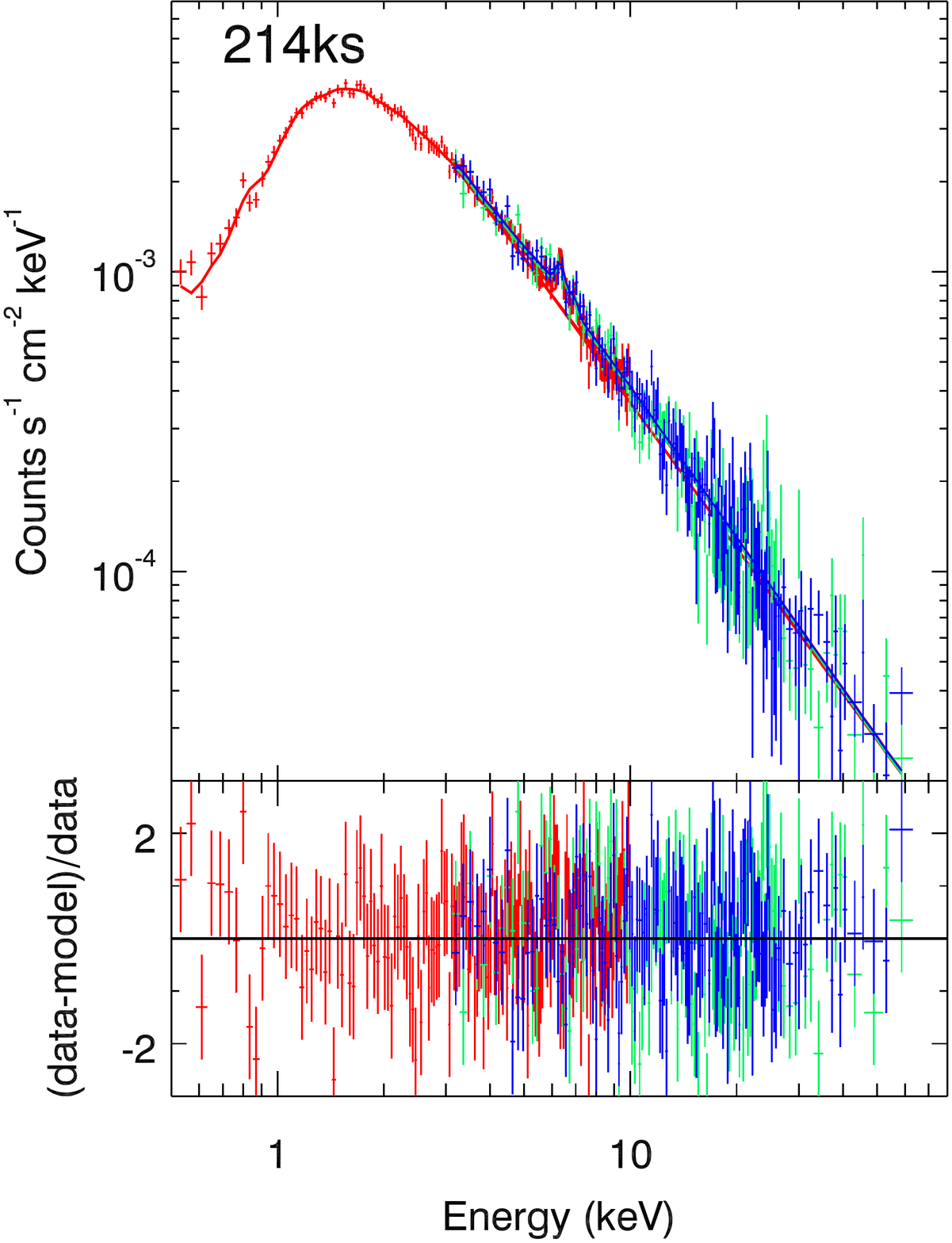}}
  \subfigure{\includegraphics[width=0.19\textwidth]{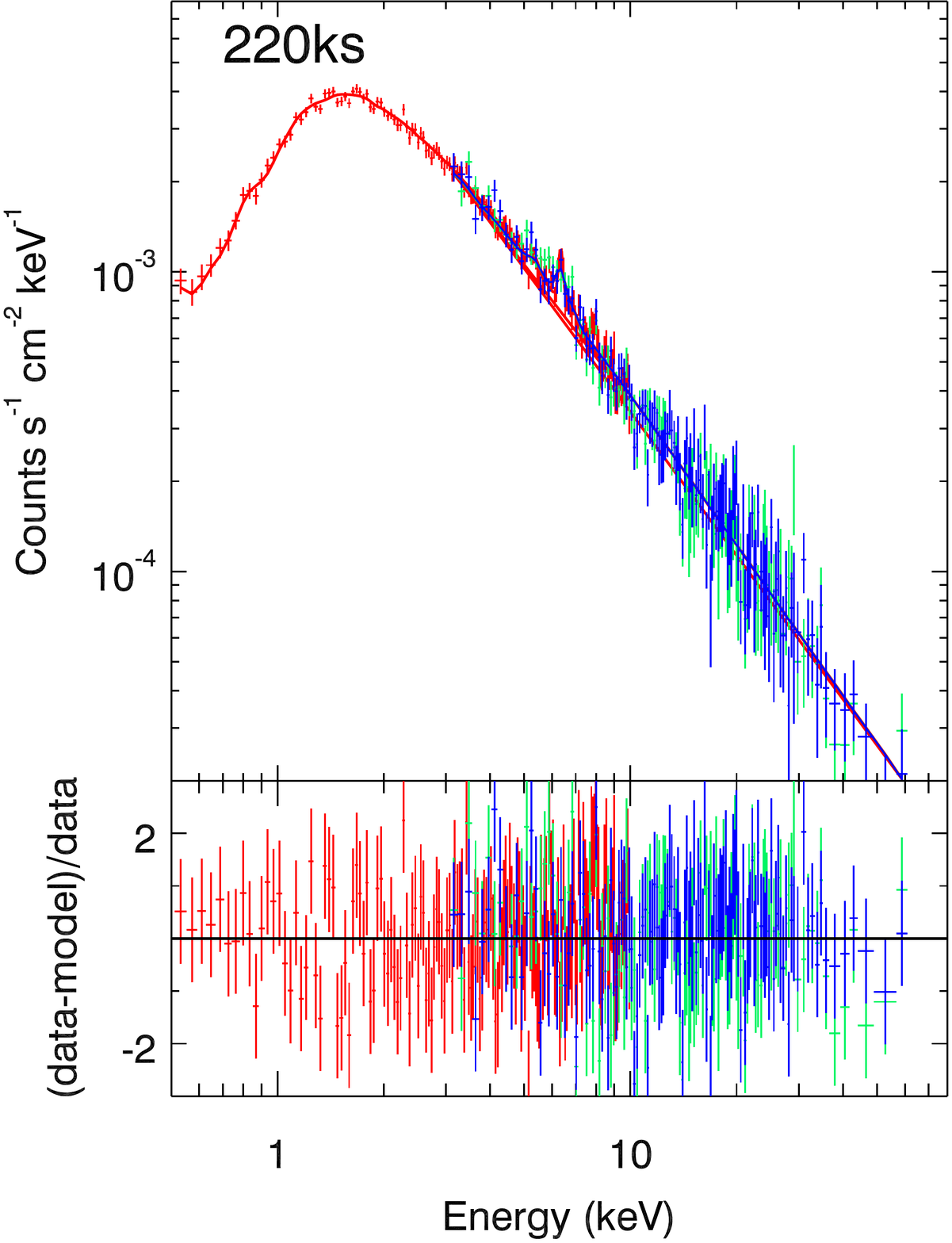}}

  \subfigure{\includegraphics[width=0.19\textwidth]{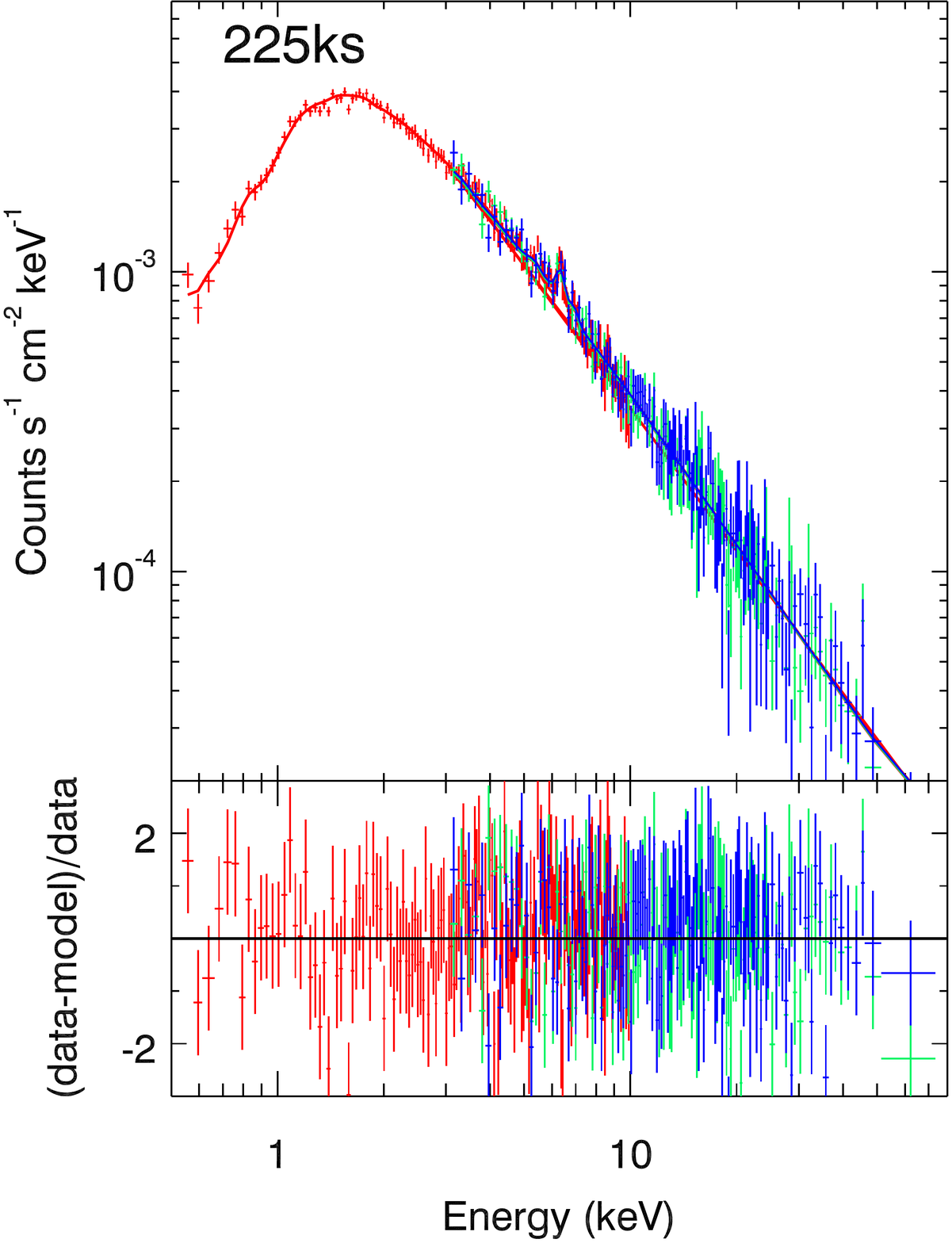}}
  \subfigure{\includegraphics[width=0.19\textwidth]{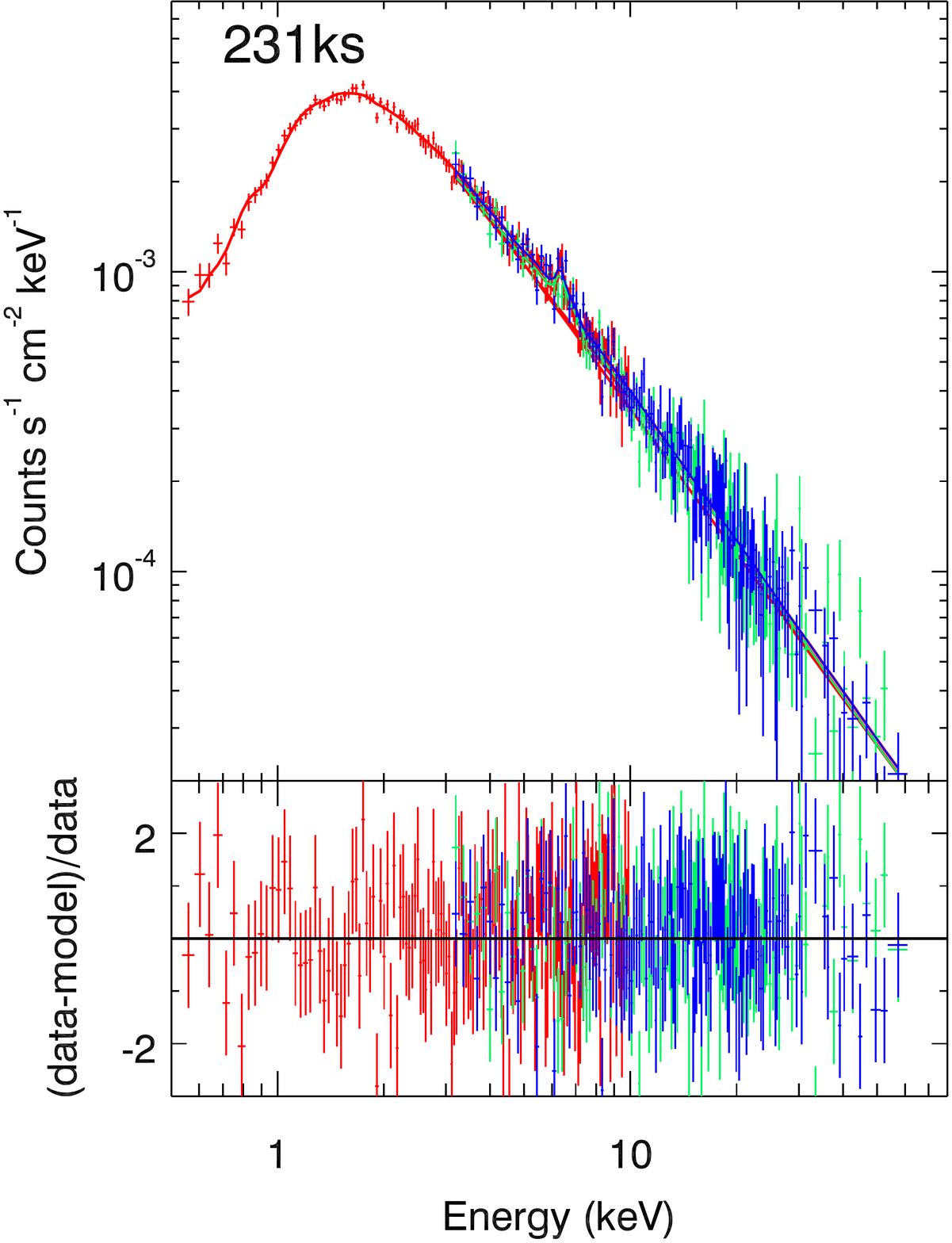}}
  \subfigure{\includegraphics[width=0.19\textwidth]{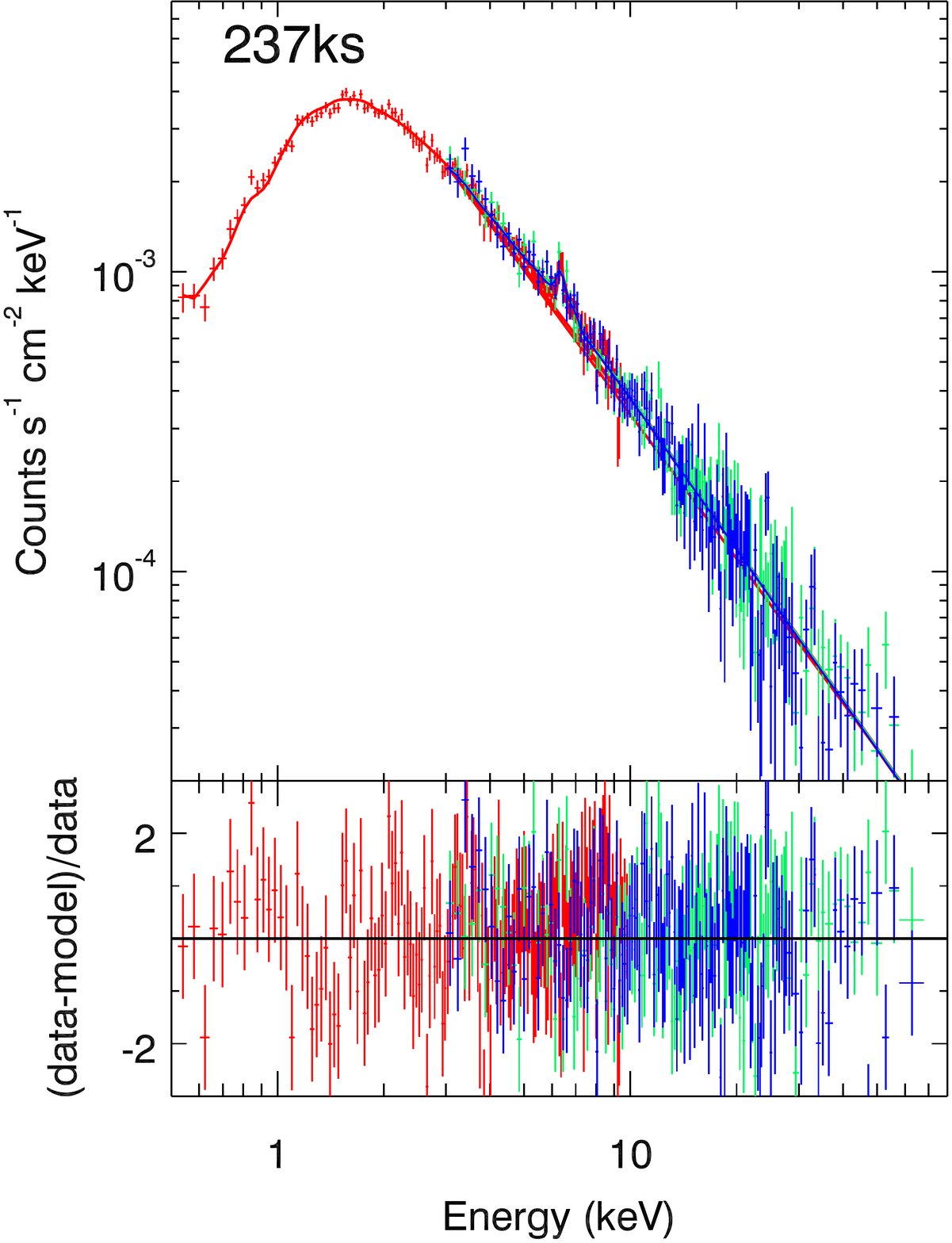}}
  \subfigure{\includegraphics[width=0.19\textwidth]{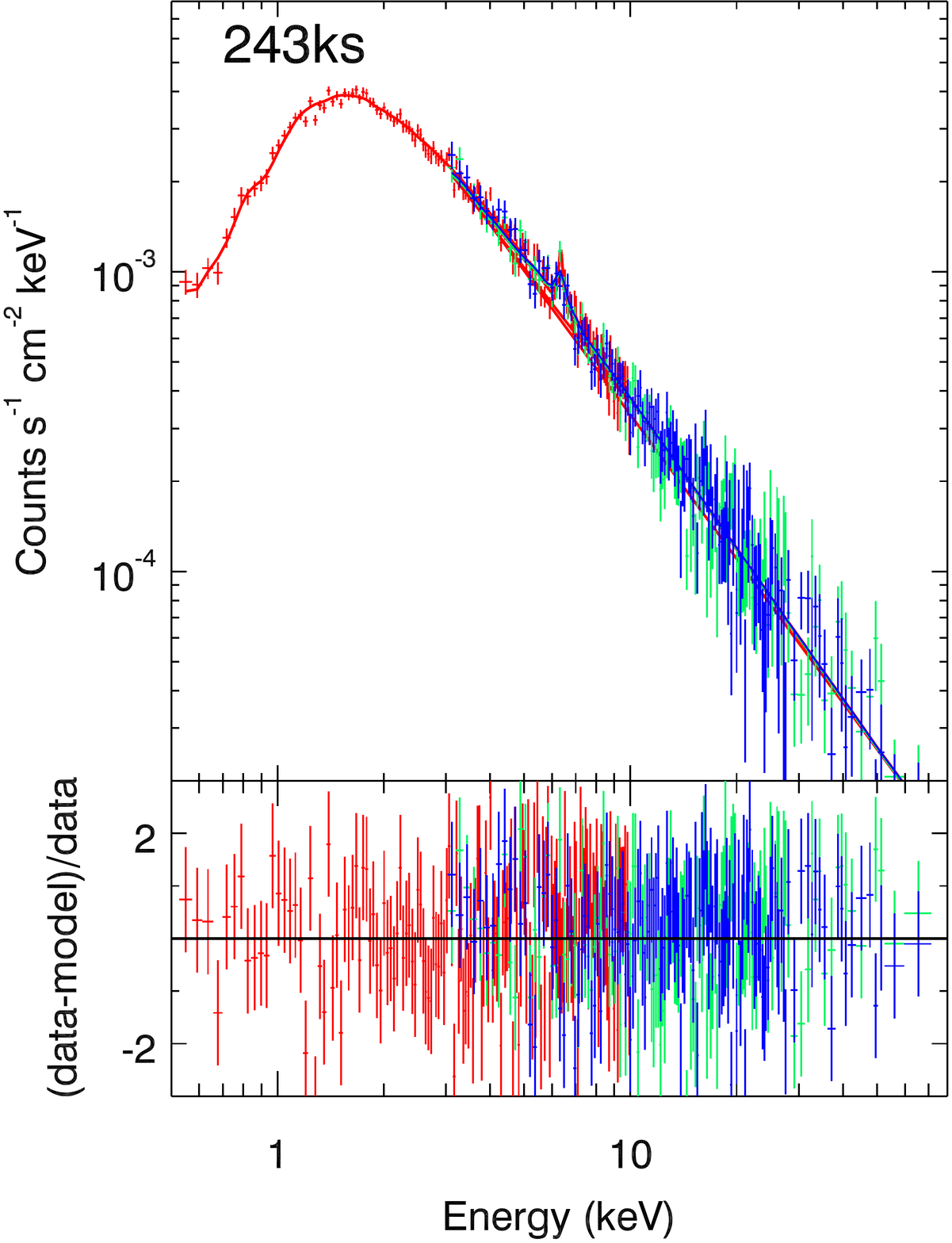}}
  \subfigure{\includegraphics[width=0.19\textwidth]{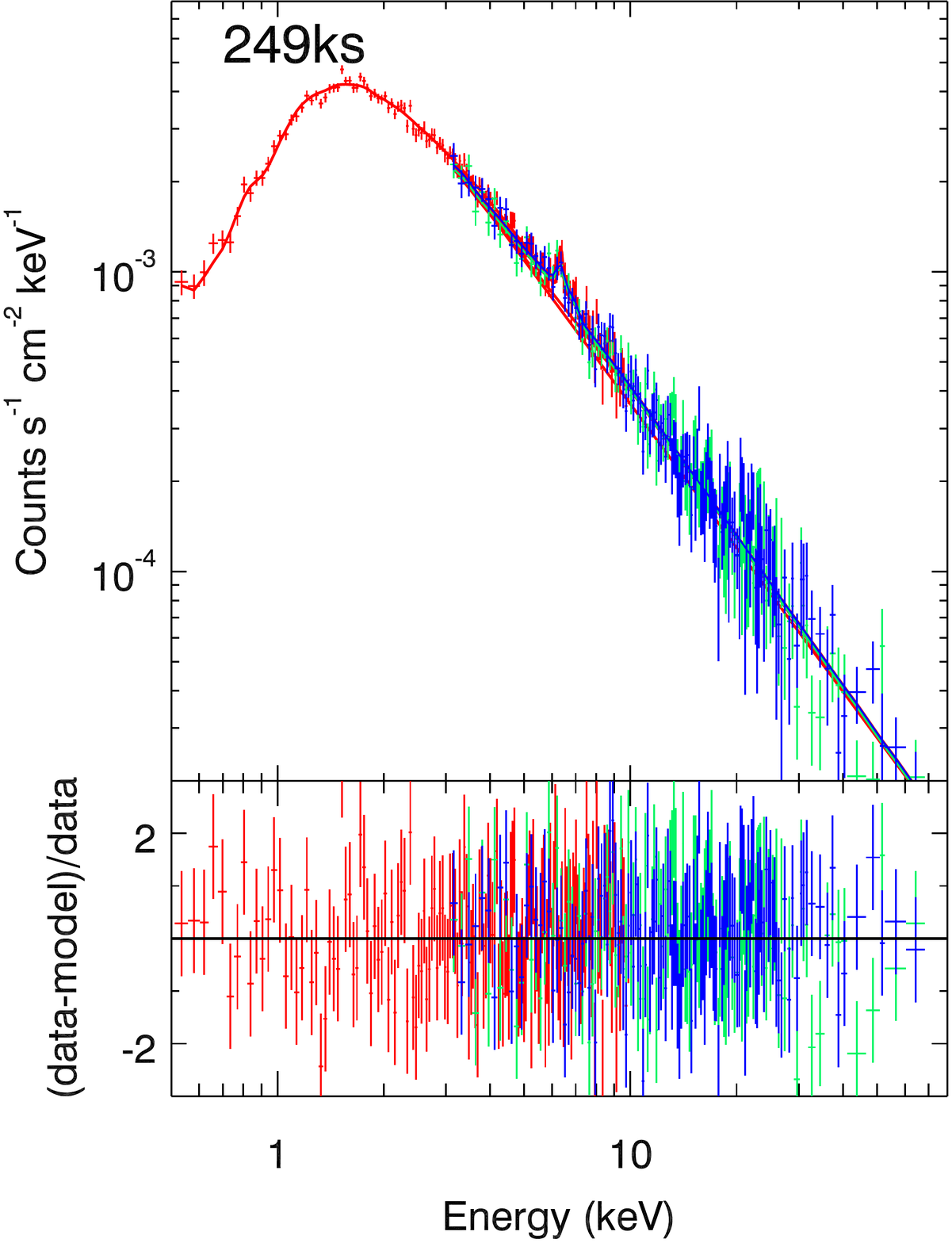}}
  
  \subfigure{\includegraphics[width=0.19\textwidth]{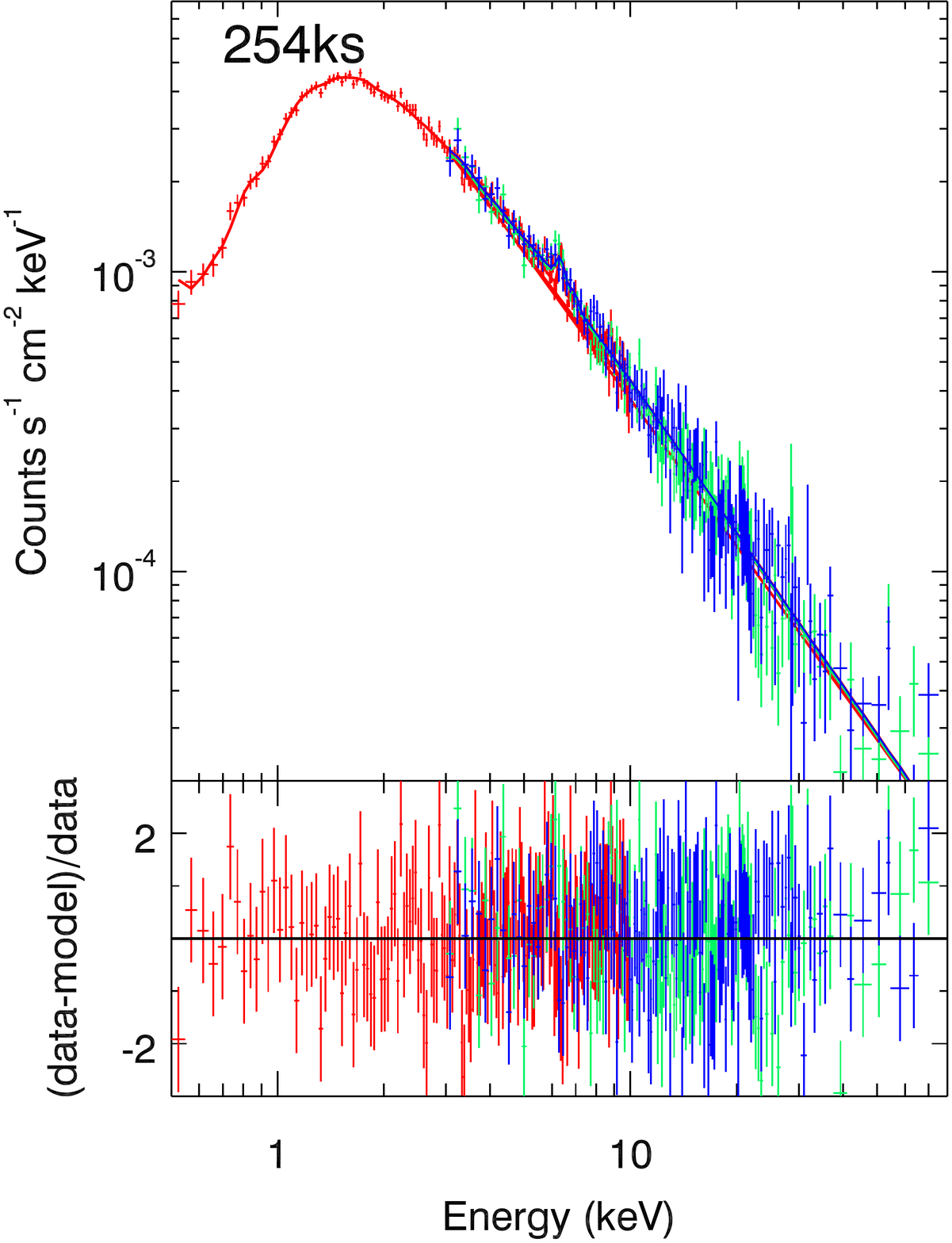}}
  \subfigure{\includegraphics[width=0.19\textwidth]{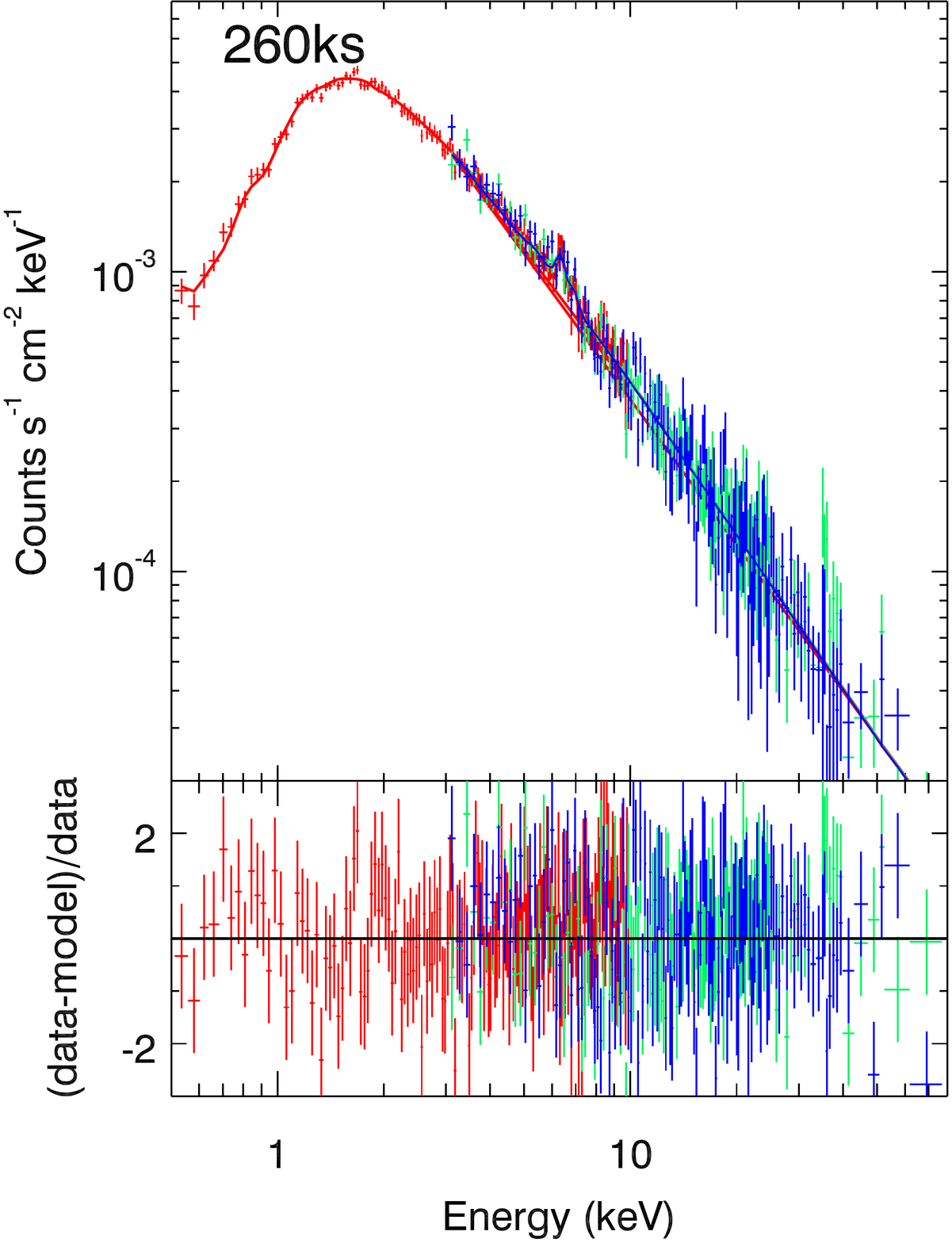}}
  \subfigure{\includegraphics[width=0.19\textwidth]{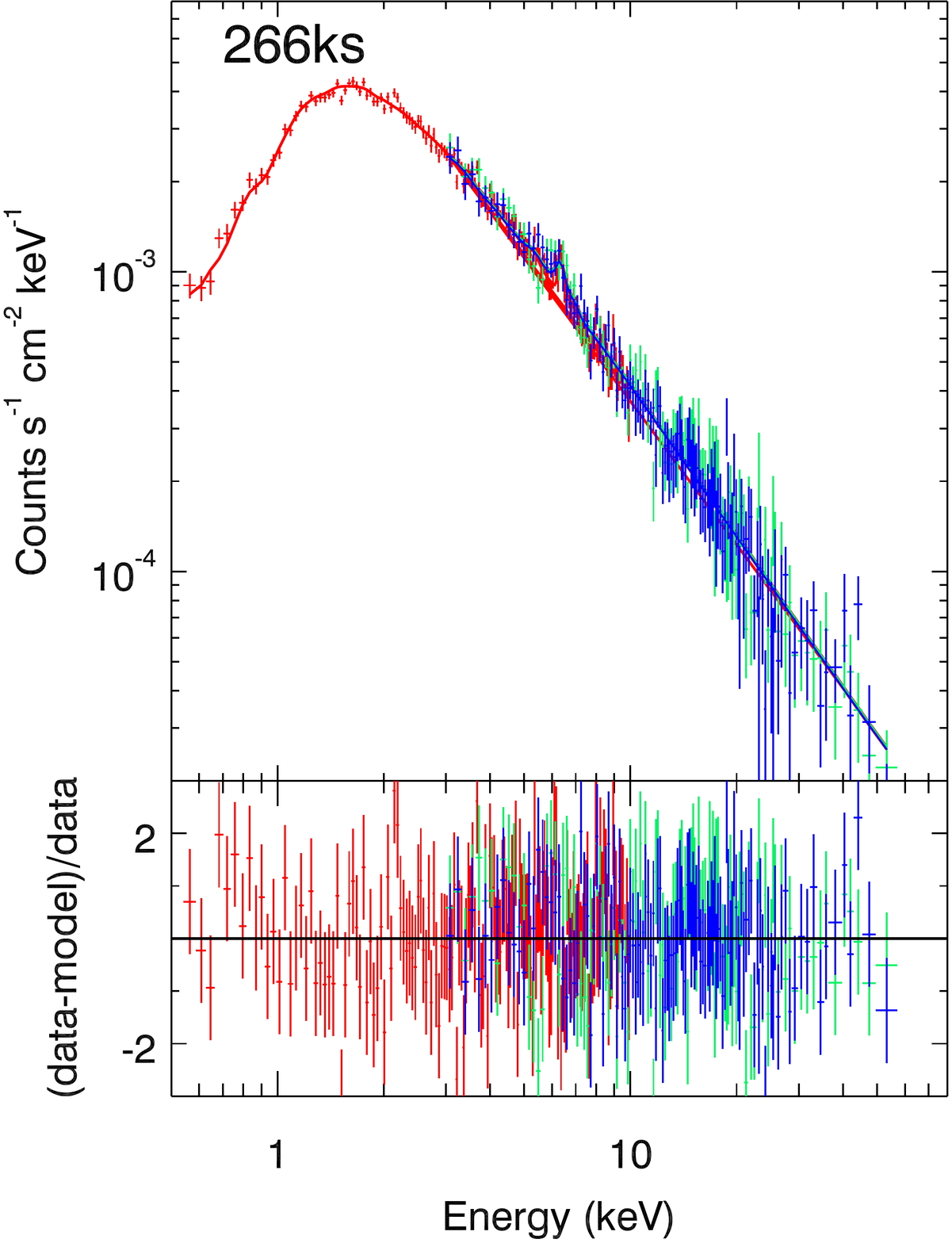}}
  \subfigure{\includegraphics[width=0.19\textwidth]{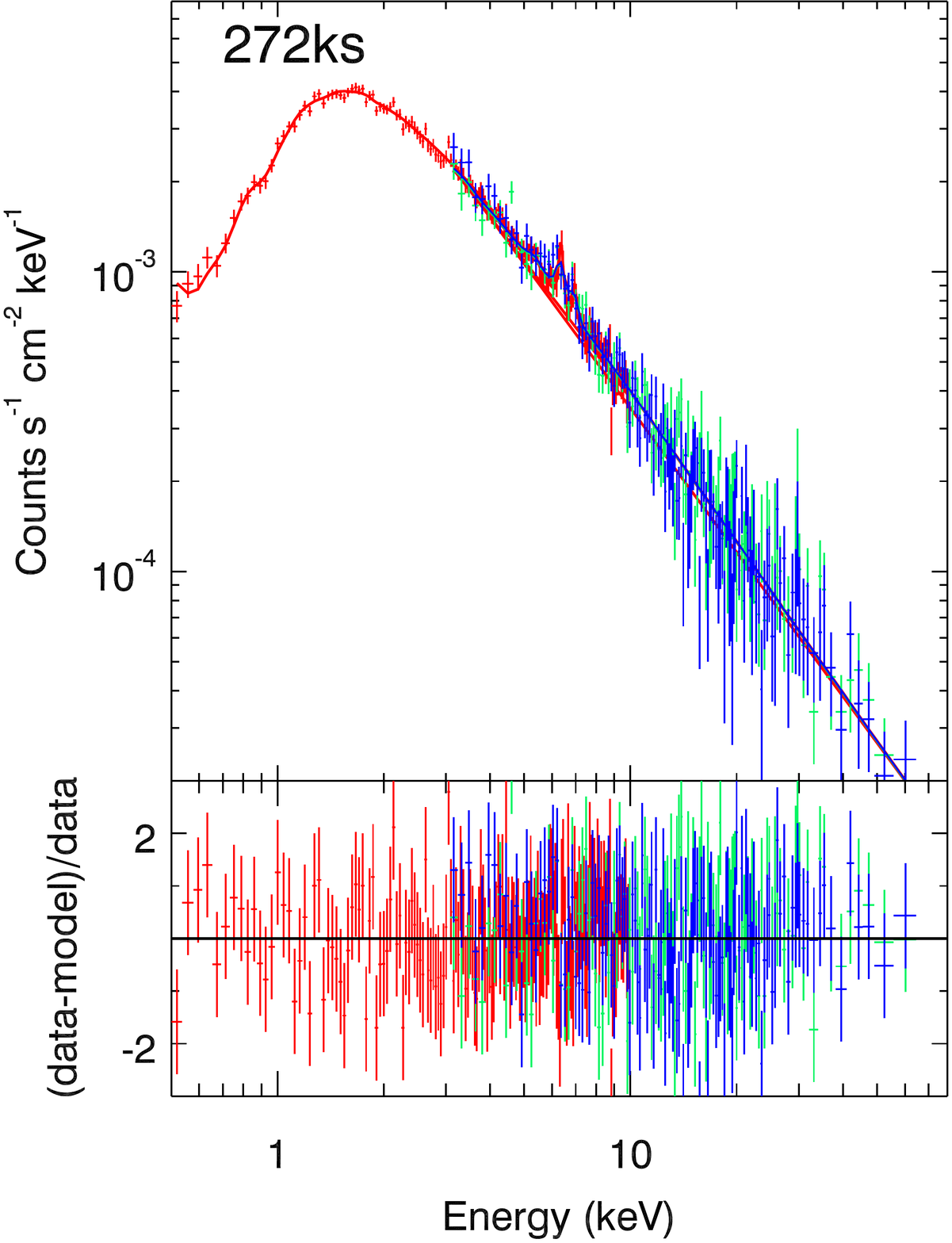}}
  \subfigure{\includegraphics[width=0.19\textwidth]{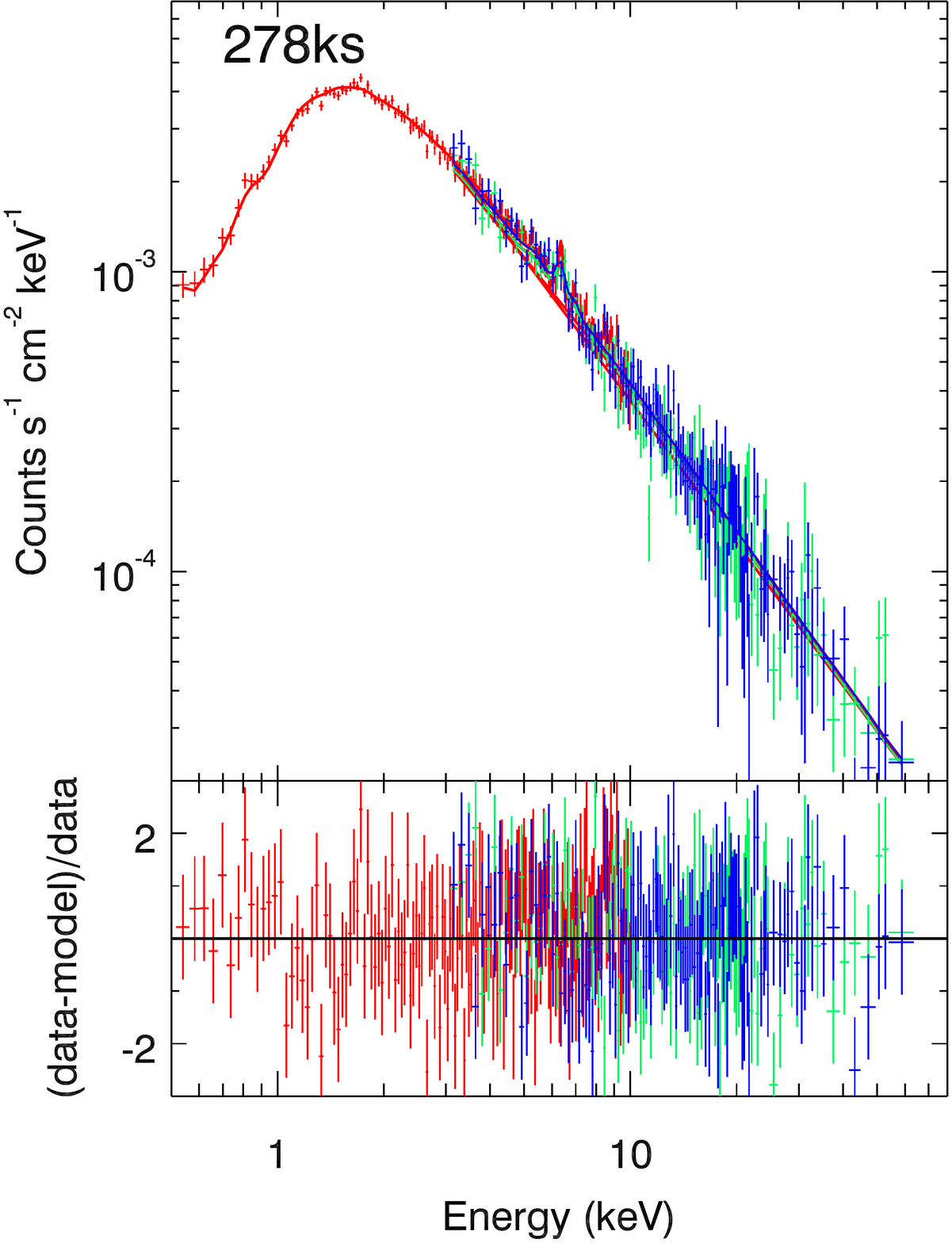}}
  
  \subfigure{\includegraphics[width=0.19\textwidth]{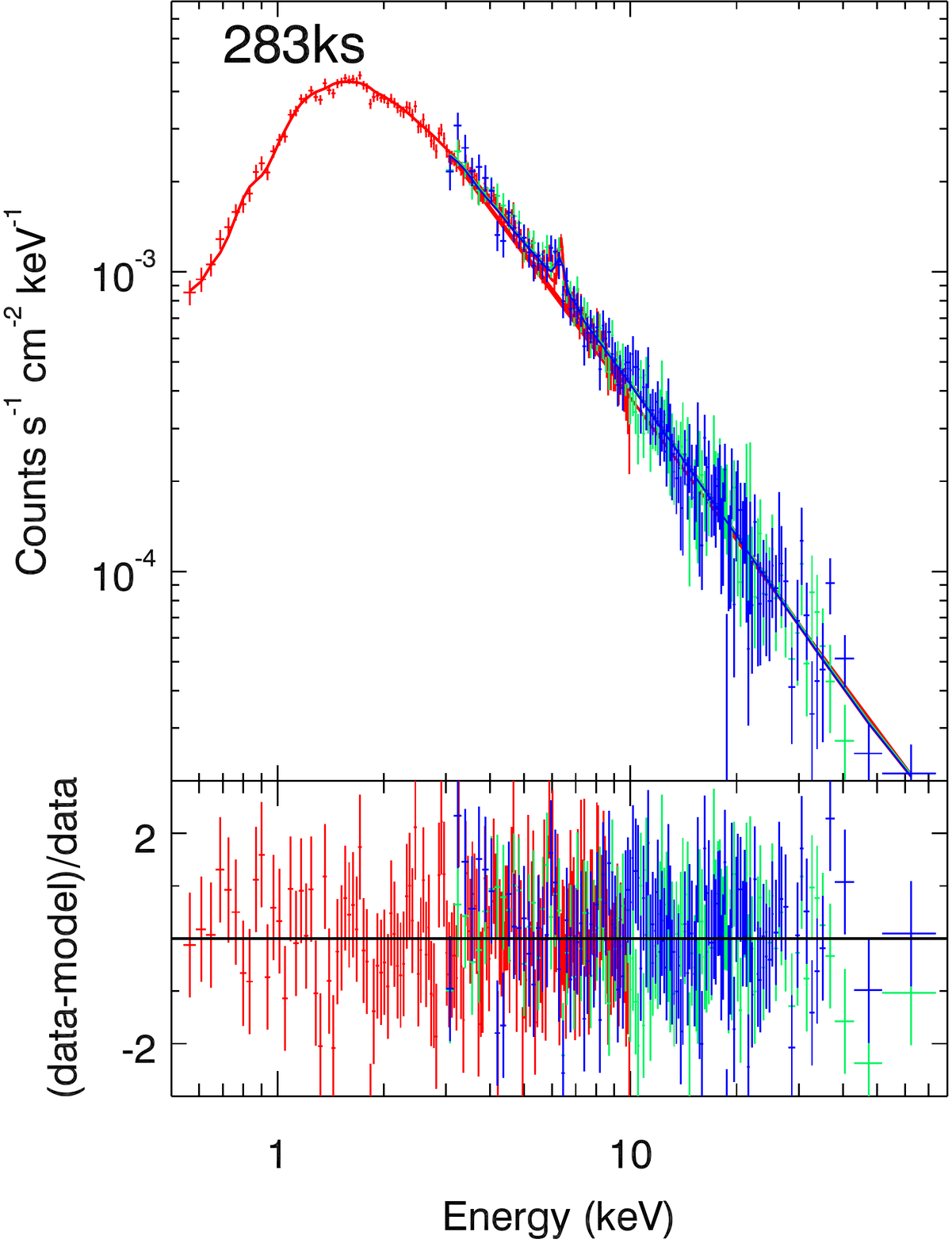}}
  \subfigure{\includegraphics[width=0.19\textwidth]{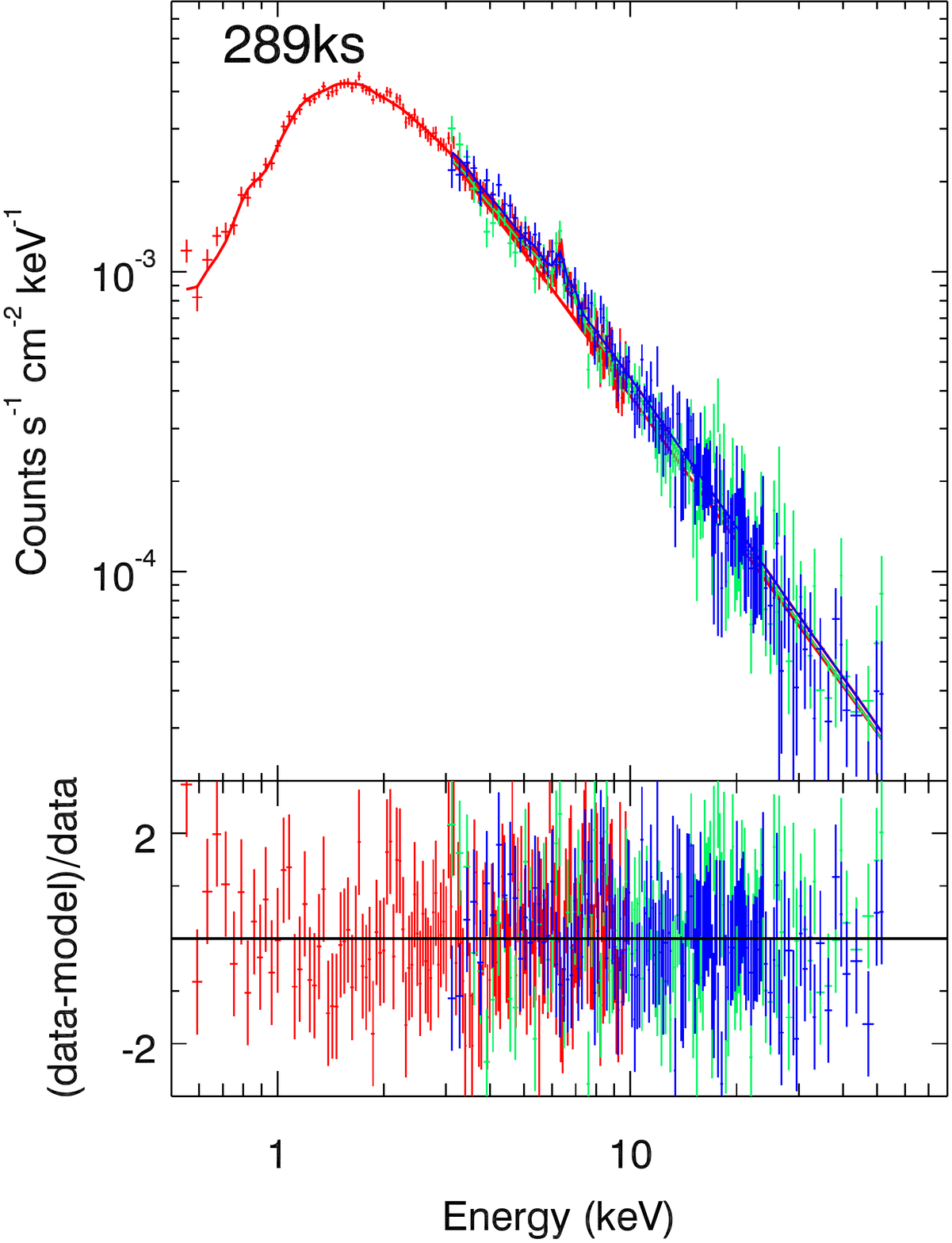}}
  \subfigure{\includegraphics[width=0.19\textwidth]{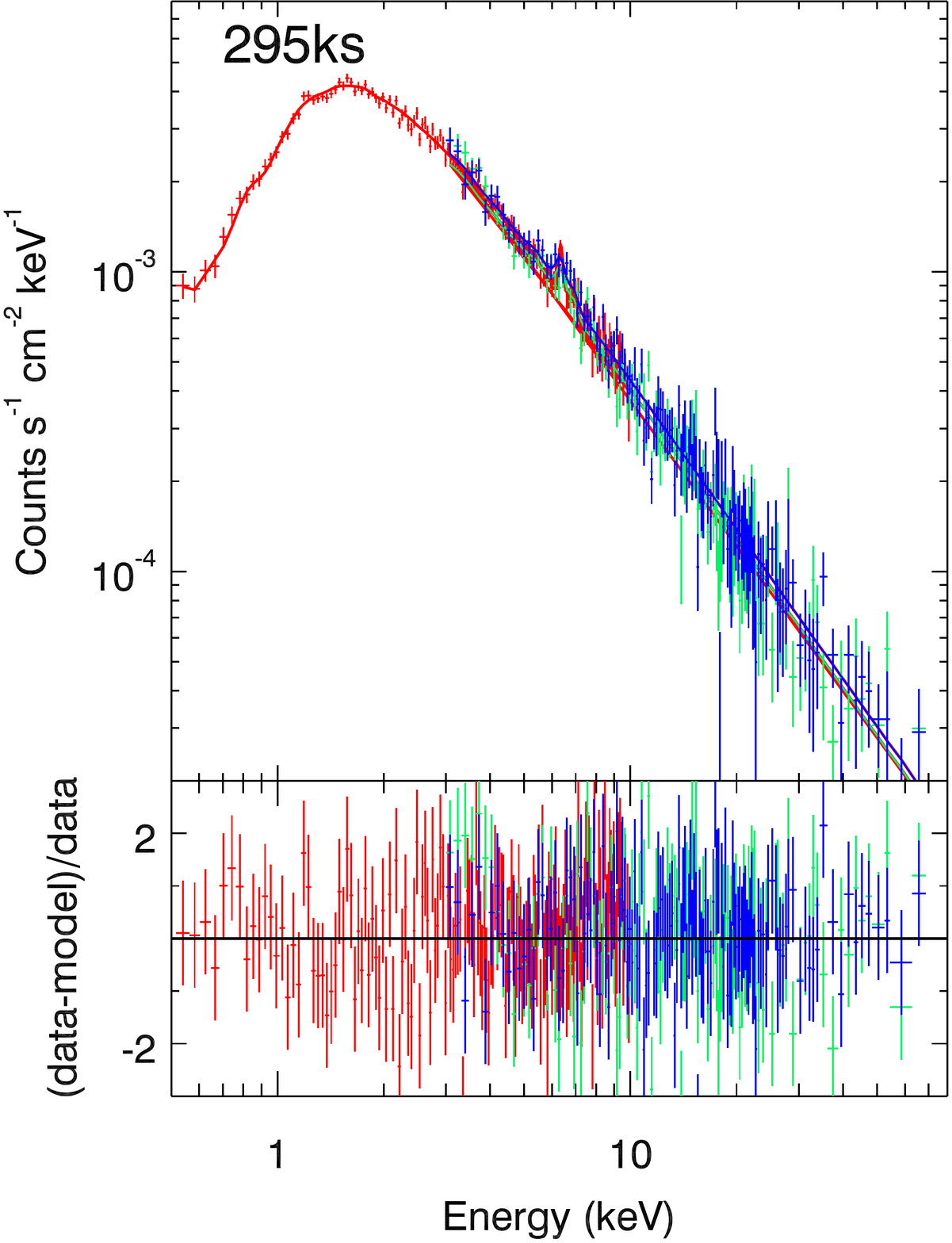}}
  \subfigure{\includegraphics[width=0.19\textwidth]{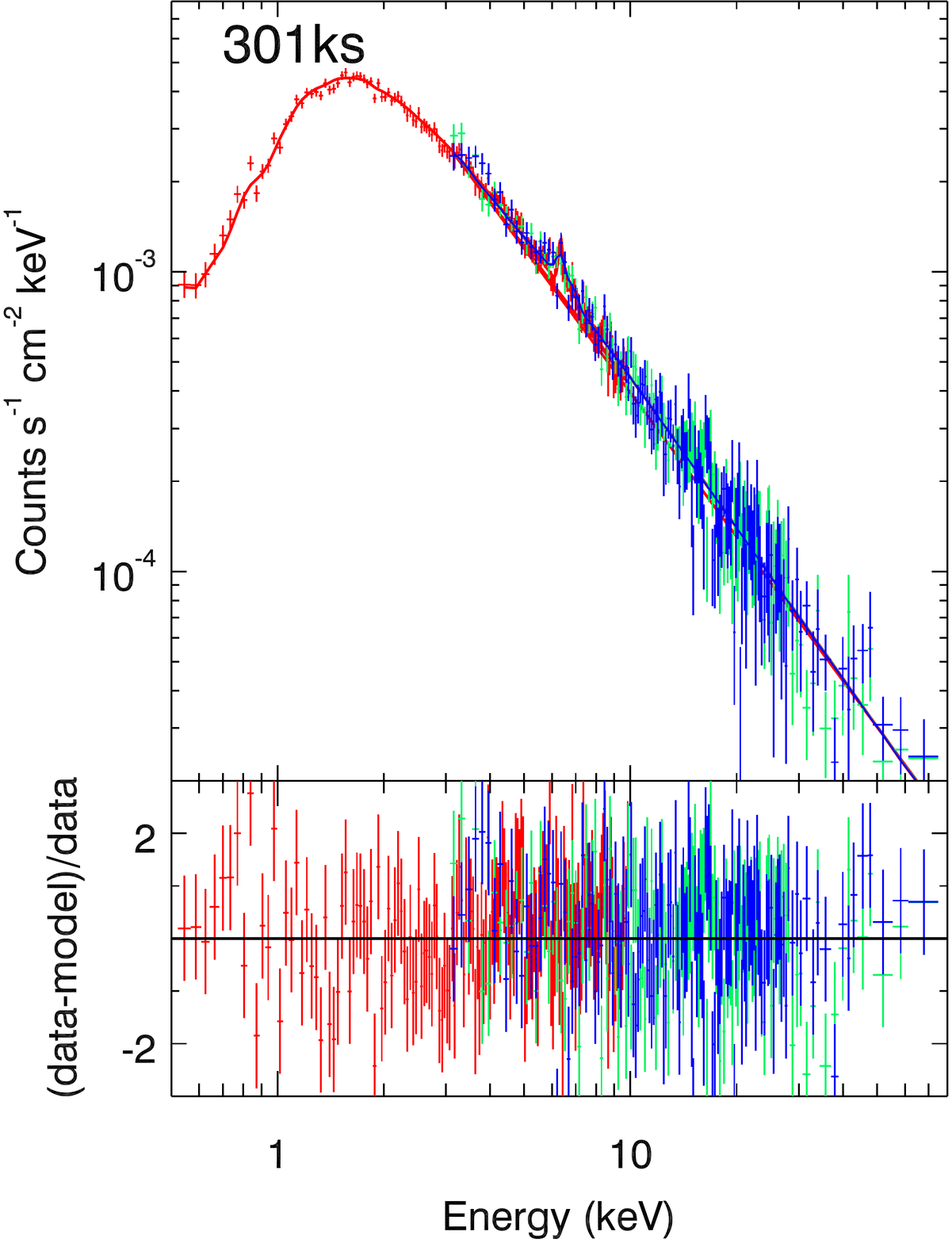}}
  \subfigure{\includegraphics[width=0.19\textwidth]{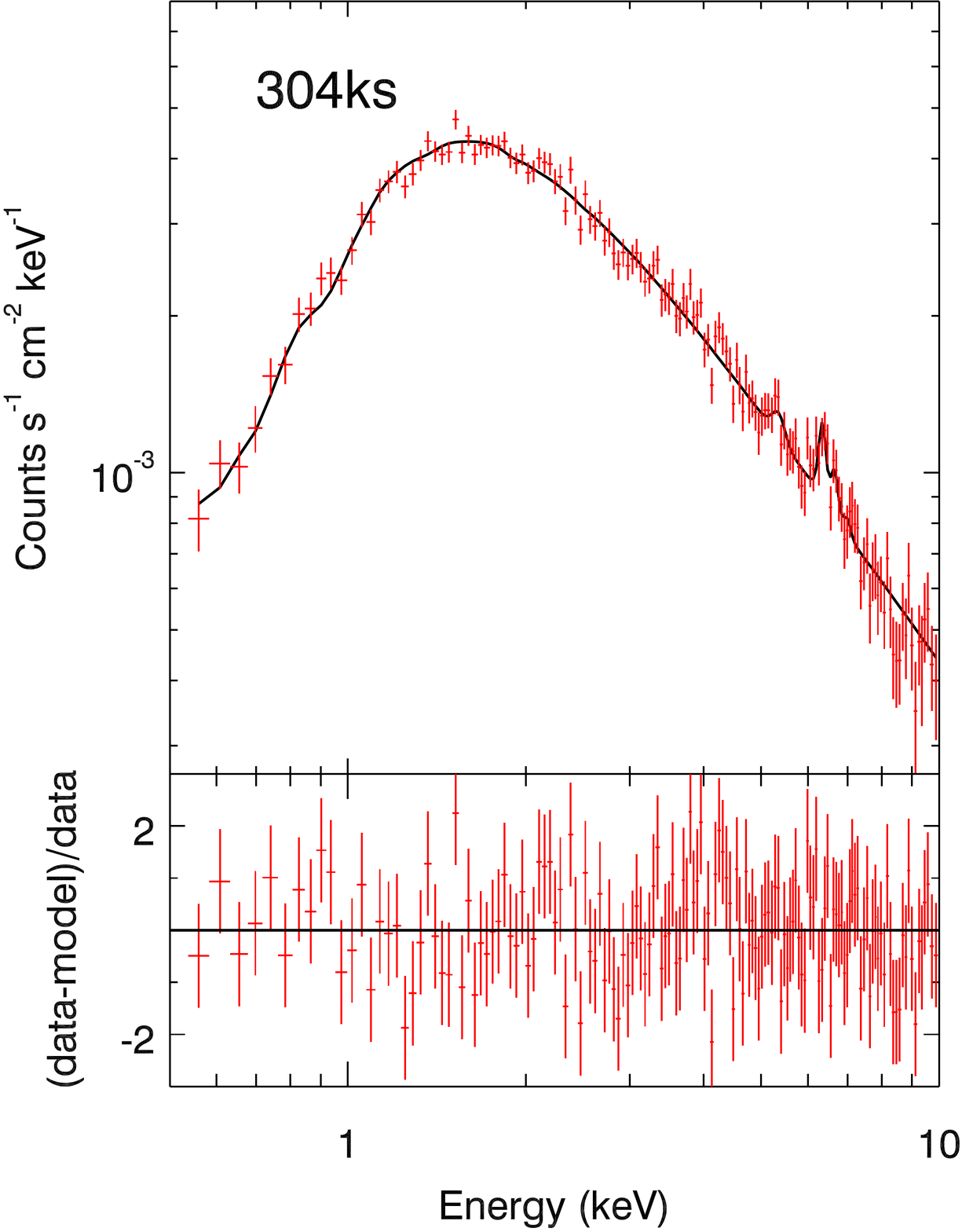}}

\caption{Continues Fig.~\ref{slices1}}
\label{slices2}
\end{figure*}
%
%
%
%
%
\bsp	
\label{lastpage}
\end{document}